\documentclass[useAMS,usenatbib]{mn2e}

\usepackage{graphicx}
\usepackage{natbib}
\usepackage{subfigure}
   \usepackage{amsmath}
   \usepackage{amsfonts}   
   \usepackage{amssymb}    
\usepackage{longtable,lscape}
\def\aap{A\&A}
\def\apjl{ApJL}
\def\apjs{ApJS}

\title[Lyman break and UV-selected galaxies at $z \sim 1$]{Lyman break and UV-selected galaxies at $z \sim 1$ \\ I. Stellar populations from ALHAMBRA survey}
\author[I. Oteo et al.]
{\parbox{\textwidth}{I. Oteo,$^{1,2}$\thanks{E-mail: \texttt{ioteo@iac.es}}
\'A. Bongiovanni$^{1,2}$, 
J. Cepa$^{1,2}$, 
A.M. P\'erez-Garc\'ia$^{1,2,3}$, 
A. Ederoclite$^{4}$, 
M. S\'anchez-Portal$^{3,5}$, 
I. Pintos-Castro$^{1,2,15}$, 
R. P\' erez-Mart\' inez$^{6}$,
J. Polednikova$^{1,2}$, 
J. A. L. Aguerri$^{6}$, 
E. J. Alfaro$^{7}$, 
T. Aparicio-Villegas$^{7,16}$, 
N. Ben\' itez$^{7}$, 
T. Broadhurst$^{8}$, 
J. Cabrera-Ca\~no$^{9}$, 
F. J. Castander$^{10}$, 
M. Cervi\~no$^{7}$, 
D. Cristobal-Hornillos$^{7,4}$, 
A. Fernandez-Soto$^{11,17}$, 
R. M. Gonzalez-Delgado$^{7}$,
C., Husillos$^{7}$, 
L. Infante$^{12}$, 
V. J. Mart\' inez$^{13,14}$, 
I. M\' arquez$^{7}$, 
J. Masegosa$^{7}$, 
I. Matute$^{7}$, 
M. Moles$^{7,4}$, 
A. Molino$^{7}$, 
A. del Olmo$^{7}$, 
J. Perea$^{7}$,
M. Povi\'c$^{7}$, 
F. Prada$^{7}$,
J. M. Quintana$^{7}$, and
K. Viironen$^{4}$}\vspace{0.4cm}\\
$^{1}$Instituto de Astrof{\'i}sica de Canarias (IAC), E-38200 La Laguna, Tenerife, Spain\\
$^{2}$Departamento de Astrof{\'i}sica, Universidad de La Laguna (ULL), E-38205 La Laguna, Tenerife, Spain\\
$^{3}$Asociaci\' on ASPID. Apartado de Correos 412, La Laguna, Tenerife, Spain\\
$^{4}$Centro de Estudios de F\'isica del Cosmos de Arag\' on, Plaza San Juan 1, Planta 2, Teruel, 44001, Spain\\
$^{5}$Herschel Science Centre (ESAC). Villafranca del Castillo, Spain\\
$^{6}$XMM/Newton Science Operations Centre (ESAC). Villafranca del Castillo. Spain\\
$^{7}$Instituto de Astrof\'isica de Andaluc\' ia (CSIC), Glorieta de la Astronom\'ia s/n, EÐ18008 Granada, Spain\\
$^{8}$School of Physics and Astronomy, Tel Aviv University, Israel\\
$^{9}$Facultad de F\'isica. Departamento de F\'isica At\'omica, Molecular y Nuclear. Universidad de Sevilla, Sevilla, Spain\\
$^{10}$Institut de Ciencies de lÕEspai, IEEC-CSIC, Barcelona, Spain\\
$^{11}$Instituto de F\' isica de Cantabria (CSIC-UC), EÐ39005, Santander, Spain\\
$^{12}$Departamento de Astronom\'ia, PontiÞcia Universidad Catolica, Santiago, Chile\\
$^{13}$Departament d' Astronom\' ia i Astrof\'isica, Universitat de Valencia, Valencia, Spain\\
$^{14}$Observatori Astronomic de la Universitat de Valencia, Valencia, Spain\\
$^{15}$Centro de Astrobiolog\'{i}a, INTA-CSIC, P.O. Box - Apdo. de correos 78, Villanueva de la Ca\~nada Madrid 28691, Spain\\
$^{16}$Observat—rio Nacional-MCT, Rua JosŽ Cristino, 77. CEP 20921-400, Rio de Janeiro-RJ, Brazil\\
$^{17}$Unidad Asociada Observatorio Astron—mico (Universitat de Valncia / IFCA-CSIC), Parc Cient'fic UV, 46980 Paterna, Spai
}

\begin{document}

\date{Accepted 1988 December 15. Received 1988 December 14; in original form 1988 October 11}

\pagerange{\pageref{firstpage}--\pageref{lastpage}} \pubyear{2002}

\maketitle

\label{firstpage}

\begin{abstract}
We take advantage of the exceptional photometric coverage provided by the combination of GALEX data in the UV and the ALHAMBRA survey in the optical and near-IR to analyze the physical properties of a sample of 1225 GALEX-selected Lyman break galaxies (LBGs) at $0.8 \lesssim z \lesssim 1.2$ located in the COSMOS field. This is the largest sample of LBGs studied at that redshift range so far. According to a spectral energy distribution (SED) fitting with synthetic stellar population templates, we find that LBGs at $z \sim 1$ are mostly young galaxies with a median age of 341 Myr and have intermediate dust attenuation, $\langle E_s (B-V) \rangle \sim 0.20$. Due to their selection criterion, LBGs at $z \sim 1$ are UV-bright galaxies and have high dust-corrected total SFR, with a median value of 16.9 $M_\odot {\rm yr}^{-1}$. Their median stellar mass is $\log{\left( M_*/M_\odot \right)} = 9.74$. We obtain that the dust-corrected total SFR of LBGs increases with stellar mass and the specific SFR is lower for more massive galaxies (downsizing scenario). Only 2\% of the galaxies selected through the Lyman break criterion have an AGN nature. LBGs at $z \sim 1$ are mostly located over the blue cloud of the color-magnitude diagram of galaxies at their redshift, with only the oldest and/or the dustiest deviating towards the green valley and red sequence. Morphologically, 69\% of LBGs are disk-like galaxies, with the fraction of interacting, compact, or irregular systems being much lower, below 12\%. LBGs have a median effective radius of 2.5 kpc and bigger galaxies have higher total SFR and stellar mass. Comparing to their high-redshift analogues, we find evidence that LBGs at lower redshifts are bigger, redder in the UV continuum, and have a major presence of older stellar populations in their SEDs. However, we do not find significant difference in the distributions of stellar mass or dust attenuation.
\end{abstract}

                


\section{Introduction}\label{intro}

Much effort has been devoted over the last decades to searching for high-redshift star-forming (SF) galaxies. Different selection criteria select distinct kinds of galaxies. Among those criteria, the most successful and used are the Lyman-alpha and the Lyman break techniques, which pick up the so-called Lyman-alpha emitters (LAEs) and Lyman break galaxies (LBGs), respectively. The Lyman-alpha technique is based on looking for a Lyman-alpha emission in the redshifted optical spectrum of galaxies by employing a combination of narrow and broad-band filters. Specifically, the narrow-band filter is used to isolate the Ly$\alpha$ line and the broad band one(s) to constrain its nearby continuum \citep{Cowiehu1998,Gronwall2007,Gawiser2006,Ouchi2008,Ouchi2010,Bongiovanni2010,Shioya2009}. The choice of the central wavelength of the narrow-band filter determines the redshift of the selected LAEs, which have been searched, found, and analyzed from $z \sim 2.0$ up to $z \sim 7$, and beyond \citep{Guaita2011,Nilsson2007,Nilsson2009,Ouchi2008,Gawiser2006,Finkelstein2009_45,Murayama2007,Pirzkal2007,Hibon2011,Hibon2012,Oteo2011,Oteo2012a,Oteo2012b}. LBGs are found by employing a combination of broad-band filters that sample the red-ward and blue-ward zones of the redshifted Lyman break of galaxies, located in 912 \AA\ in the rest-frame \citep{Madau1996,Steidel1996,Steidel2003}. The choice of the red-ward and blue-ward filters determines the location in wavelength of the Lyman break and, consequently, the redshift of the selected galaxies. Many samples of LBGs have been found and examined at different redshifts, mostly at $z \gtrsim 3$ \citep{Madau1996,Steidel1996,Steidel1999,Steidel2003,Stanway2003,Giavalisco2004LBGs,Bunker2004,Verma2007,Iwata2007}. At $z \lesssim 3$ the number of LBGs reported and studied is much lower than that at higher redshifts \citep{Burgarella2006,Burgarella2007,Ly2009,Ly2011,Basu2011,Hathi2010,Nilsson2011_LBG,Hathi2013,Haberzettl2012,Chen2013} despite this redshift range is quite important since it is thought that the peak of the cosmic star formation of the universe took place in that epoch.

Apart from the Ly$\alpha$ and Lyman break techniques, some other methods have been used for selecting high-redshift galaxies in the literature. \cite{Adelberger2004} defined some color selection criteria that employs several combinations of optical colors to select galaxies at different redshifts: $GRi$ for sources within $0.85 \lesssim z \lesssim 1.15$, $GRz$ for $1.0 \lesssim z \lesssim 1.5$, and $U_nGR$ for $1.4 \lesssim z \lesssim 2.1$ and $1.9 \lesssim z \lesssim 2.7$. The galaxies selected in this way have been traditionally called BM/BX galaxies. Another ground-based optical color selection criterion is the $BzK$ method, which is aimed at finding galaxies in the redshift range $1.4 \lesssim z \lesssim 2.5$ and classifying them as SF or passively evolving systems \citep{Daddi2004}. Both kinds of galaxies are often associated to LBGs. \cite{Haberzettl2012} find that NUV data  provide greater efficiency for selecting SF galaxies. Furthermore, they report that, although the BM/MX and $BzK$ techniques are very efficient for detecting sources within $1.0 \lesssim z \lesssim 3.0$, they are biased against those SF galaxies which are more massive and contain a noticeable amount of red stellar populations. \cite{Haberzettl2012} argue that, therefore, a NUV-based LBG selection criterion is more adequate to compare with the populations found at $z \gtrsim 3.0$.

The physical properties of high-redshift SF galaxies have been traditionally analyzed by fitting their observed spectral
energy distributions (SED) built from their photometric data\footnote{In this work we apply the term SED to refer to a set of photometric points. However, it should be noted that SED is also applied to spectroscopic data in many works. Some of the limitations quoted here for the SED-fitting technique only apply to photometric SEDs but not to spectroscopic one} to SED templates obtained from stellar population models (such as \cite{Bruzual2003}, hereafter BC03) \citep{Lai2008,GAwiser2007,Nilsson2007,Nilsson2009,Yabe2009,Finkelstein2009,Finkelstein2008,Finkelstein2009d,Finkelstein2010,Magdis2010_IRAC}. This procedure, in principle, enables the determination of age, dust attenuation, star formation rate (and histories), metallicity, and stellar mass. This is because the SED obtained in stellar population models depends (among others) on all those parameters. However, in practice, this procedure has several limitations. For example, metallicity does not have a strong influence on the shape of the rest-frame optical SEDs and, therefore, its determination from SED-fitting suffers from large uncertainties. On the other hand, the degeneracy between dust attenuation and age or star-formation history (SFH) and age produces that both parameters are difficult to constrain accurately at the same time. With a good wavelength coverage of the UV continuum and the 4000 \AA\ Balmer break it is feasible to improve the determination of the SED-derived dust attenuation and stellar age. However, dust attenuation would still suffer from uncertainties and the only way to obtain accurate values is by employing direct FIR detections around the dust emission peak \citep{Burgarella2011,Oteo2013_LBGsz3}. Despite these caveats, many previous works have analyzed the physical properties of LBGs at different redshifts by employing an SED-fitting method since it is the only way to analyze their properties with large samples of galaxies. At $z \sim 5$, LBGs have been reported to be much younger ($<$100Myr) and to have lower stellar masses (10$^{9}$M$_{\odot}$) than their analogs at $z \sim 2.0-3.0$ in a similar rest-frame UV luminosity range \citep{Verma2007,Yabe2009,Haberzettl2012}. 

Most previous works focus their studies on LBGs which are located at $z \gtrsim 3$, where the Lyman break is shifted to the optical and can be sampled with filters in ground-based telescopes. At $z \lesssim 2$, the Lyman Break is located in the UV and LBGs can only be found with observations from space, for example with GALEX \citep{Burgarella2006,Burgarella2007,Haberzettl2012}, HST and UVIS filters \citep{Hathi2013}, or \emph{Swift} satellite \citep{Basu2011}. In this work, we aim at analyzing the physical properties of a sample of 1225 GALEX-selected LBGs at $z \sim 1$ located in the COSMOS field by using data coming from the Advanced Large, Homogeneous Area Medium Band Redshift Astronomical (ALHAMBRA) survey \citep{Moles2008}, which covers the optical range with 20 medium-band filters (width about 300\AA) and the near-IR with the traditional JHKs broad-band filters. The combination of the ALHAMBRA survey with observations in other wavelengths (GALEX and IRAC) allows an unprecedented coverage of the UV continuum and optical Balmer break. This can disrupt some of the degeneracies outlined before and provide more accurate results for the SED-derived physical properties. Since we study LBGs at $z \sim 1$, their observed fluxes are high enough so that the photometry has a good signal-to-noise ratio. These two facts (exceptional coverage of the SED and the high signal-to-noise ratio) are not usually achieved at higher redshifts. This emphasizes the importance of studying intermediate-redshift LBGs.

The paper is organized as follows: In Section \ref{data} we present the data sets employed both in the UV with GALEX and in the optical-to-near-IR with the ALHAMBRA survey. In Section \ref{combination} we combine these two data sets to build a general sample of UV-selected galaxies at $0 \lesssim z \lesssim 2$. In Section \ref{combination} we also explain how we carry out the SED fits using BC03 templates for these UV-selected sources with the aim of obtaining their photometric redshift, rest-frame UV luminosity, and other physical properties such as dust attenuation, age, and stellar mass. In Section \ref{UV_08} we define the selection criterion adopted in this work to look for LBGs at $z \sim 1$. The SED-derived physical properties of the selected LBGs are discussed in Section \ref{stellar}. The morphology and physical sizes of the LBGs studied are analyzed in Section \ref{morfo}, and in Section \ref{color_mag} we show their location in the color-magnitude (CMD) diagram. In Section \ref{high_z} we compare the properties of our GALEX-selected LBGs to those reported in previous works for LBGs at higher redshifts. Finally,  we summarize the main conclusions of the work in Section \ref{conclu}.

Throughout this paper we assume a flat universe with $(\Omega_m, \Omega_\Lambda, h_0)=(0.3, 0.7, 0.7)$, and all magnitudes are listed in the AB system \citep{Oke1983}.

\section{Data sets}\label{data}

On the UV side we use data coming from observations of the COSMOS field with the GALEX satellite \citep{Martin2005} in both NUV and FUV bands as part of the Deep Imaging Survey (PI: D. Schiminovich). GALEX catalogs were created by using the EM-algorithm \citep{Gillaume2006}, aimed at resolving the blended objects in the far and the near UV using the information (position and shape) available from existing, well resolved catalogs on the visible range. In the concrete case of the COSMOS field, the prior optical photometric informations corresponds to a u*-band mosaic (and its SExtractor-derived catalog) based on CFHT-u* observations. With a list of optical prior positions, the algorithm measures their UV fluxes on the GALEX  images by adjusting a GALEX PSF's model. The algorithm was run on the four NUV and the four FUV GALEX images covering the COSMOS field obtained as a product of the GALEX pipeline processing, version 1.61.

On the optical and near-IR side we use the ALHAMBRA survey \citep{Moles2005,Moles2008}, which employs a set of 20 equal-width ($\sim 300$ \AA) medium-band filters covering the optical range from 3500 to 9700\AA\ plus the traditional JHKs broad-bands near-IR filters to observe a region of 4 square degrees distributed into 8 different fields. Among them, we focus our work in the COSMOS field due to the wealth of photometric and spectroscopic data in the same and other wavelengths than UV, optical, and near-IR. The observations were carried out with the 3.5m telescope of the Calar Alto observatory using the wide-field camera LAICA in the optical and the OMEGA-2000 camera in the near-IR. The optical filters system adopted in ALHAMBRA was set with the aim of optimize the output of the survey in terms of the photo-z accuracy \citep{Benitez2009}. The simulations performed in \cite{Benitez2009} relating the image depth, photo-z accuracy, and number of filters indicate that the filter system of ALHAMBRA enables to get a photo-z precision, for normal SF galaxies, that is three times better than that for traditional 4-5 broadband filter sets. Additionally, the complementary usage of near-IR data improves the determination of photometric redshifts. In this work we have used the catalogs coming from the Internal Data Release 3. The data reduction and the construction of the catalogs were carried out by the ALHAMBRA team. Since the ALHAMBRA survey performs observations in 23 filters, one must work with care when defining the detection of the objects. With the aim of not biasing the detection to any special kind of object as a consequence of the selection in a single band, a special technique was employed. It is based on creating a deep detection image built as the sum of the individual frames with the highest efficiencies. This includes the filters centered between 457 and 829 nm. The photometry of the sources is done by running SExtractor in its dual mode. The deep image is used for source detection and the photometry is then extracted in each individual frame. As result, the average depth (for $3\sigma$ detections) is 24.5 and 22 magnitudes in the optical and near-IR, respectively. In this work we employ the SExtractor-derived \verb+AUTO_MAG+ as the best approximation to the total magnitude for all the calculations. The characterization in the optical range of the ALHAMBRA photometric system can be found in \cite{AparicioVillegas2010} and the near-IR number counts of one of the fields is presented in \cite{CristobalHornillos2209}. More details on the quality of the data, the reduction process, the depth, etc. will be published in Husillos et al. (in prep). We note that the characterization of ALHAMBRA filters (complete wavelength coverage with almost no overlapping filters) provides an SED that can be considered as a low resolution optical spectra of the observed sources. It should be noted that the Lyman break selection that will be employed in this work is purely based on UV GALEX data, whereas the ALHAMBRA survey is used in the analysis of their SED-derived physical properties.

\section{GALEX and ALHAMBRA data: UV-selected galaxies and SED fitting}\label{combination}

\begin{figure*}
\centering
\includegraphics[width=0.3\textwidth]{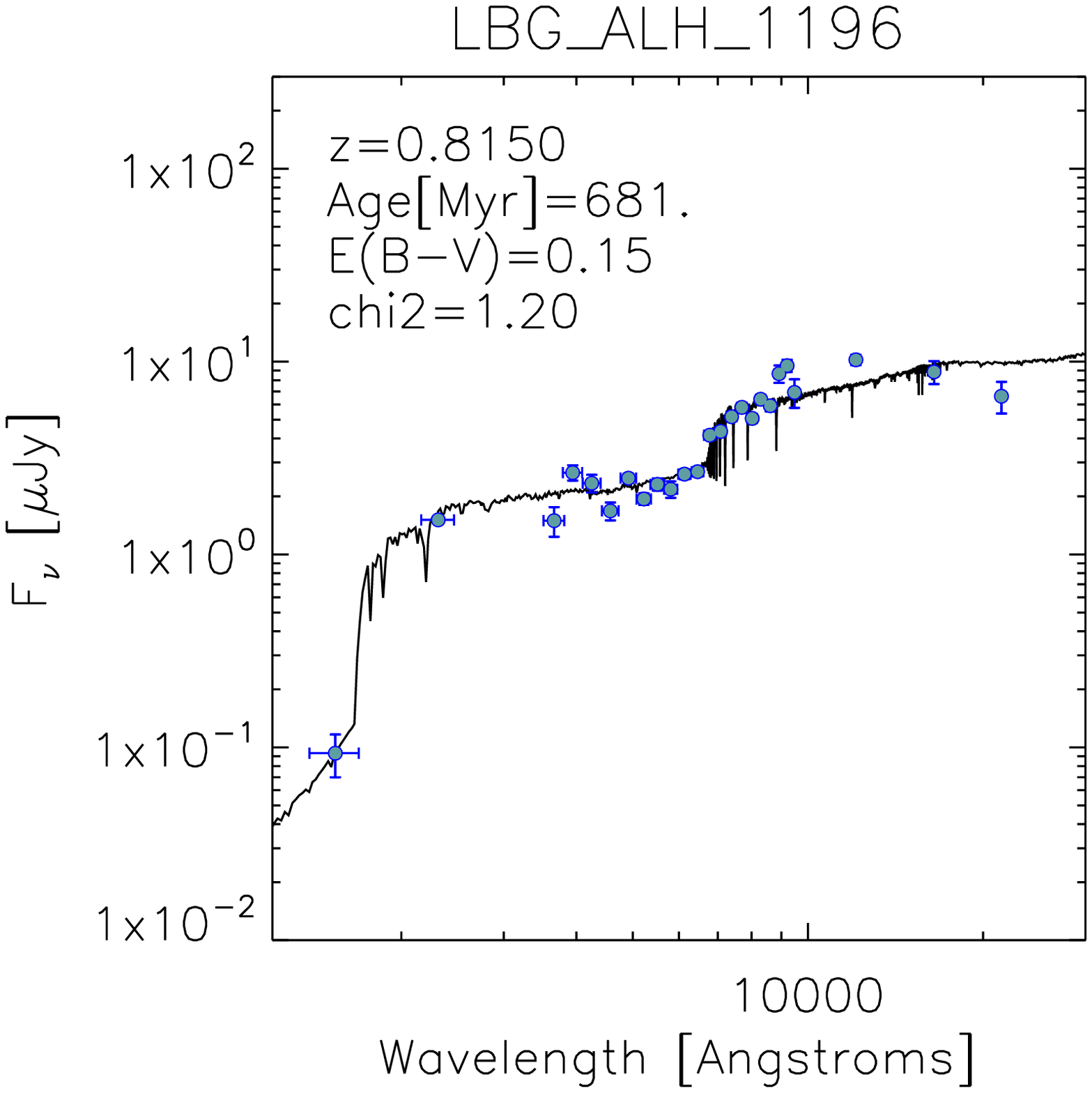}
\includegraphics[width=0.3\textwidth]{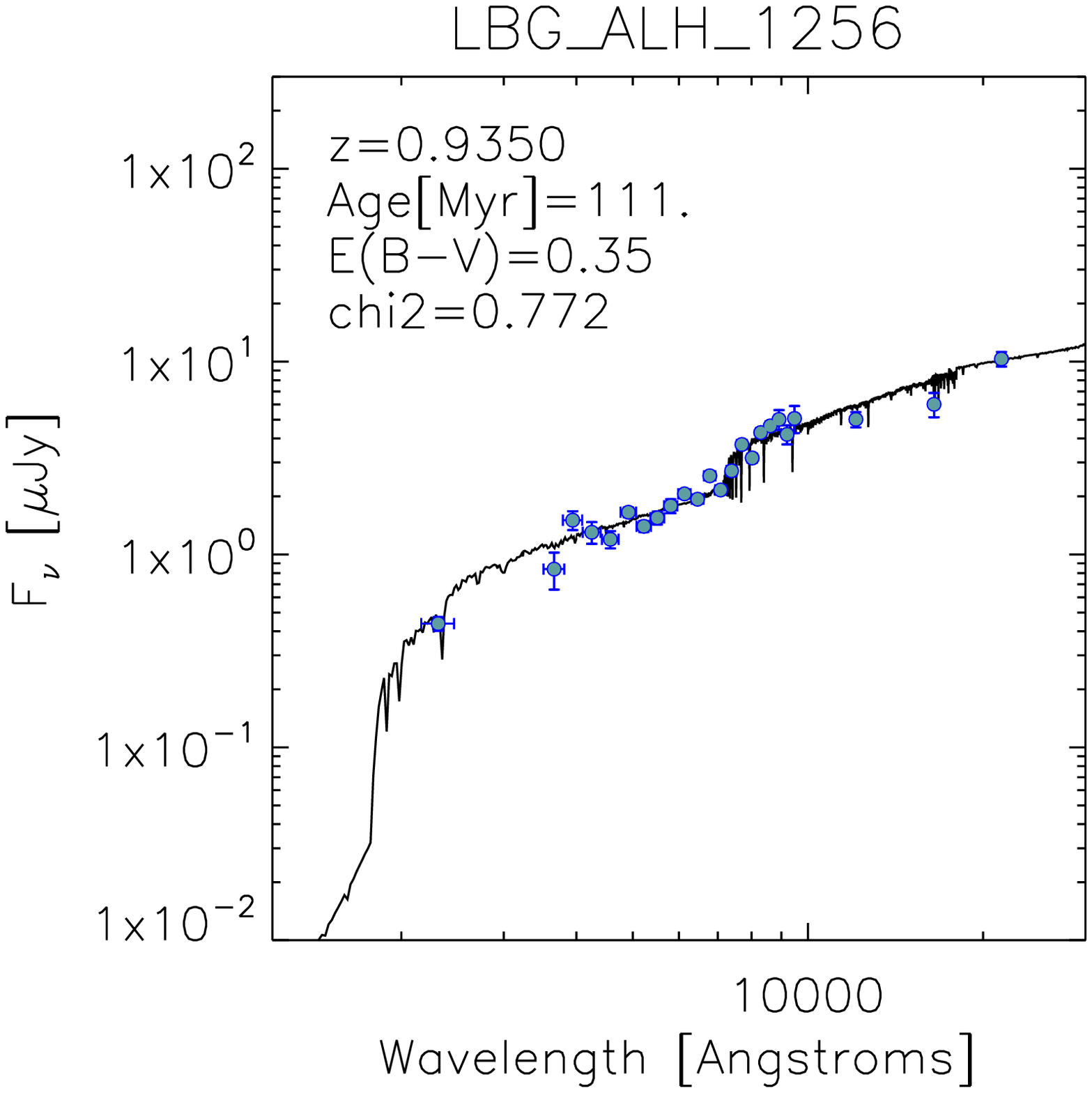}
\includegraphics[width=0.3\textwidth]{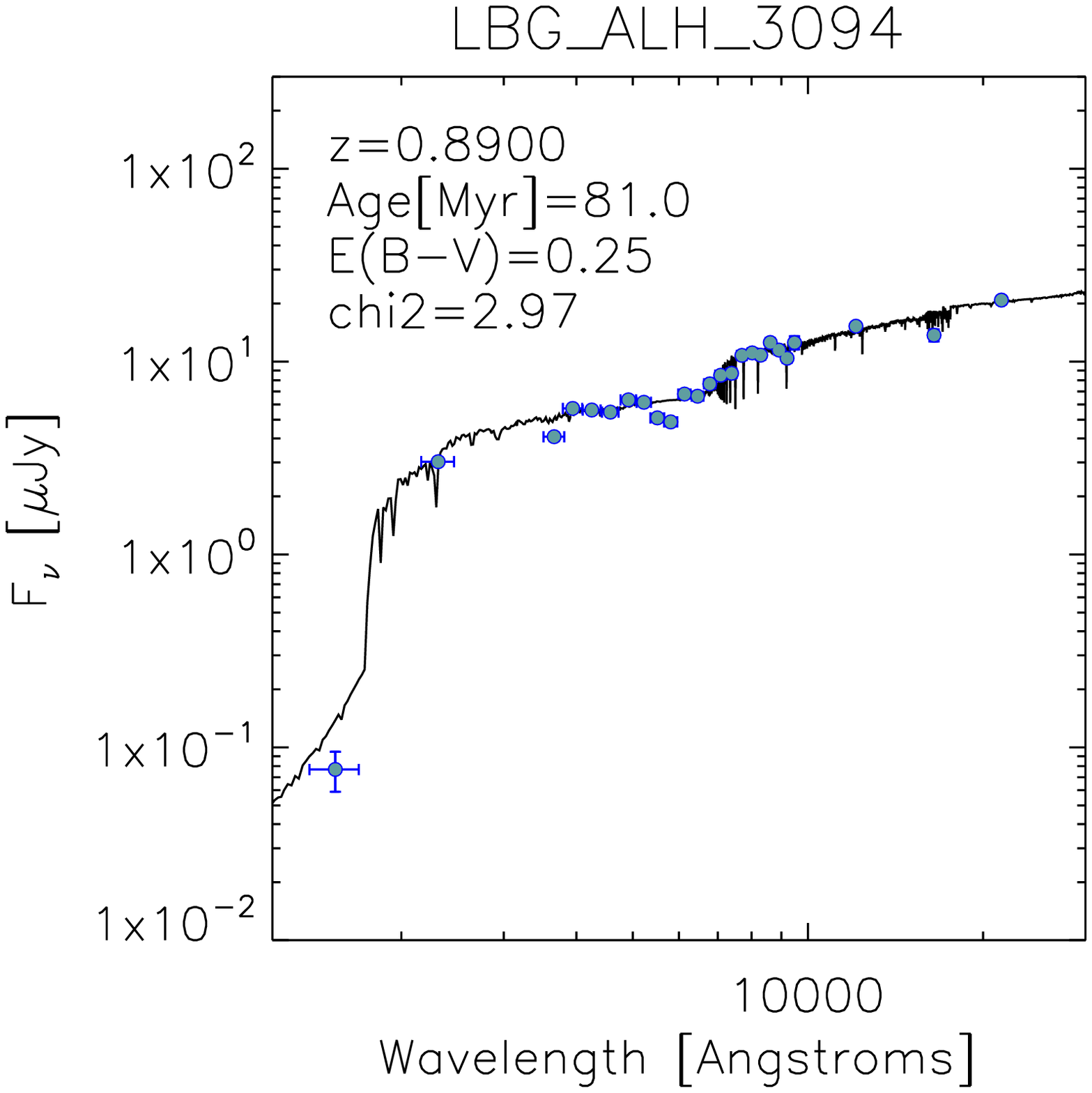}
\includegraphics[width=0.3\textwidth]{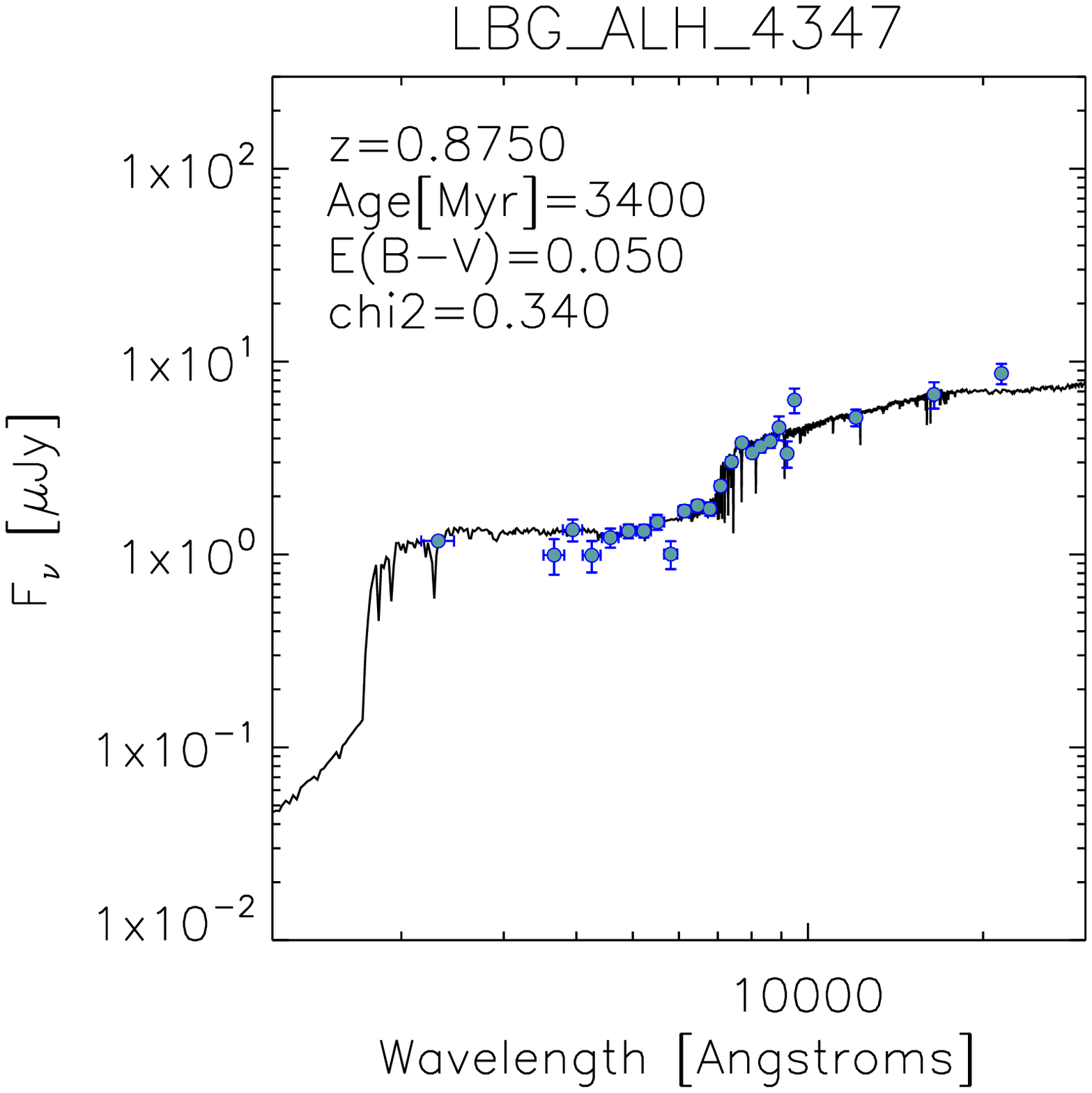}
\includegraphics[width=0.3\textwidth]{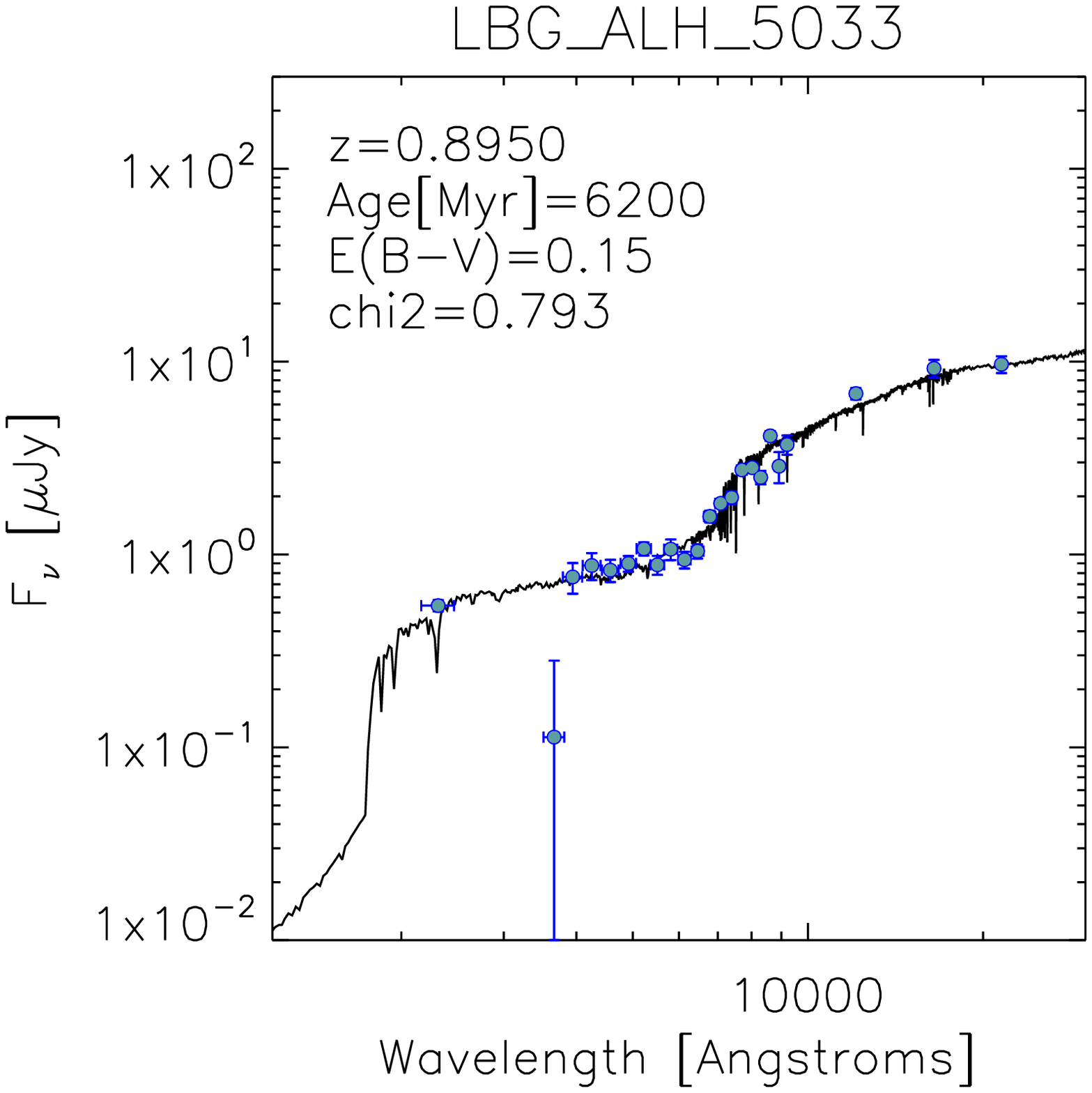}
\includegraphics[width=0.3\textwidth]{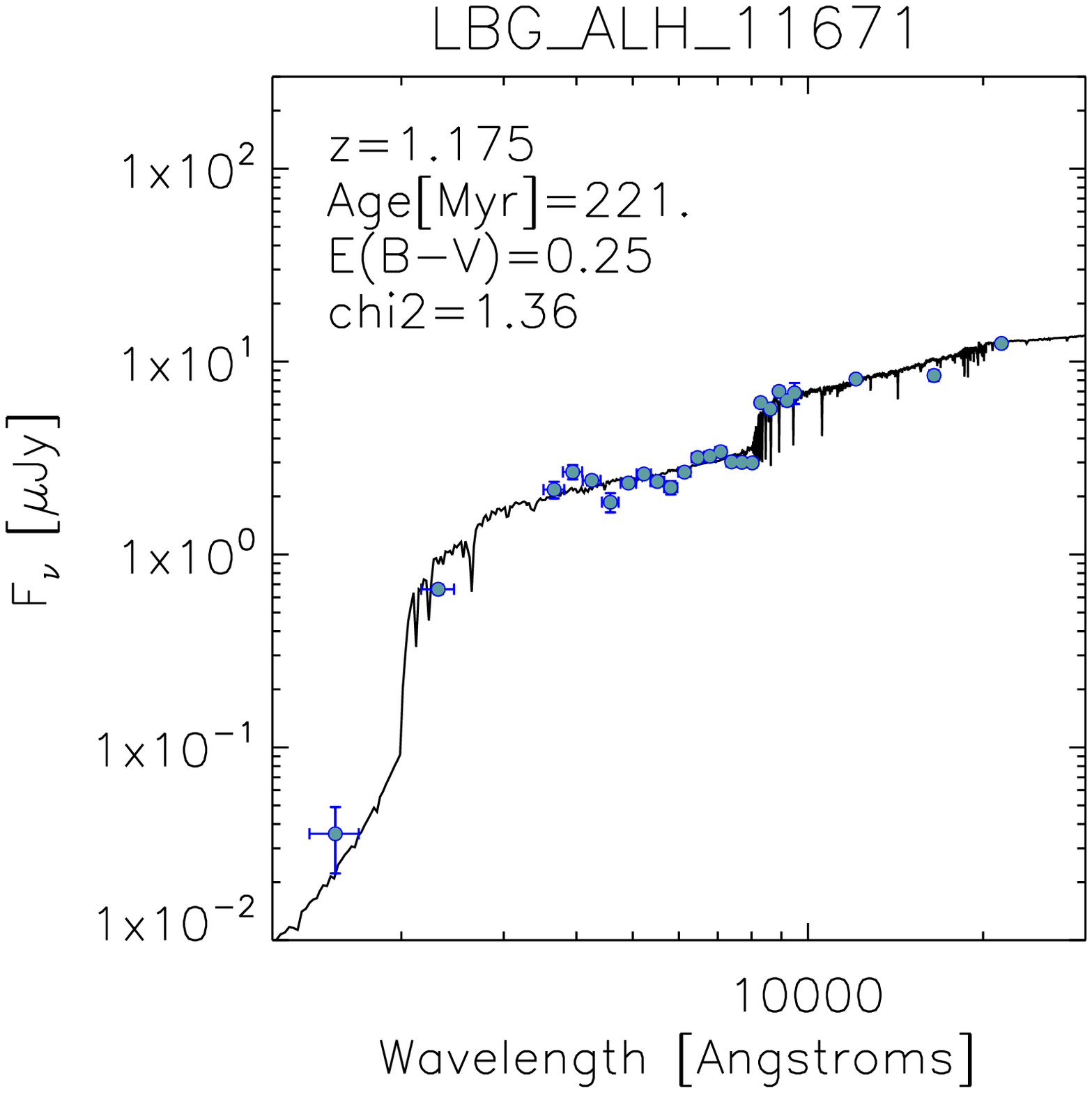}
\includegraphics[width=0.3\textwidth]{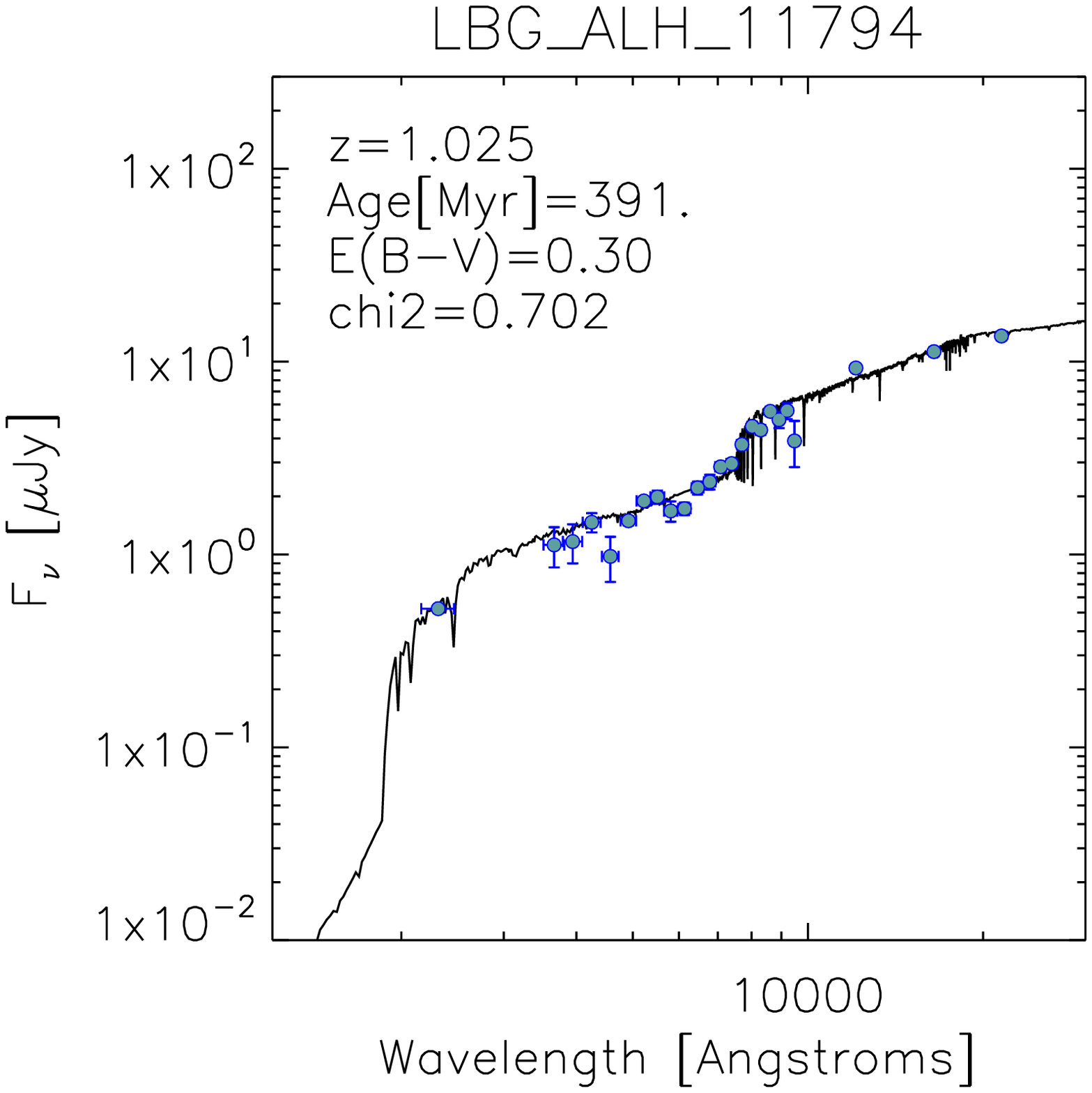}
\includegraphics[width=0.3\textwidth]{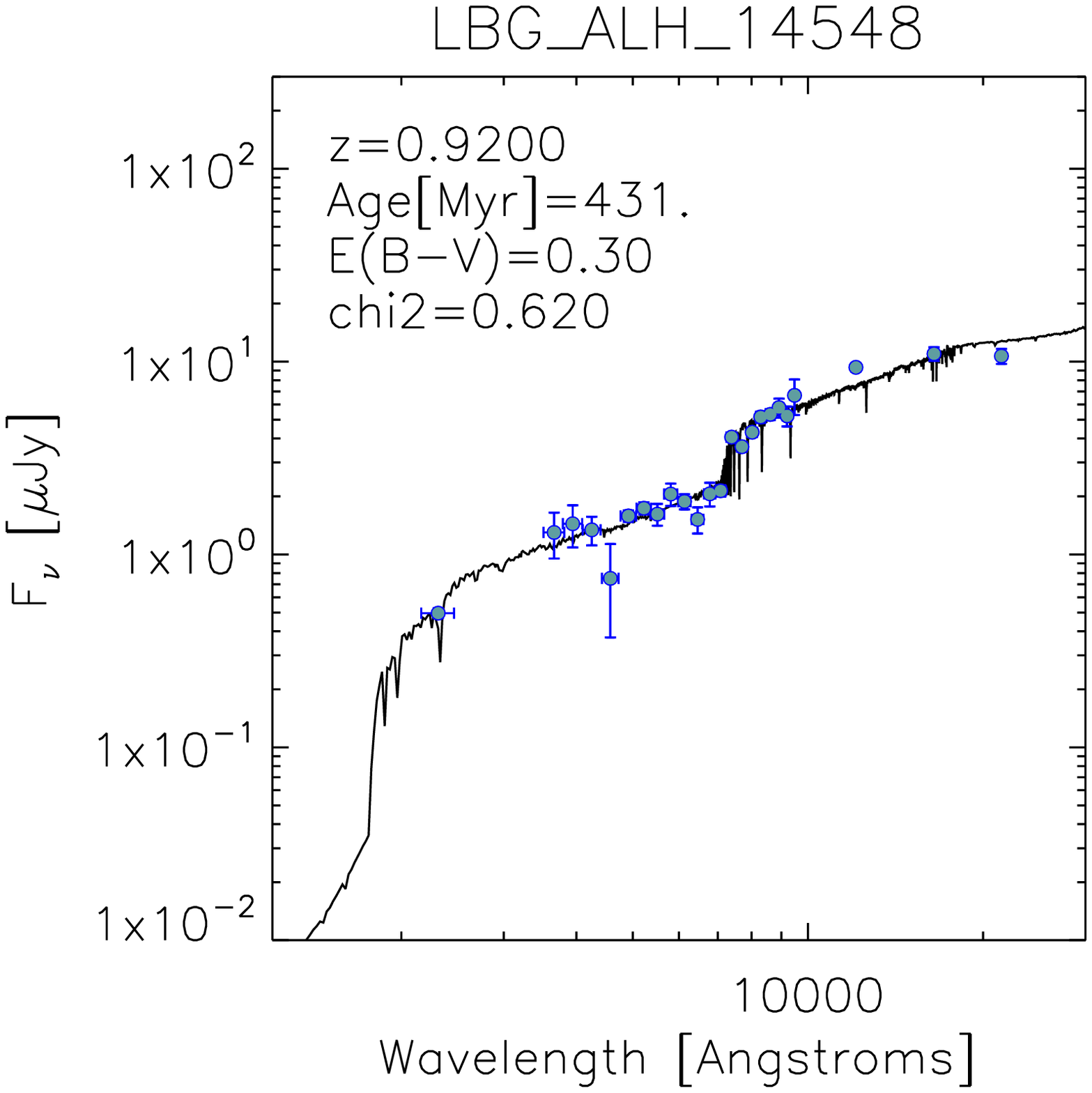}
\includegraphics[width=0.3\textwidth]{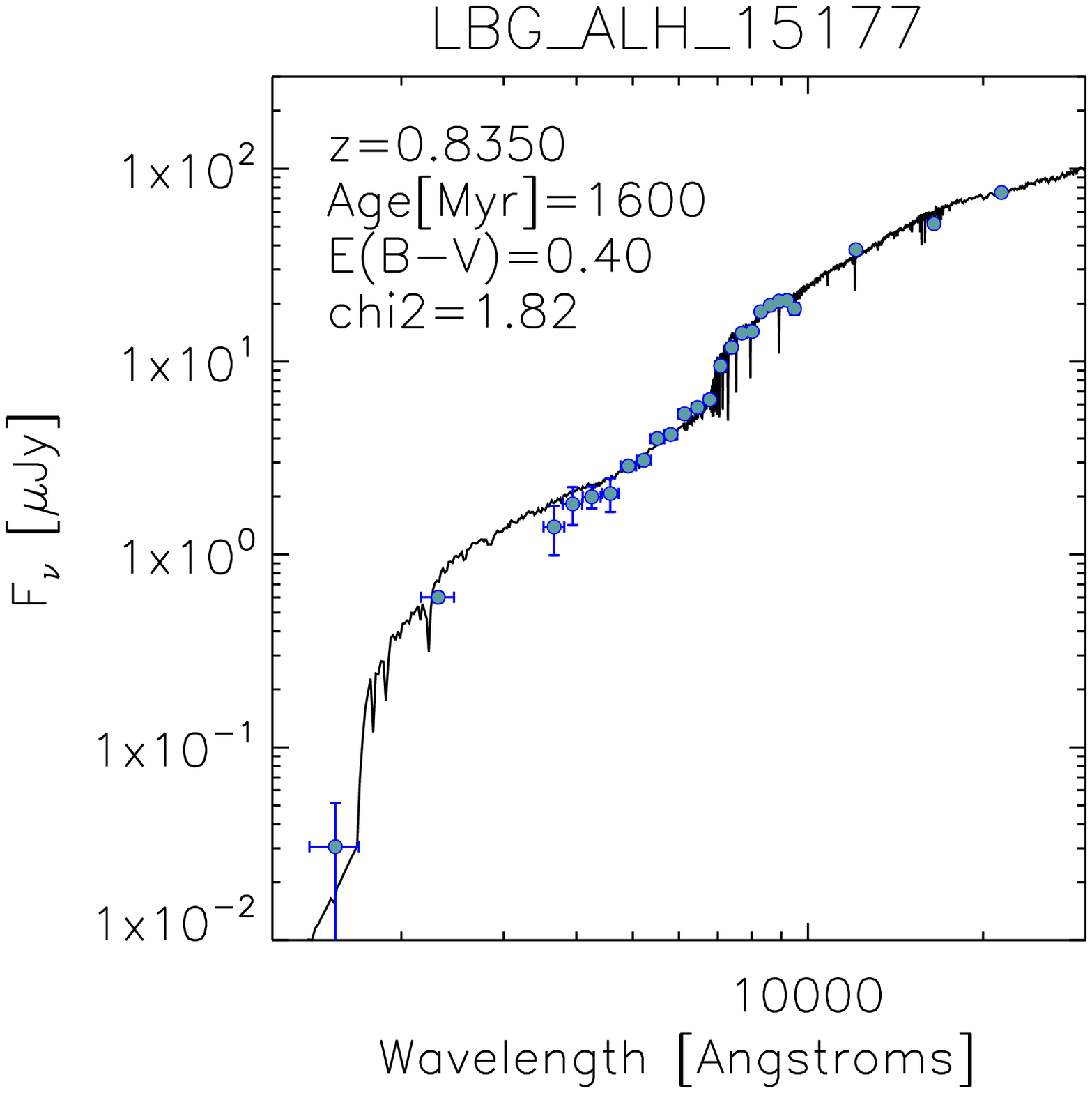}
\caption{Examples of SED-fitting results with GALEX+ALHAMBRA data for nine GALEX-selected LBGs randomly selected within the whole sample. Blue points are the observed GALEX and ALHAMBRA fluxes and the black curves are the best-fitted \citet{Bruzual2003} (BC03) templates of each object. The BC03 templates were build by assuming a constant SFR, salpeter IMF, and a fixed metallicity of $Z=0.2Z_\odot$. In each panel we indicate the SED-derived redshift, age, dust attenuation, and reduced $\chi^2$ associated to each best-fitted template. It can be seen that the combination of GALEX and ALHAMBRA data provides a very good sampling of the rest-frame UV continuum and 4000 \AA\ Balmer break of our UV-selected galaxies.
}
\label{SEDs_IRAC}
\end{figure*}

The LBGs that will be studied in this work are taken from a multi-wavelength catalog of UV-selected SF galaxies that we build by combining the GALEX observations with the data coming from the ALHAMBRA survey: we look for GALEX detections around 2'' of the optical position of the sources in the ALHAMBRA survey and retain only those galaxies which have, at least, a detection in the NUV band. This produces a sample of 39734 UV-selected sources for which we have photometric information from the UV to the near-IR.

With the aim of obtaining the photometric redshifts and the physical properties of those UV-selected sources we fit their observed fluxes to BC03 templates with the Zurich Extragalactic Bayesian Redshift Analyzer \citep[ZEBRA,][]{Feldmann2006} code which, in its maximum-likelihood mode, employs a $\chi^2$ minimization algorithm over the templates to find the one which fits the observed SED of each input object best. We build a set of BC03 templates associated to different physical properties of galaxies by using the software \verb+GALAXEV+. In this process we adopt a \cite{Salpeter1955} initial mass function (IMF) distributing stars from 0.1 to 100 M$_\odot$ and select a fixed value for metallicity of $Z=0.2Z_{\odot}$. We consider values of age from 1 Myr to 7 Gyr, in steps of 10 Myr from 1 Myr to 1 Gyr and in steps of 100 Myr from 1 Gyr to 7 Gyr. Dust attenuation is included in the templates via the \cite{Calzetti2000} law and parametrized through the color excess in the stellar continuum, $E_s(B-V)$. We select values for $E_s(B-V)$ ranging from 0 to 0.7 in steps of 0.05. We include intergalactic medium absorption adopting the\cite{Madau1995} prescription. Regarding SFR, we adopt time-constant models. In this case, different values of the SFR does not change the shape of the templates and the SFR can be obtained by using the \cite{Kennicutt1998} calibration: 

\begin{equation}\label{SFR_UV}
\textrm{SFR}_{UV,uncorrected}[M_{\odot}\textrm{yr}^{-1}] = 1.4 \times 10^{-28}L_{1500}
\end{equation}

\noindent where $L_{1500}$ is the rest-frame UV luminosity in 1500\AA. The $L_{1500}$ is obtained for each galaxy by convolving its best-fitted template with a top-hat filter (300 \AA\ width) centered in rest-frame 1500 \AA. It should be noted that, throughout the work, we distinguish between L$_{UV}$ defined in a $\nu$L$_{\nu}$ way (units of $\textrm{erg} \, \textrm{s}^{-1}$) and L$_{1500}$ considered in L$_{\nu}$ units ($\textrm{erg} \, \textrm{s}^{-1} \, \textrm{Hz}^{-1}$). The SFR derived from Equation \ref{SFR_UV} is uncorrected for the attenuation that dust produces in SEDs of galaxies. In order to obtain an estimation of the dust-corrected total SFR we have to introduce into Equation \ref{SFR_UV} the dust-corrected $L_{1500}$. It is obtained from $L_{1500}$ by multiplying it by the dust correction factor 10$^{0.4A_{1500}}$, where $A_{1500}$ is the dust attenuation in 1500\AA. The values of $A_{1500}$ are obtained from the SED-derived $E_s(B-V)$ assuming the \cite{Calzetti2000} law. Once both age and dust-corrected total SFR are known for each source, and according to the assumed time-independent SFH, the stellar mass can be obtained from the product of both quantities.

In this work we also analyze the UV continuum slope, $\beta$, of our UV-selected galaxies \citep[see for example][]{Calzetti1994}. This parameter is important in the study of how galaxies built up since it is related to age, metallicity, stellar IMF, and most importantly, dust attenuation. Furthermore, UV colors seem to present a relation with UV luminosities of SF galaxies \citep{Bouwens2009,Bouwens2010,Bouwens2012} and are relatively easy to measure than optical rest-frame colors in high-redshift galaxies, for which IRAC detections would be mandatory. The combination of the GALEX photometry and the bluest optical bands of the ALHAMBRA survey provides a good sampling of the UV continuum at $z \sim 1$, the redshift range where our LBGs are located, as it can be seen in the SED fits shown in Figure \ref{SEDs_IRAC}. Different works employ different methods to obtain the UV continuum slope of galaxies at different redshifts ranges, being the most popular and traditionally used that in where $\beta$ is quantified by using two broadband filters which sample two zones of the observed UV continuum \citep{Kong2004,Hathi2008,Meurer1997,Overzier2008_beta,Bouwens2010_beta,Finkelstein2010_beta,Dunlop2012}. In other works $\beta$ is obtained by using a power-law fit to the observed fluxes of the studied galaxies, using filters that sample the same zone of the SED at different redshifts \citep{Bouwens2011}. In our work, we obtain $\beta$ for each galaxy by fitting the UV continuum of its best-fitted template with a power law in the form $f_\lambda \sim \lambda^\beta$ \citep{Calzetti1994}. In this process we employ the rest-frame wavelength range between 1300\AA\ $\lesssim \lambda \lesssim$ 3000\AA. This range contains all the windows defined in \cite{Calzetti1994} in their definition of the UV continuum slope. This approach has the advantage of using all the available fluxes of each source, resulting in more robust S/N determinations. The method employed here is similar to that used by \cite{Finkelstein2012} in their study of the redshift evolution of the UV continuum slope from $z \sim 8$ to $z \sim 4$. In that work, they present some illustrative examples showing the differences in the UV continuum slope when using the different techniques.

\begin{figure*}
\centering
\includegraphics[width=0.49\textwidth]{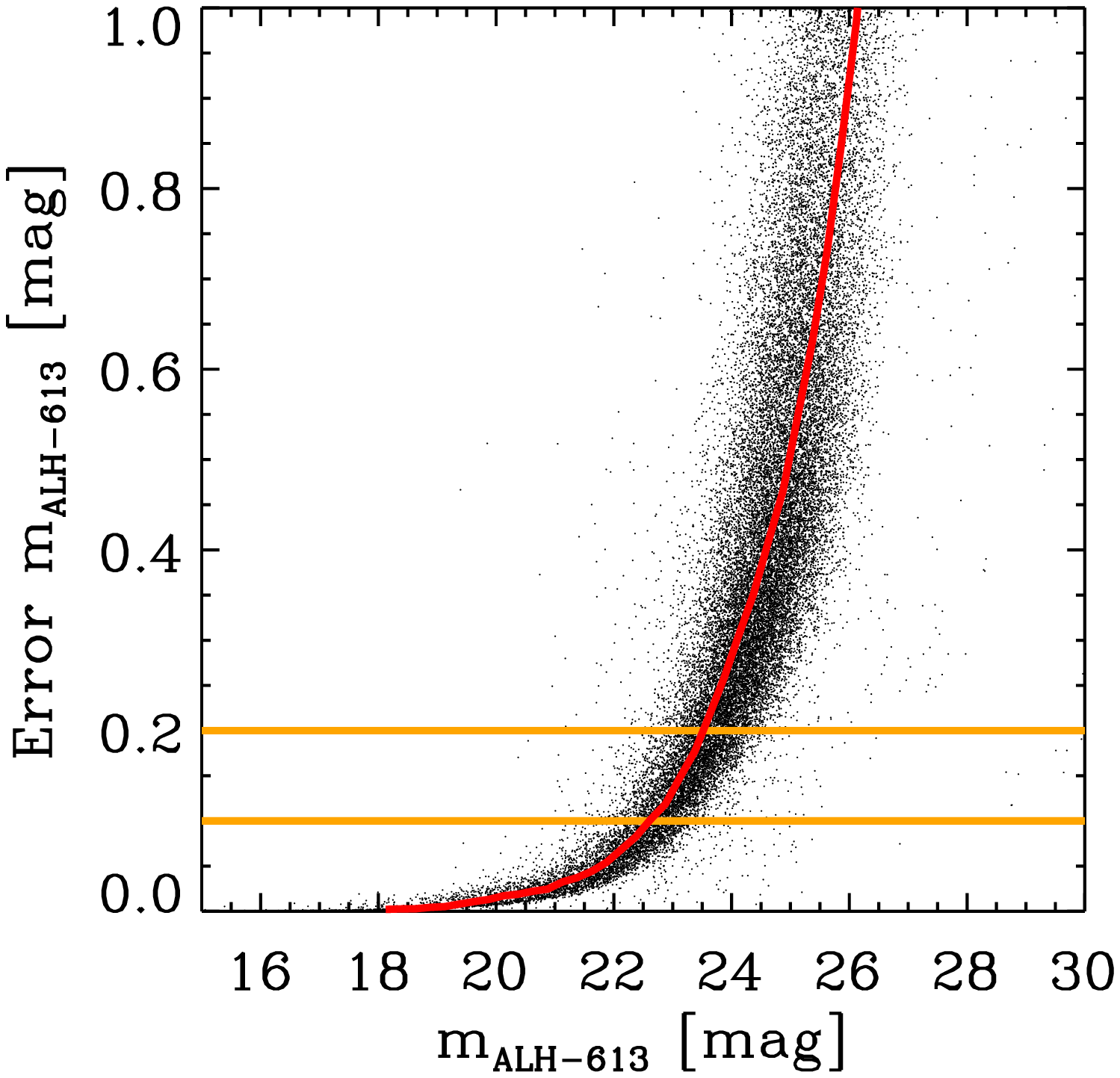}
\includegraphics[width=0.49\textwidth]{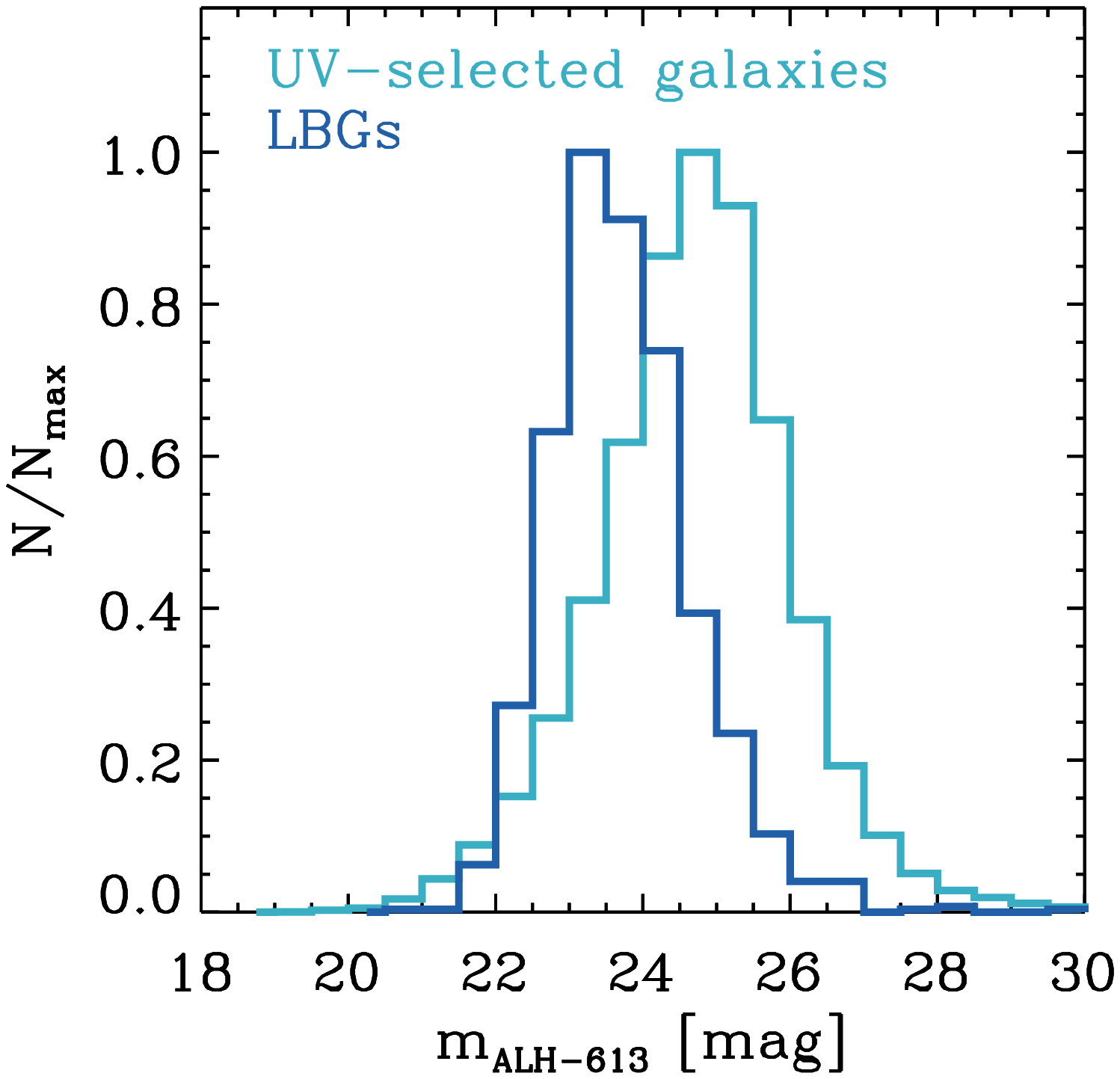}
\caption{\emph{Left:} Photometric errors against the observed magnitude in the ALHAMBRA filter centered in 613 nm for our sample of UV-selected galaxies with ALHAMBRA measurements and $\chi^2_r < 10$ in the SED-fitting results. Red curve corresponds to the median value of the error distribution for each value of the observed magnitude. Horizontal lines represent photometric errors of 0.1 and 0.2 mag. \emph{Right:} Distribution of the observed magnitude in the ALHAMBRA filter centered in 613 nm for LBGs and a sample of UV-selected galaxies with ALHAMBRA counterparts at the same redshift range than LBGs. Histograms have been normalized to their maxima in order to clarify the representation.
}
\label{errores}
\end{figure*}

In an SED-fitting procedure, the reliability of the results i.e. the similarity between the observed SED and the represented by its best-fitted template, is related to the $\chi^2$ value of the fits. Here we define the reduced $\chi^2$, $\chi^2_r$ of each best-fitted template as the ratio between its $\chi^2$ and the number of filters minus one employed in the fit, $\chi^2_r = \chi^2/(N-1)$ \citep[see for example][]{Barros2012}. From a visual inspection of the SED-fitting results, we consider that the BC03 templates truly represent the observed SED for each galaxy when $\chi^2_r  < 10$ (see some examples of $\chi^2$ values and the quality of the fittings in Figure \ref{SEDs_IRAC}). Imposing $\chi_r^2 < 10$ to the fittings of the whole sample of 39754 UV-selected galaxies, we end up with a robust sample of 35810 galaxies. From now on, only sources with $\chi^2_r < 10$ are considered.

\begin{figure*}
\centering
\includegraphics[width=0.49\textwidth]{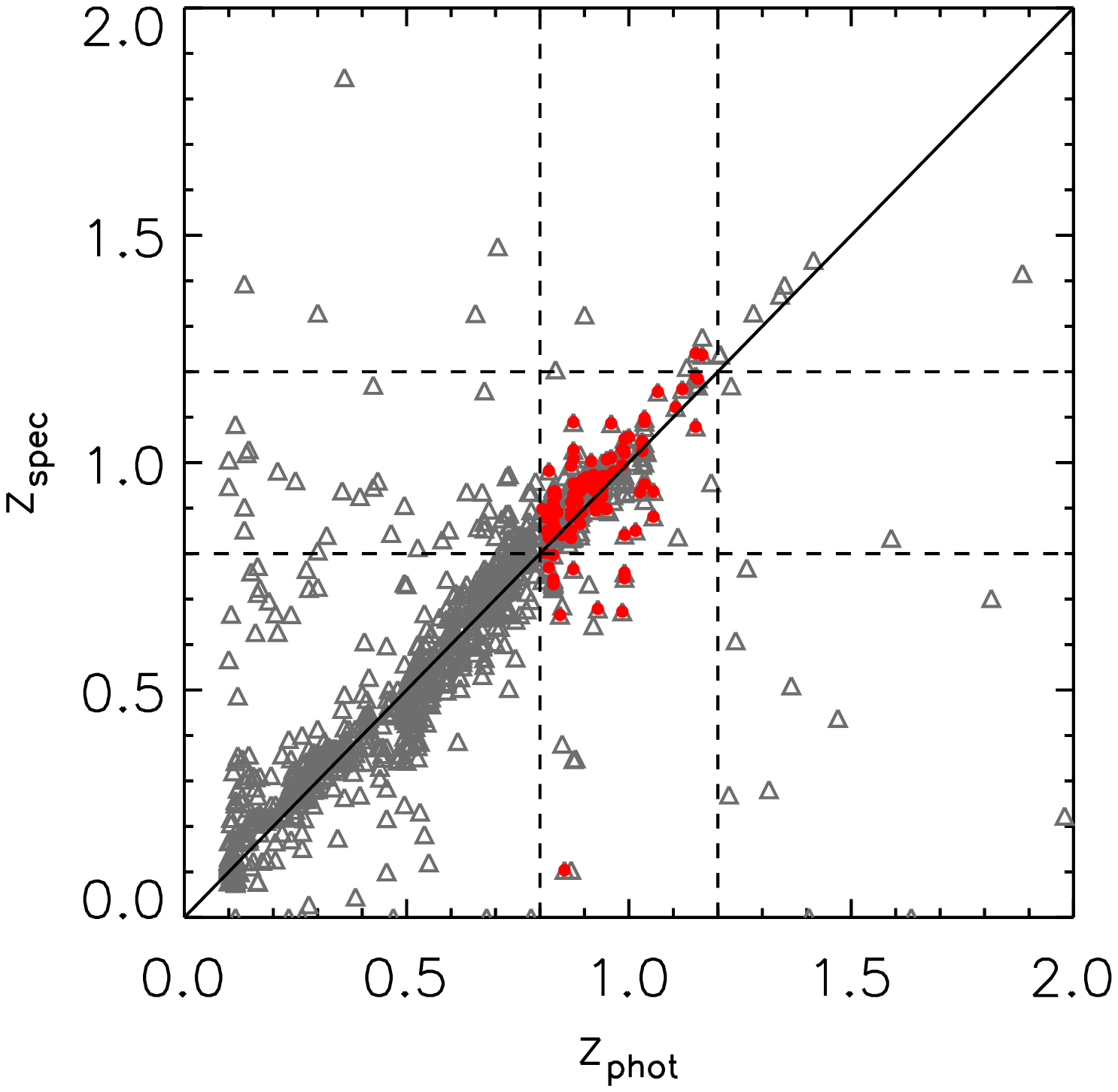}
\includegraphics[width=0.49\textwidth]{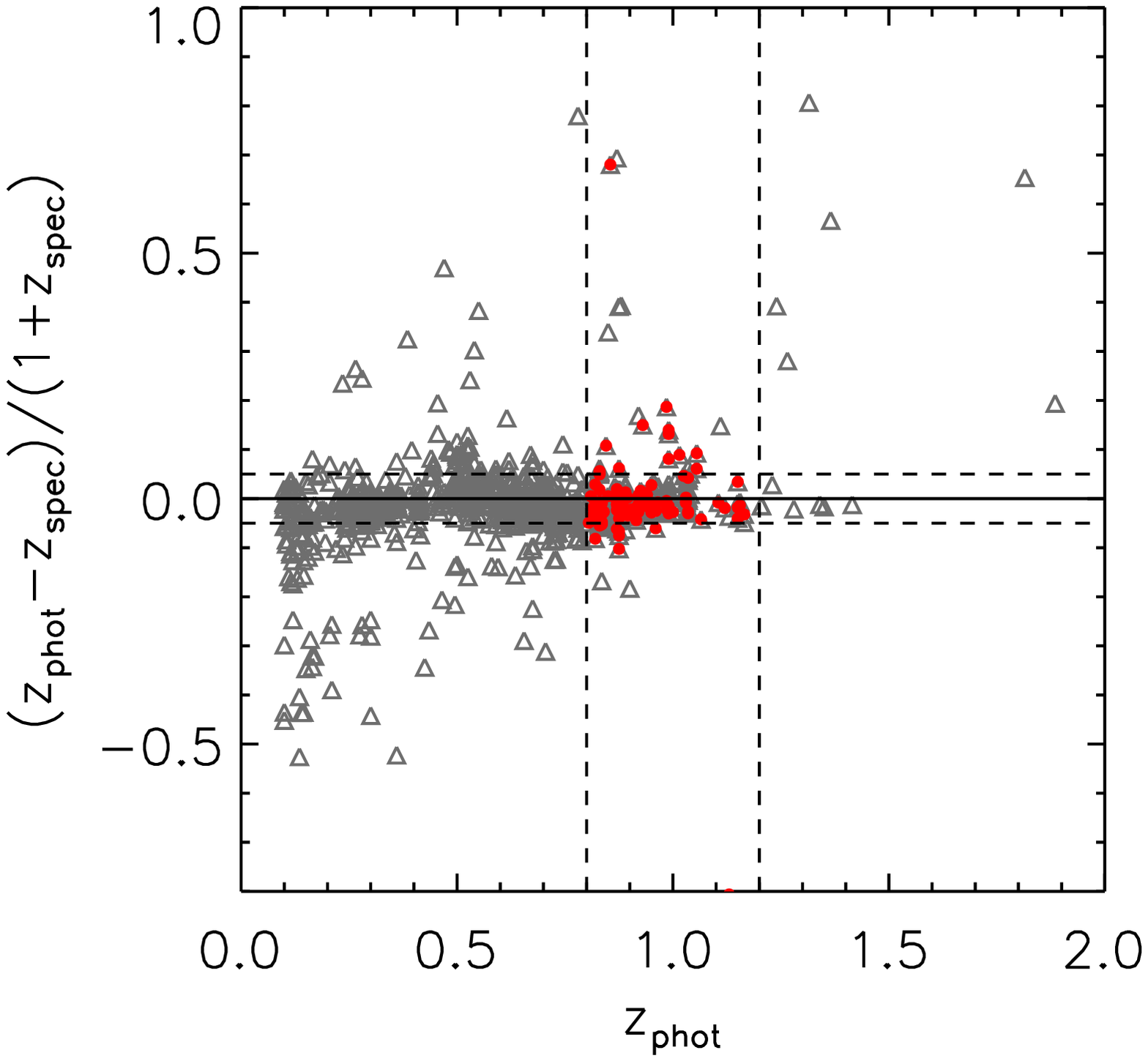}
\caption{Accuracy of the photometric redshift determination with the combination of GALEX and ALHAMBRA data for the whole sample of 35810 UV-selected galaxies with ALHAMBRA measurements and $\chi^2_r < 10$ in the SED-fitting results. In these plots, only galaxies with available spectroscopic redshift from zCOSMOS survey \citep{Lilly2007} are considered. Red dots represent our GALEX-selected LBGs and grey open triangles are the remaining galaxies in the sample. In the left panel, vertical and horizontal dashed straight lines represent the photometric redshift locus where most GALEX-selected LBGs are expected to be located according to their UV color selection, $0.8 \lesssim z \lesssim 1.2$. In the right panel, vertical dashed straight lines represent the photometric redshift locus where most GALEX-selected LBGs are expected to be located according to their UV color selection. The horizontal dashed straight lines indicates the values of the photometric redshift accuracy, $\sigma_{\Delta z} = \Delta z/(1+z_{spec} )$, equal to $\pm$0.05. The horizontal continuous straight line represents where spectroscopic and photometric redshifts would agree.
              }
\label{reliability}
\end{figure*}

The $\chi^2_r$ values depend upon the observed fluxes and also upon their photometric uncertainties. In this way, if a galaxy has a photometry with high photometric errors, the $\chi^2_r$ might be low even when its best-fitted template does not represent its observed SED properly. Therefore, a low value of the $\chi^2_r$ can be due to either a good SED-fitting or to a bad SED fit with a photometry with high uncertainties. Thus, we should check the typical photometric errors of the ALHAMBRA photometry of our sources to analyze whether the low values of $\chi^2_r$ are due to truly accurate fits or are the consequence of high photometric errors. As an example, we show in the left panel of Figure \ref{errores} the photometric errors in the ALHAMBRA filter centered in 613 nm of our sample of UV-selected sources at $0 \leq z \leq 2$ as a function of their observed magnitude in the same band. As expected, the photometric errors increase with the observed magnitude. If we consider that a fit is reliable for galaxies with typical photometric errors below 0.4 mag, we can only trust those SED-fitting results for galaxies typically brighter than about 25 mag. In the right panel of Figure \ref{errores} we represent the observed magnitudes of the GALEX-selected LBGs that will be studied in this work. It can be seen that most of them meet the previous criterion and, therefore, we can consider that the low values of $\chi^2_r$ are statistically due to good SED fits rather than to high photometric errors.

It should be noted that in this work we employ BC03 templates associated to a time-independent SFR. Other kinds of SFHs can be used, such as exponentially declining or composed by different bursts of star formation. In the first case, the SFR is characterized by the decaying time-scale, $\tau_{\rm SFR}$, which would be another parameter to obtain in the SED fitting, increasing the degrees of freedom in the process. Distinguishing between different kind of SFHs is very challenging even with a good photometric coverage of the SED of galaxies. Therefore, the results reported in this work should be understood as those derived with that choice of the SFR, but different values of the SED-derived parameters might be obtained if other temporal variations of the SFH are assumed. The analysis of the differences in the SED-fitting results depending on the assumption of the SFH will be studied in Section \ref{SFH}.

\begin{figure*}
\centering
\includegraphics[width=0.49\textwidth]{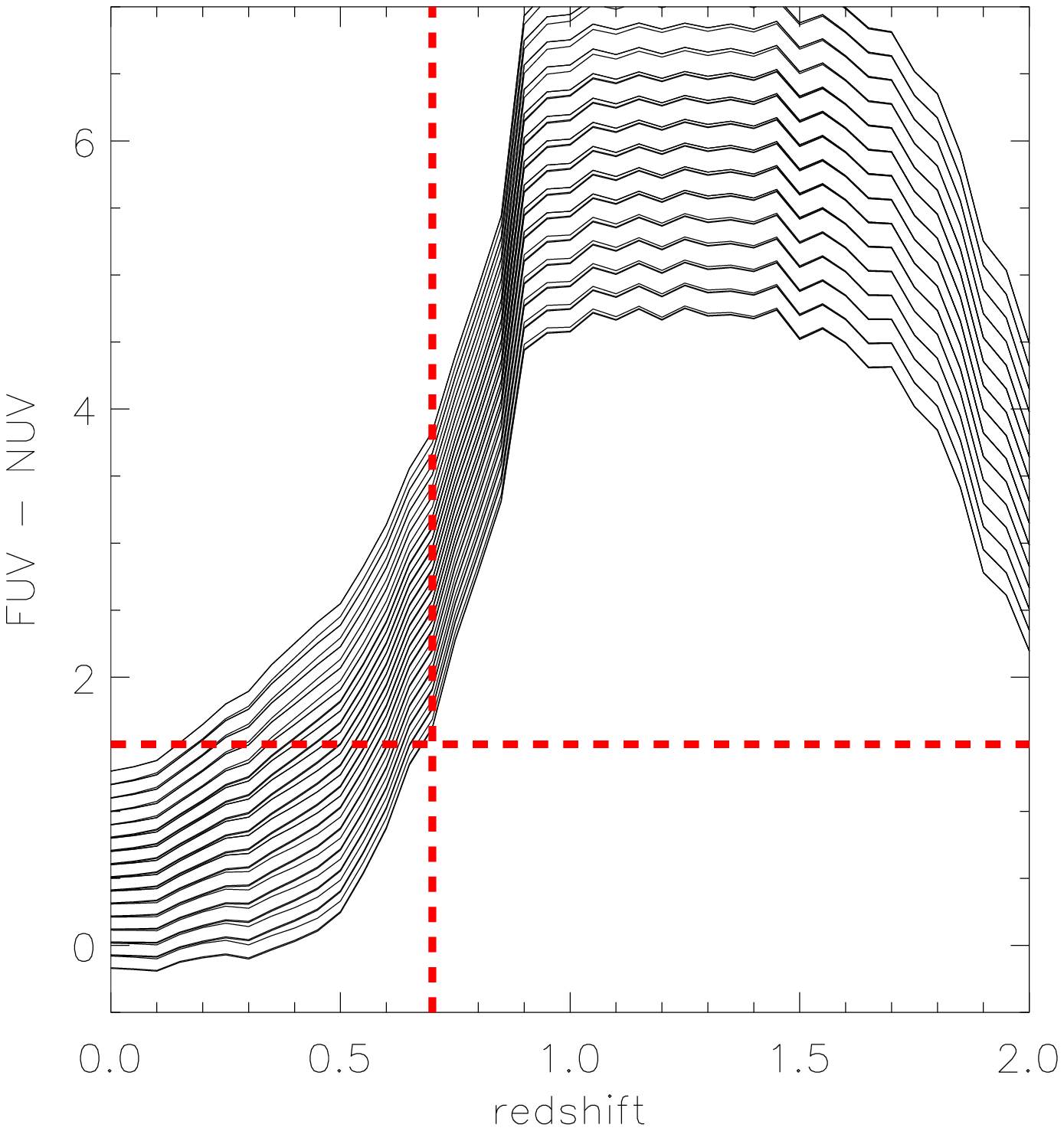}
\includegraphics[width=0.49\textwidth]{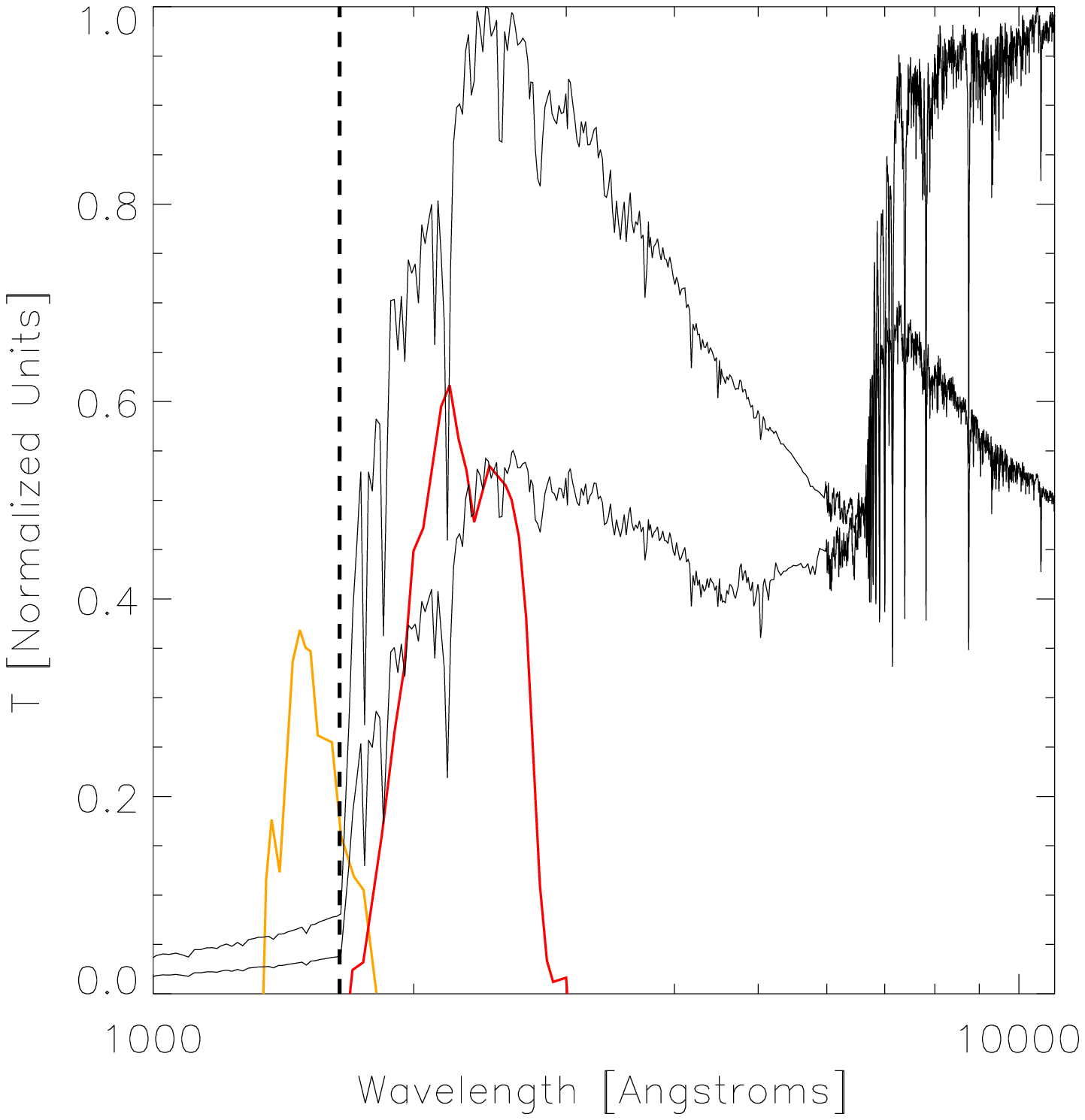}
\caption{Color selection of the GALEX-selected LBGs studied in this work. \emph{Left}: synthetic FUV-NUV color tracks as a function of redshift according to a set of BC03 stellar population templates associated to different values of age and dust attenuation. The horizontal dashed red line indicates the color cut employed in this work, which according to the tracks shown in black is expected to select galaxies at $z \geq 0.7$ (vertical red dashed line). \emph{Right}: Transmission curves of the FUV (orange) and NUV (red) GALEX filters. We also show the location of the Lyman break at $z = 0.8$ with a vertical dashed black line and the \citet{Bruzual2003} stellar population templates associated to two different values of age and dust attenuation (black curves). It can be seen that, although the color selection shown in the left panel is expected to segregate galaxies at $z \geq 0.7$, the Lyman break has not almost completely passed the FUV channel until $z \sim 0.8$.
}
\label{GALEX}
\end{figure*}

\subsection{Photometric redshifts and their accuracy}\label{zphot}

The good coverage of the observed UV-to-near-IR SED of galaxies provided by the ALHAMBRA survey in combination with GALEX data is expected to give accurate determination of photometric redshifts ($z_{\rm phot}$) at the expected redshift range of GALEX-selected LBGs, i.e. $z \sim 1$. This is due to the fact that at $z \sim 1$, GALEX+ALHAMBRA data cover the rest-frame UV continuum and the Balmer break, which are two of the most important features to fit in the SED of galaxies for determining photometric redshifts. We compare in Figure \ref{reliability} the obtained z$_{\rm phot}$ with spectroscopic redshifts ($z_{\rm spec}$) for those sources in the whole sample of 35810 UV-selected galaxies with $\chi^2_r < 10$ which have available spectra from the zCOSMOS survey \citep{Lilly2007}. Here we define the accuracy of the $z_{\rm phot}$ as $\sigma_{\Delta z} = \Delta z/(1+z_{\rm spec} )$, with $\Delta z = |z_{\rm phot} - z_{\rm spec}|$. See also \cite{Matute2012} for a discussion of the photometric redshift accuracy of ALHAMBRA survey. It can be seen in Figure \ref{reliability} that within the redshift range $0.8 \lesssim z \lesssim 1.2$ there is a good agreement between the photometric and spectroscopic redshifts, being $\sigma_{\Delta z}$ less than 0.05 for most galaxies (see horizontal dashed lines). It should be noted that this accuracy only applies to galaxies which are as bright as the sources in the spectroscopic survey. All the galaxies with spectroscopic redshift from zCOSMOS survey shown in Figure \ref{reliability} have $r$-band observed magnitudes typically brighter than 23.5 mag and, therefore, the reliability of the photometric redshifts can be guaranteed up to that limit. From now on, photometric redshifts are used for the sources without zCOSMOS counterpart. For those sources with a zCOSMOS spectrum, redo the SED fits and employ the results based on $z_{\rm spec}$.

\section{UV-selected galaxies at $z \sim 1$}\label{UV_08}

As it was commented in Section \ref{intro}, the choice of the red-ward and blue-ward filters determines the wavelength where the Lyman break is located and, therefore, the redshifts of the selected galaxies. In this work we aim at analyzing the physical properties of those LBGs whose Lyman break is located between the GALEX FUV and NUV filters, which are centered at 1528\AA\ and 2271\AA\ (in terms of their effective wavelengths) and have bandwidths of 1344-1786 \AA\ and 1771-2831 \AA, respectively. These wavelengths imply that the redshifts of these GALEX-selected LBGs are expected to be around $z \sim 0.95$ considering an intermediate wavelength of 1780 \AA\ between the two filters. 

\begin{figure}
\centering
\includegraphics[width=0.49\textwidth]{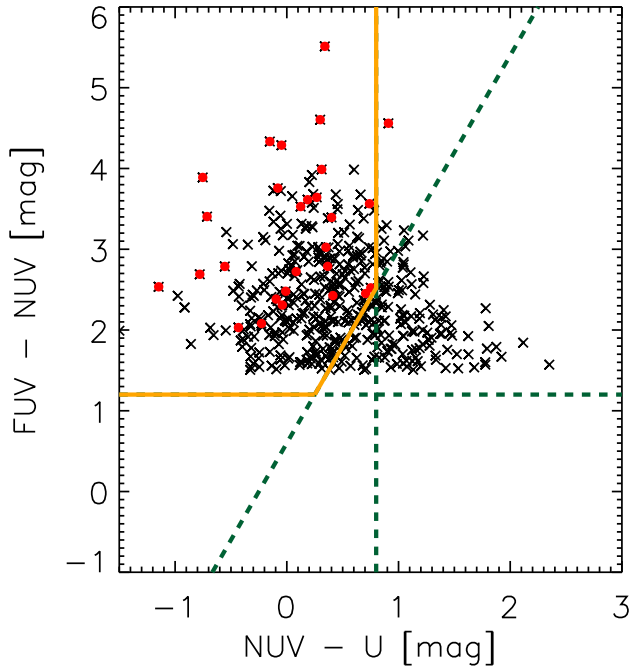}
\caption{Location of our GALEX-selected LBGs in a color-color diagram. The window enclosed by the orange solid lines is the selection region for LBGs at $0.6 < z < 1.4$ with $NUV < 23.75$. The subsample of our GALEX-selected LBGs which satisfies $NUV < 23.75$ are represented with red filled dots, while the remaining fainter LGBs are plotted with black symbols.
}
\label{barger}
\end{figure}

In order to formulate an analytic selection criterion to segregate our LBGs we use a large set of BC03 templates associated to a metallicity of $Z=0.2Z_\odot$, constant star formation rate, and different values of age and dust attenuation. We study the location of those templates in a color-color diagram as a function of redshift. To do that, each template is redshifted from $z = 0$ up to $z = 2$, then we apply the corresponding absorption in the IGM following the \cite{Madau1996} prescription, and obtain the $FUV-NUV$ synthetic observed colors by convolving the templates with the transmission curves of the GALEX filters. The result is shown in Figure \ref{GALEX}. As a general trend and as it could be expected by the location of the Lyman break as a function of redshift, the $FUV-NUV$ color increases with redshift up to $z \sim 1$. Looking at the different tracks represented in Figure \ref{GALEX}, we decide to impose a color cut of 1.5 (red dashed horizontal line) and, therefore, our color selection for GALEX-selected LBGs is:

\begin{equation}\label{criterion}
FUV-NUV>1.5
\end{equation}

\noindent It is important to note that the application of this criterion requires the detection of each galaxy in both FUV and NUV channels. Left panel of Figure \ref{GALEX} indicates that imposing such a color selection criterion we segregate galaxies located at $z \geq 0.7$ (this threshold is represented by the red dashed vertical line). However, at $0.7 \leq z \leq 0.8$, the FUV flux is importantly affected by the photons of the Lyman continuum (Lyc) i.e. those UV photons whose wavelengths are lower than the wavelength of the Lyman break. Therefore, if we really want to sample the Lyman break between the FUV and NUV without significant contamination of Lyc photons in the FUV filter we have to limit the redshift of the galaxies to $z \geq 0.8$. This situation is schematized in the right panel of Figure \ref{GALEX}. Then, we define, as a first approximation, GALEX-selected LBGs as those galaxies which are detected in both FUV and NUV channels and whose fluxes in each band meet the Equation \ref{criterion} and are located at $z \geq 0.8$. This sample is formed by 475 galaxies. 

It is worth remarking that there is a difference between the selection criterion that we apply here, and those applied to look for high-redshift LBGs. At $z \gtrsim 2$, LBGs are usually found employing not only the difference in color which characterizes the Lyman Break (Equation \ref{criterion}) but also a difference in color in redder wavelengths \citep[see for example][]{Steidel2003,Madau1996}. This is done for ruling out lower-redshift interlopers. This has to the done because at high redshift the photometric redshifts might suffer from large uncertainties and then, it is not always possible to select galaxies at a specific redshift range basing on $z_{\rm phot}$. However, in our case, as it was commented in Section \ref{zphot} and shown in Figure \ref{reliability}, we have accurate values of the photometric redshift for our UV-selected galaxies at $z \sim 1$ and, therefore, that supplementary condition is not needed. \cite{Barger2008} select LBGs at $0.6 \leq z \leq 1.4$ by employing a double color selection criterion combining $FUV-NUV$ and $NUV-U$. If we limit our sample in NUV magnitude to their same limit, $NUV < 23.75$, all but one of our GALEX-selected LBGs satisfy the double color selection criterion of \cite{Barger2008}. This is schematized in Figure \ref{barger}. The U-band data for the galaxies in the panel have been taken from the broad-band photometric catalog in the COSMOS field \citep{Capak2007}. Conversely, if we trust the photometric redshifts obtained from the combination of GALEX and ALHAMBRA data we find that with the selection criterion of \cite{Barger2008} we would miss a population of GALEX-selected LBGs fainter that $NUV=23.5$ mag whose $NUV-U$ color are typically redder that those GALEX-selected LBGs with $NUV < 23.5$ mag.

\begin{figure}
   \centering
   \includegraphics[width=0.45\textwidth]{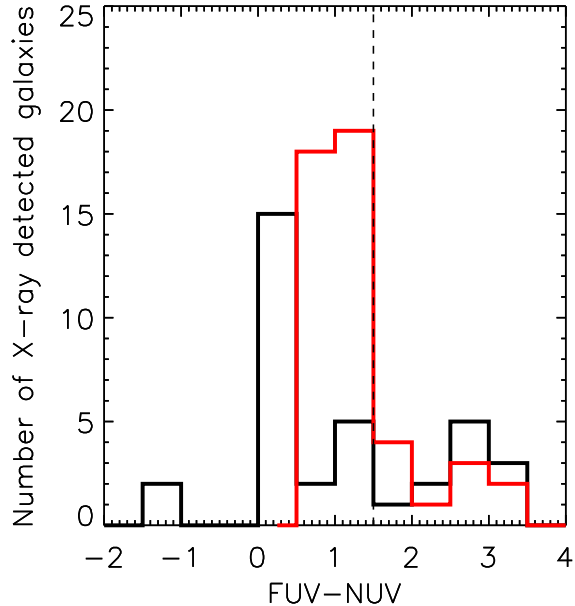}
\caption{Distribution of the $FUV-NUV$ color for Xray-detected galaxies at $0.8 \leq z \leq 1.2$ with GALEX and ALHAMBRA counterparts (black histogram). Red histogram represents the distribution of the $FUV-NUV$ color of the galaxies at $0.8 \leq z \leq 1.2$ spectroscopically classified as AGNs in \citet{Cowie2010} via emission line diagnosis. The vertical dashed line indicates the color threshold for selecting LBGs in this work. In this plot, only galaxies with detection in both $FUV$ and $NUV$ channels are included.
              }
\label{xray}
\end{figure}

\begin{figure*}
   \centering
   \includegraphics[width=0.22\textwidth]{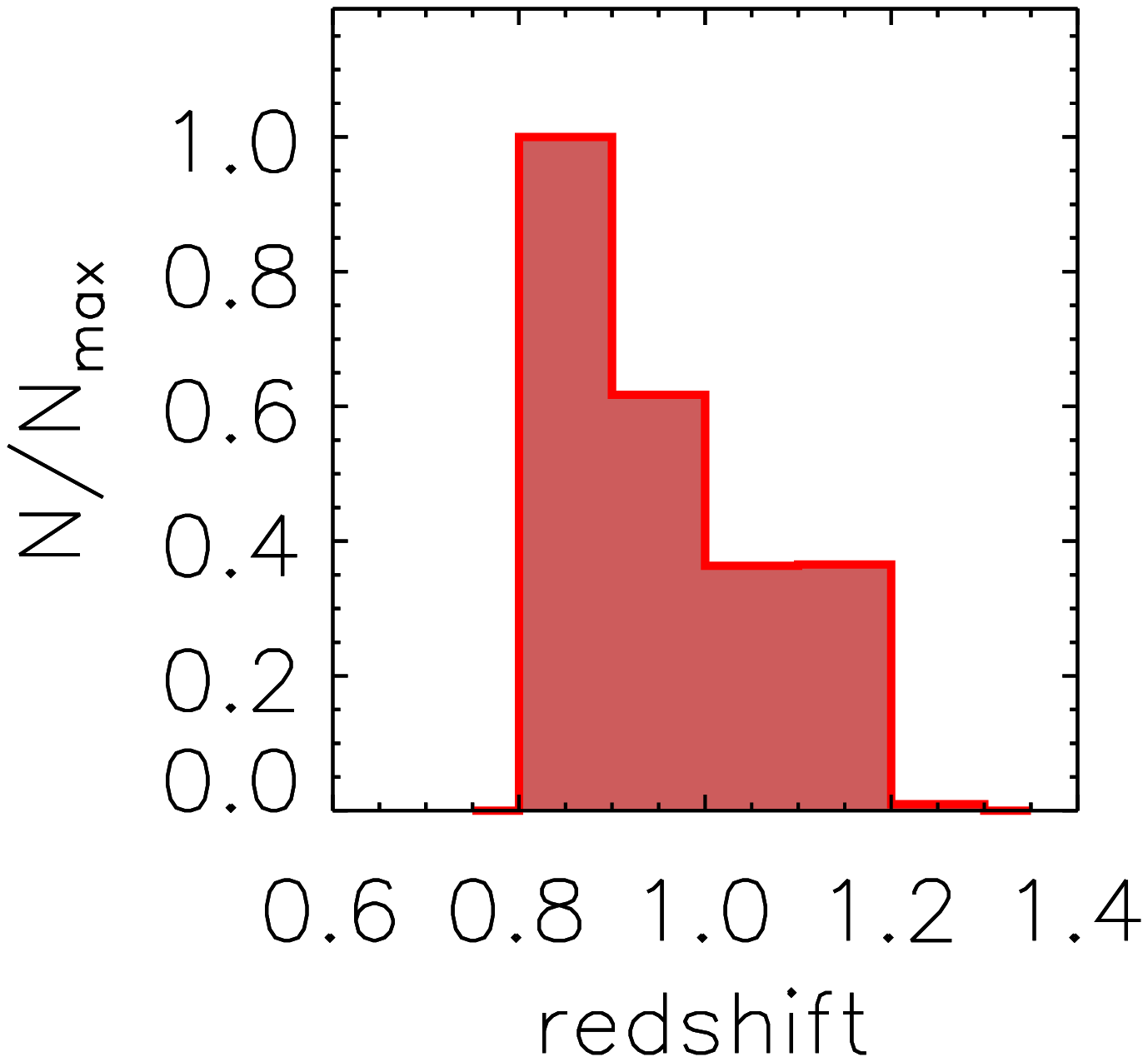}
   \includegraphics[width=0.22\textwidth]{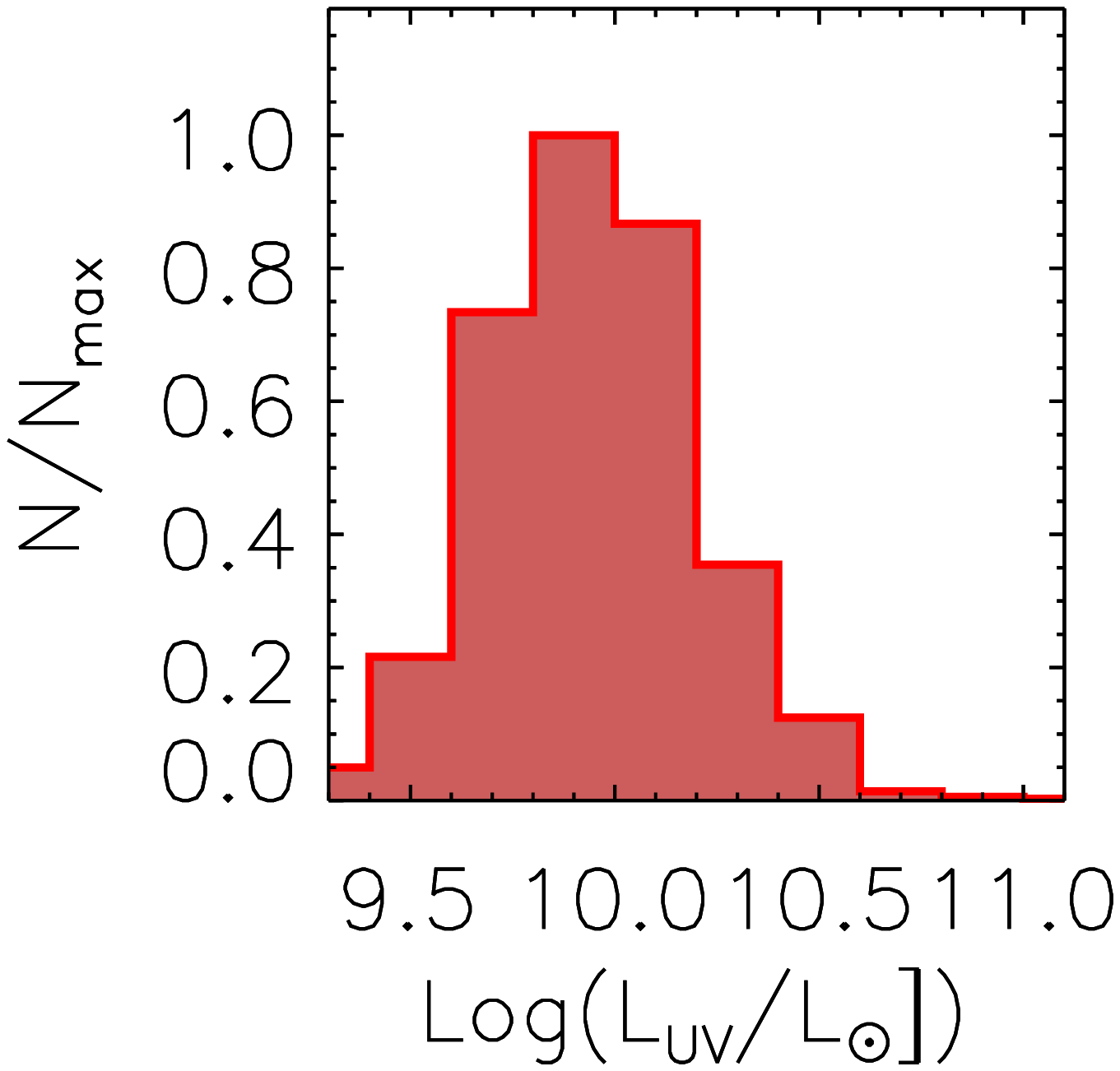}
   \includegraphics[width=0.22\textwidth]{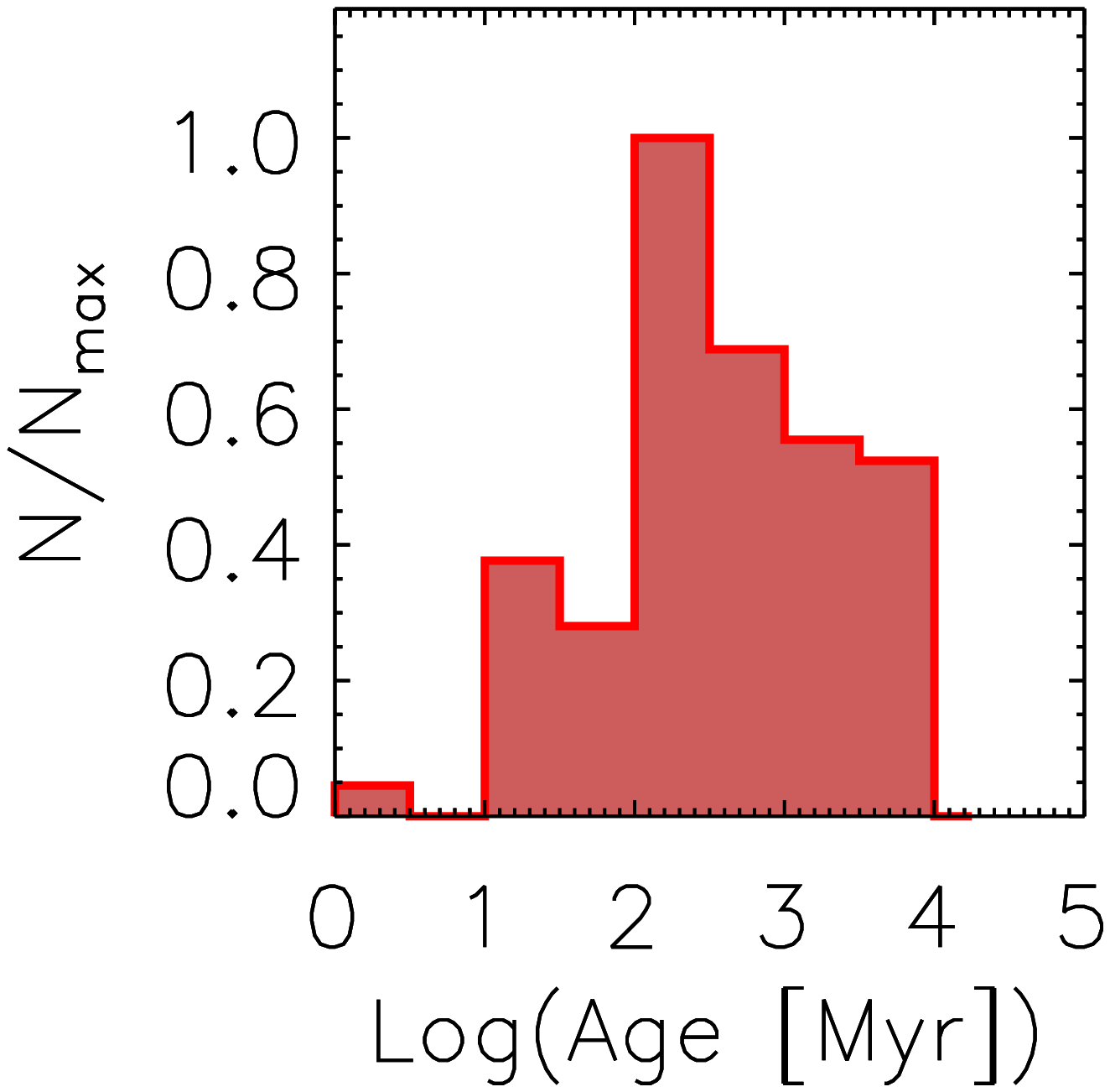}\\ 
   \includegraphics[width=0.22\textwidth]{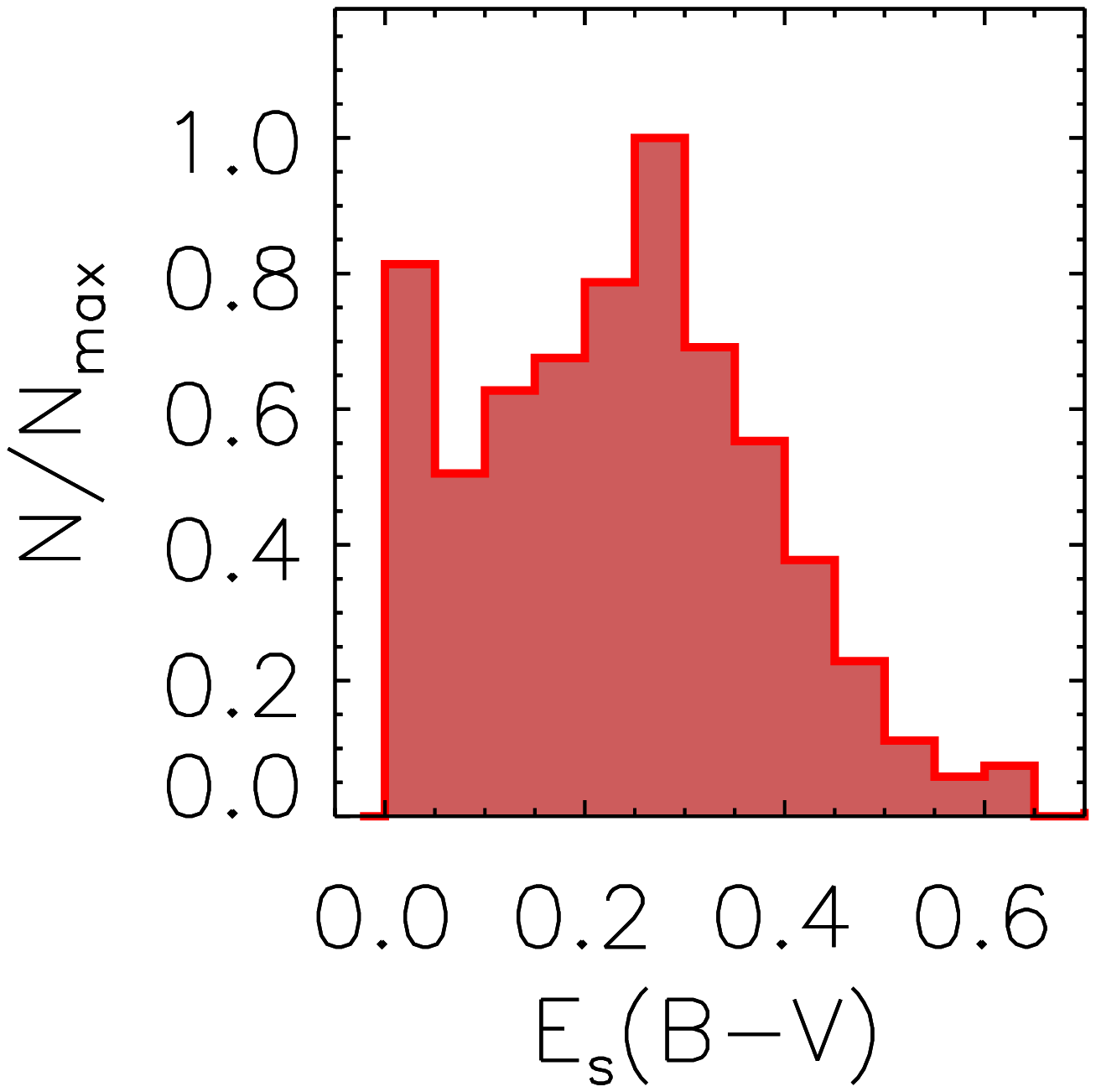}
    \includegraphics[width=0.22\textwidth]{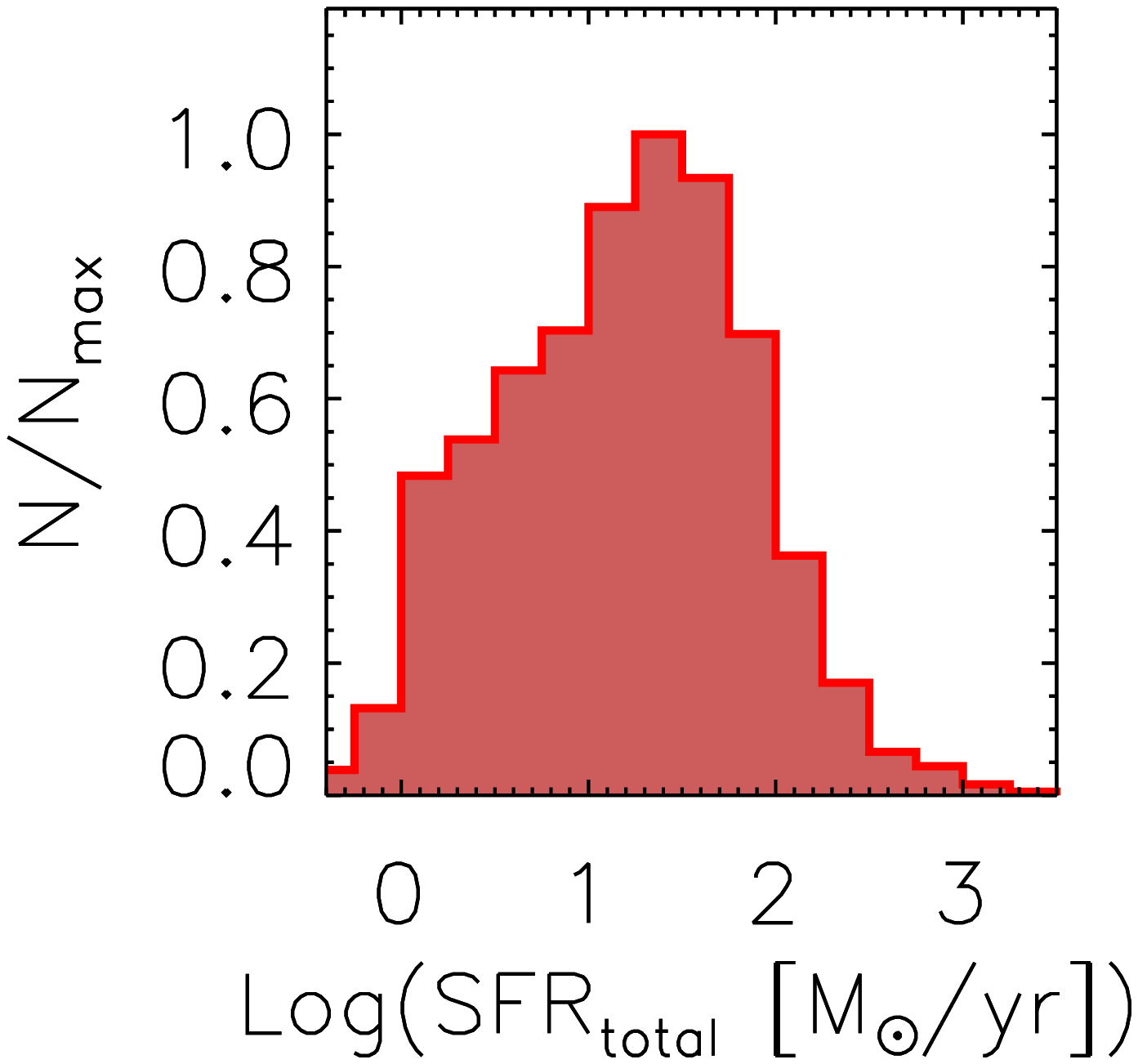}
    \includegraphics[width=0.22\textwidth]{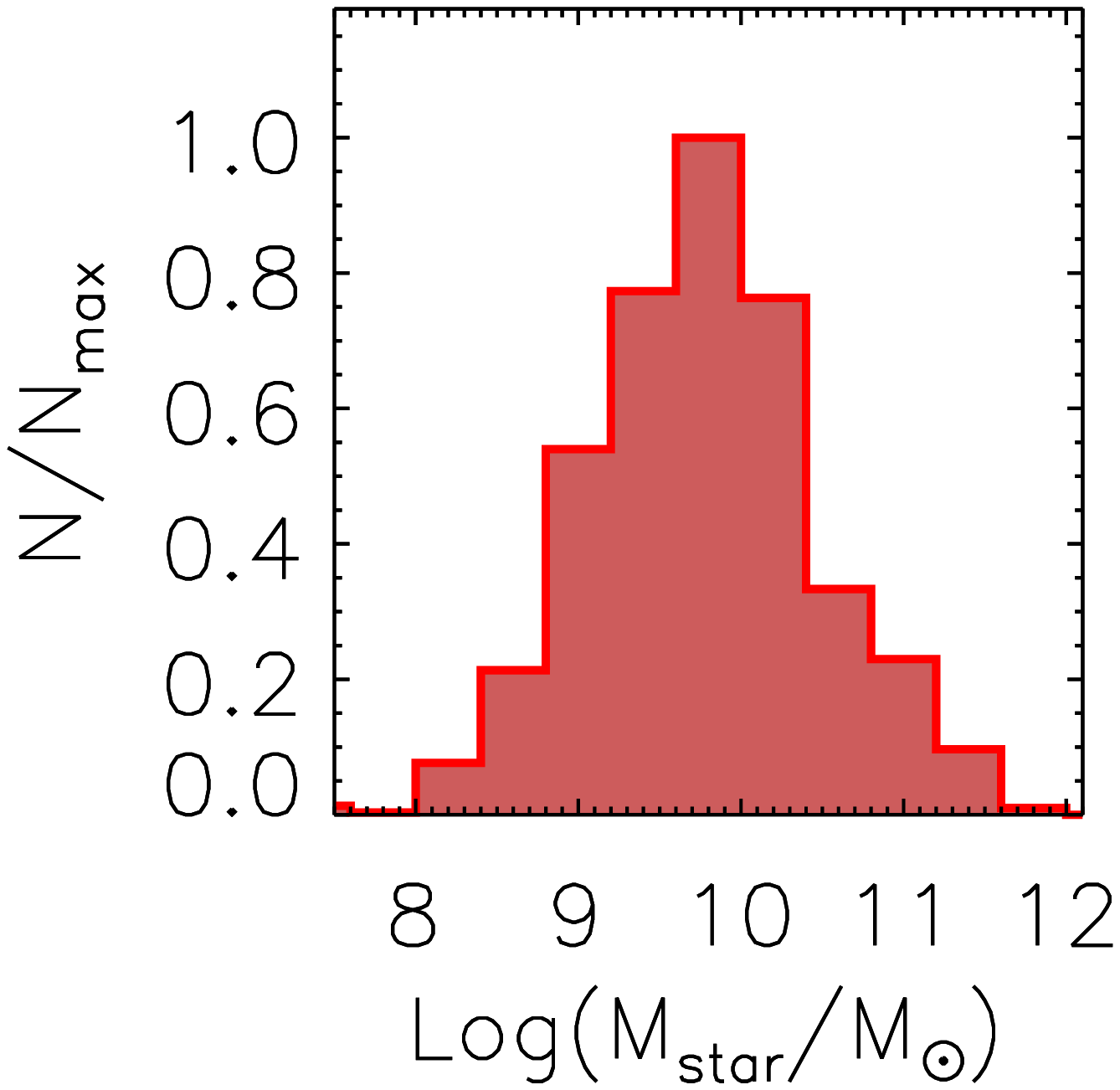}
    \includegraphics[width=0.22\textwidth]{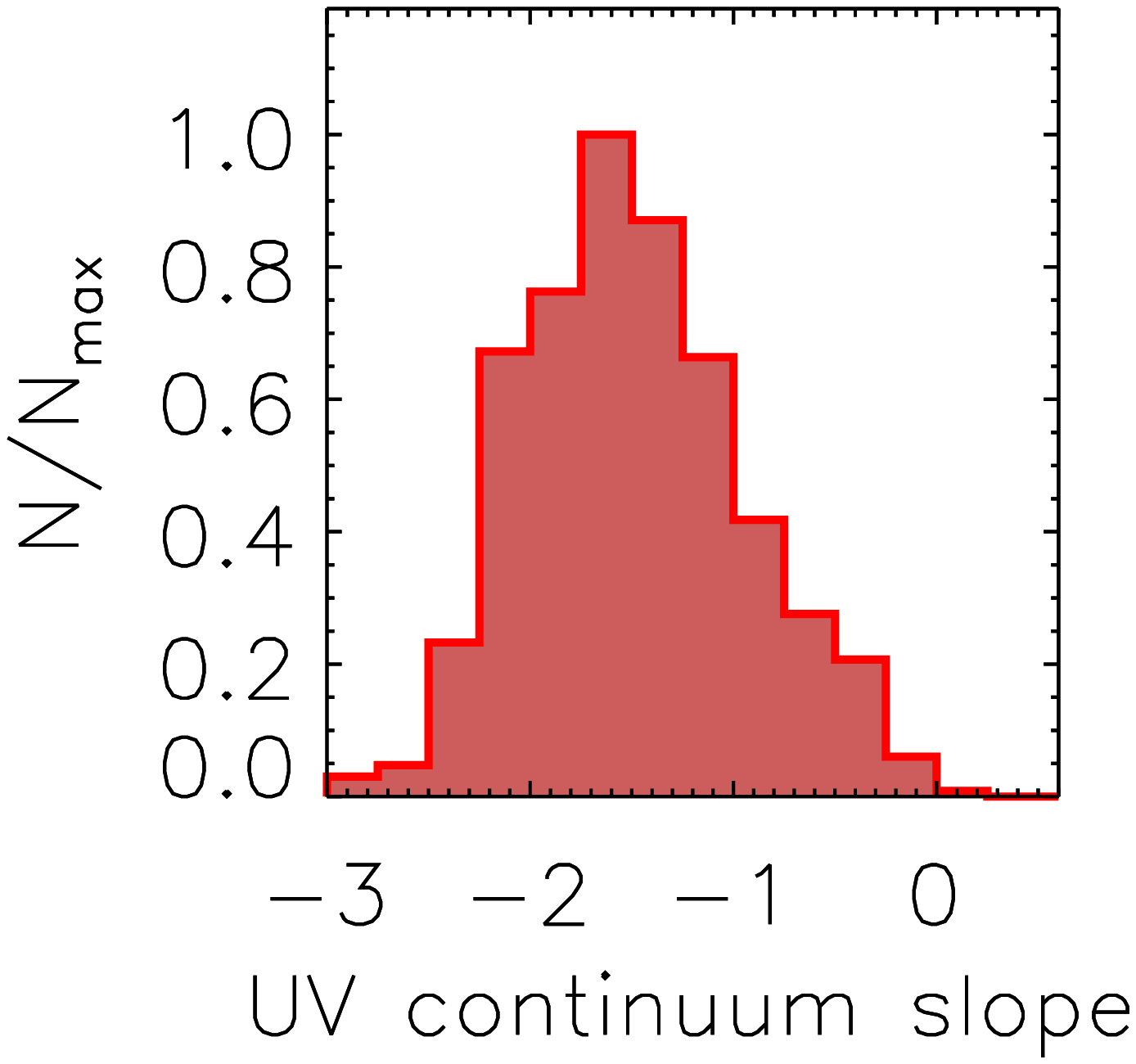}
\caption{From left to right and from top to bottom, distributions of redshift (photometric or spectroscopic), rest-frame UV luminosity, and SED-derived age, dust attenuation, dust-corrected total SFR, stellar mass, and UV continuum slope for our GALEX-selected LBGs. BC03 templates associated to a constant SFR, Salpeter IMF, and metallicity $Z=0.2Z_\odot$ are considered in the SED fits. Histograms have been normalized to their maxima.
              }
\label{stellar_populations}
\end{figure*}


The application of Equation \ref{criterion} requires the detection of the galaxies in both the FUV and NUV GALEX channels so that the amplitude of the break can be measured. However, it can also occur that a galaxy has such a strong Lyman break that is is detected in the NUV but undetected in the FUV channel. In order to include these FUV-undetected galaxies into the sample of GALEX-selected LBGs we have to ensure that the non-detection in FUV is caused by a strong Lyman Break. FUV observations in the COSMOS field have a limiting magnitude of $FUV \sim 26.5 \,{\rm mag}$. Galaxies brighter than that value in the wavelength range covered by the FUV filters should have been detected. Since we select LBGs with a $FUV-NUV$ color cut of 1.5, that limiting magnitude would imply a limit of 25 mag in the NUV channel. This way, we include in the previous sample of LBGs those galaxies that are brighter than 25 mag in the NUV channel, are at $0.8 \leq z \leq 1.2$, and are undetected in the FUV channel. This method for selecting FUV-undetected LBGs has been also applied, for example, in \cite{Burgarella2007}. With these galaxies included, we end up with an initial sample of 1246 GALEX-selected LBGs. A visual inspection of the galaxies with available ACS information (see Section \ref{morfo} for more details) reveals that the contamination due to the low spatial resolution of the GALEX images is lower than 5\%. Additionally, from this visual inspection we check that there is no stellar contamination in the derived LBG sample.

\subsection{X-ray counterparts and AGN contamination}

In this work we are only interested in those LBGs which are SF galaxies, we need to rule out the AGN contribution. To this aim, we look for CHANDRA X-ray detections \citep{Elvis2009} around 3'' \citep{Povic2009,Povic2012} of the ALHAMBRA-based spatial coordinates of our GALEX-selected LBGs. The area where our LBGs are located is almost totally covered by the CHANDRA footprint. We find that only 21 LBGs are detected in X-rays and, therefore, likely have an AGN nature. By using the catalog of AGNs in the COSMOS field \citep{Salvato2011} we do not find any extra AGN identification. The only AGN-LBGs represents an AGN contamination of about 2\%. 

Figure \ref{xray} represents (black histogram) the distribution of $FUV-NUV$ colors of the galaxies in the whole sample of UV-selected sources with measurements in $FUV$ and $NUV$ channels which are detected in X-rays, have GALEX and ALHAMBRA counterparts, and are at $0.8 \leq z_{\rm phot} \leq 1.2$ (the redshift range where our GALEX-selected LBGs are located). It can be seen that most X-ray-detected galaxies and, therefore, galaxies with an AGN nature, have $FUV-NUV$ colors below the color threshold utilized in this work for selecting LBGs (see Equation \ref{criterion}). This UV color distribution for AGNs at $z \sim 1$ explains the low percentage of AGNs among the GALEX-selected LBGs. We also plot in Figure \ref{xray} the $FUV-NUV$ color distribution of the galaxies spectroscopically classified as AGNs via emission line diagnosis in \cite{Cowie2010} and which are located at the same redshift range of our GALEX-selected LBGs. It can be seen again that most AGNs at $z \sim 1$ have $FUV-NUV$ colors below the color threshold considered in this work for selecting LBGs, reinforcing the fact that the LBG color selection does not tend to segregate galaxies with an AGN nature.

\begin{figure}
\centering
\includegraphics[width=0.49\textwidth]{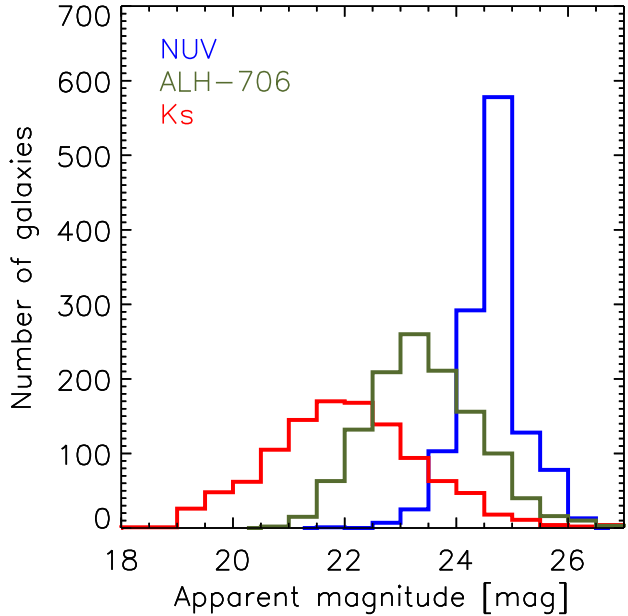}
\caption{Distribution of the NUV (blue histogram), ALH-706 (red histogram), and Ks (red histogram) apparent magnitudes for our GALEX-selected LBGs at $z \sim 1$. 
}
\label{magnitudes}
\end{figure}

Low values of the AGN contribution in samples of LBGs at different redshifts have been also reported. \cite{Lehmer2005} found AGN fractions of 1.2\%, 0.4\%, 0.3\% and 0.4\% in their sample of $U$-, $B_{435}$-, $V_{606}$-, and $i_{775}$-dropouts, respectively. \cite{Basu2011} reported an AGN fraction for their sample of LBGs at $0.5 < z < 2.0$ of 5\%-6\%, and \cite{Nandra2002} found an AGN contribution of about 3\% in their sample of LBGs at $z \sim 3$. In the subsequent analysis we do not take into consideration the GALEX-selected LBGs with an AGN nature. With this, we end up with a sample of 1225 SF GALEX-selected LBGs. This is the largest sample of LBGs studied at $z \sim 1$ so far. Figure \ref{SEDs_IRAC} shows the UV-to-near-IR SED of nine GALEX-selected LBGs in our final sample. This small subsample is representative of the whole sample of LBGs. It can be clearly seen that the combination of GALEX and ALHAMBRA provides an excellent coverage of the rest-frame UV continuum, Balmer break, and near-IR SEDs of these galaxies. Figure \ref{magnitudes} represents the distributions of the apparent brightness of our GALEX-selected LBGs in the NUV channel, optical ALH-706 ALHAMBRA filter, and in the near-IR Ks band. These distributions should be taken into account for comparing our results for our LBGs with those reported in other published studies that employ different photometric information. Our GALEX-selected LBGs at $z \sim 1$ have NUV magnitudes around 24.5 - 25.0 mag and optical magnitudes typically between 22 and 24.5 mag. The median value of their Ks magnitude is 22 mag. Furthermore, it can be seen that the spread in the magnitudes increases with the central wavelengths of the filters. Whereas the NUV magnitudes are mostly distributed within a range of 1 mag width, the Ks band magnitude spans from 20 to 24 mag.

\subsection{High-redshift analogues}\label{huh}

In this work, we also aim at comparing the SED-derived physical properties of LBGs at different redshifts. LBGs at high redshift (i.e. $z > 3$) tend to be intrinsically more luminous than those selected in the present work due to an observational bias. If we want to compare LBGs at different redshifts and, therefore, galaxies which are selected with similar selection criteria, we must limit the rest-frame UV luminosity of our GALEX-selected LBGs at $z \sim 1$ to the same range than that for LBGs at $z > 3$, which is typically $\log{\left( L_{\rm UV}/L_\odot \right)} \geq 10.2$. In this sense, we define \emph{UV-bright LBGs} as those LBGs at $0.8 \lesssim z \lesssim 1.2$ which have $\log{\left( L_{\rm UV}/L_\odot \right)} \geq 10.2$. This sub-sample is formed by 181 galaxies.

\subsection{UV-faint galaxies}

At the redshift range of our GALEX-selected LBGs there are many SF galaxies that are not selected via the dropout technique because they do not have a strong break between the FUV and NUV filters or they are undetected in the FUV channel and are not bright enough in the NUV filter to ensure a strong Lyman break between both filters. All these galaxies  will be called \emph{UV-faint galaxies}. This sample will not be studied in the present work but it will be used in an incoming work (Oteo et al. in prep.) where FIR observations will be used to constrain the FIR SED of both GALEX-selected LBGs and UV-faint galaxies. In that case, the comparison between FIR-detected LBGs and UV-faint galaxies will help to understand the galaxies which are selected under the dropout technique in opposition to other UV-fainter SF galaxies and to place LBGs in a more general scenario of SF galaxies at $0.8 \lesssim z \lesssim 1.2$.

\section{SED-derived stellar populations}\label{stellar}

\subsection{Physical properties of LBGs at $z \sim 1$}\label{properties}


Figure \ref{stellar_populations} shows with red shaded histograms the distributions of photometric/spectroscopic redshift, rest-frame UV luminosities, age, dust attenuation, dust-corrected total SFR, stellar mass, and UV continuum slope for our GALEX-selected LBGs. As a consequence of their selection criterion, our GALEX-selected LBGs are located at $0.8 \lesssim z \lesssim 1.2$ and have rest-frame UV luminosities $\log{\left( L_{\rm UV}/L_{\odot} \right)} > 9.6$. The median values of the SED-derived physical properties of our GALEX-selected LBGs are summarized in Table \ref{properties}. It can be seen that they are blue and young galaxies with moderate dust attenuation. Due to their brightness in the rest-frame UV, they have relatively high values of the UV-derived and dust-corrected total SFRs.

\begin{table}
\caption{\label{properties}SED-derived physical properties of the studied GALEX-selected LBGs at $z \sim 1$}
\centering
\begin{tabular}{lcc}
\hline\hline
Property & Median value & Width of the distribution \\
\hline
Age [Myr]								&	341		&	2206		\\
$E_s(B-V)$							&	0.20		&	0.14		\\
$SFR_{\rm UV} \, [M_\odot \, {\rm yr}^{-1}]$	&	1.90		&	2.63		\\
$SFR_{\rm total} \, [M_\odot \, {\rm yr}^{-1}]$	&	16.94	&	112.94	\\
$\log{\left( M_*/M_\odot \right)}$ 			&	9.74		&	0.75		\\
UV slope								&	-1.53		&	0.55		\\
\hline
\end{tabular}
\end{table}

\begin{figure*}
\centering
\includegraphics[width=0.3\textwidth]{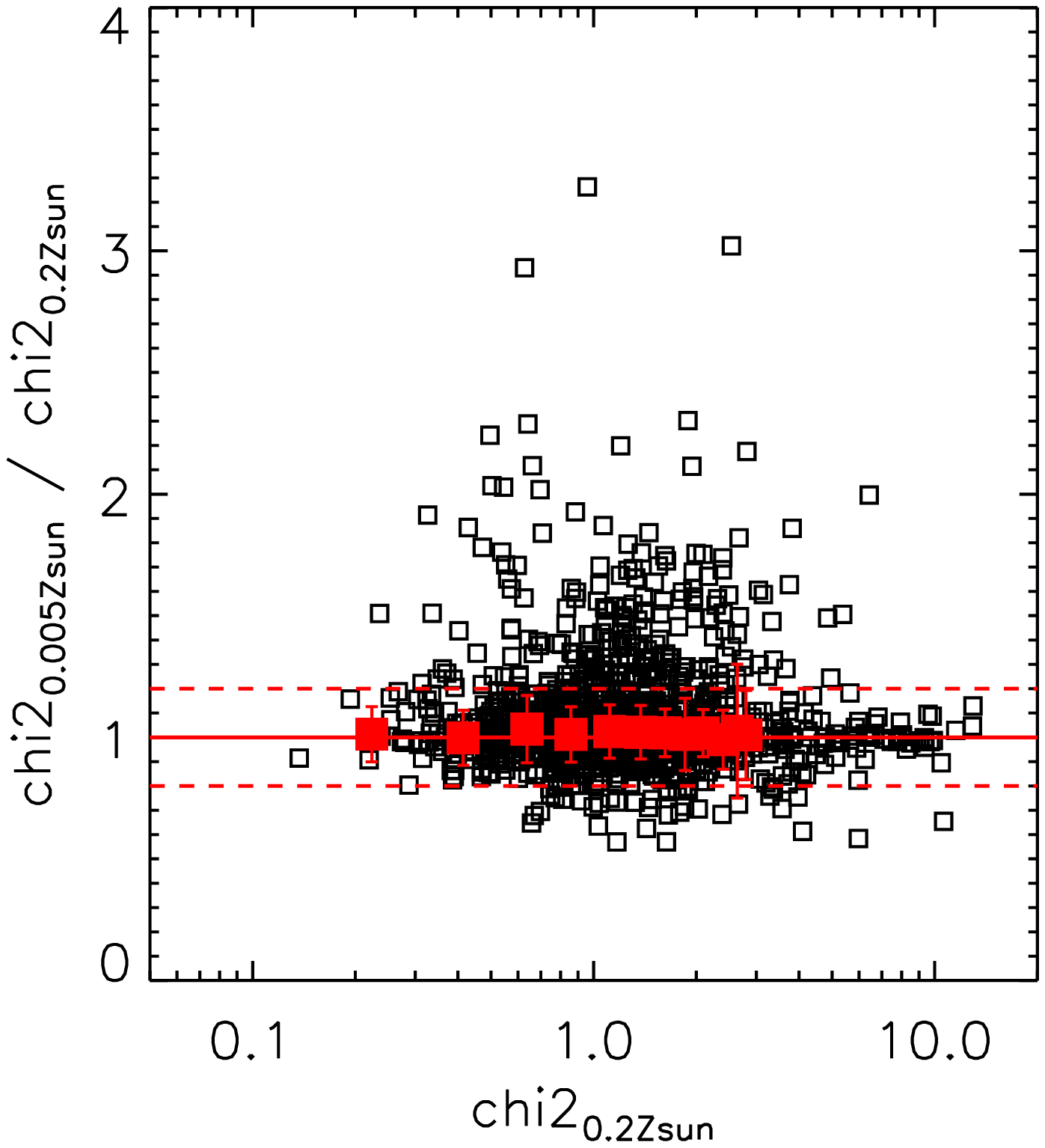}
\includegraphics[width=0.3\textwidth]{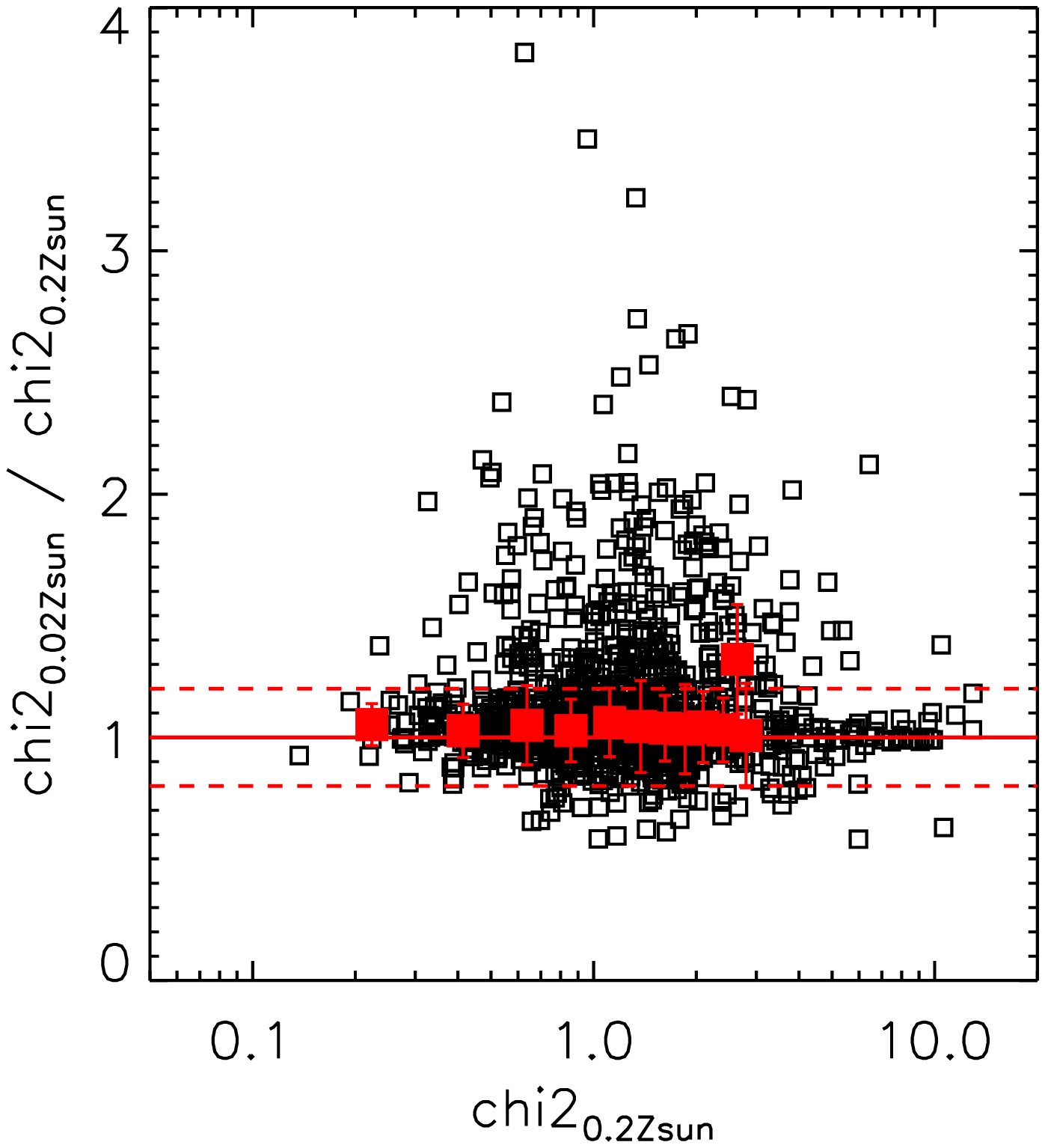}\\
\includegraphics[width=0.3\textwidth]{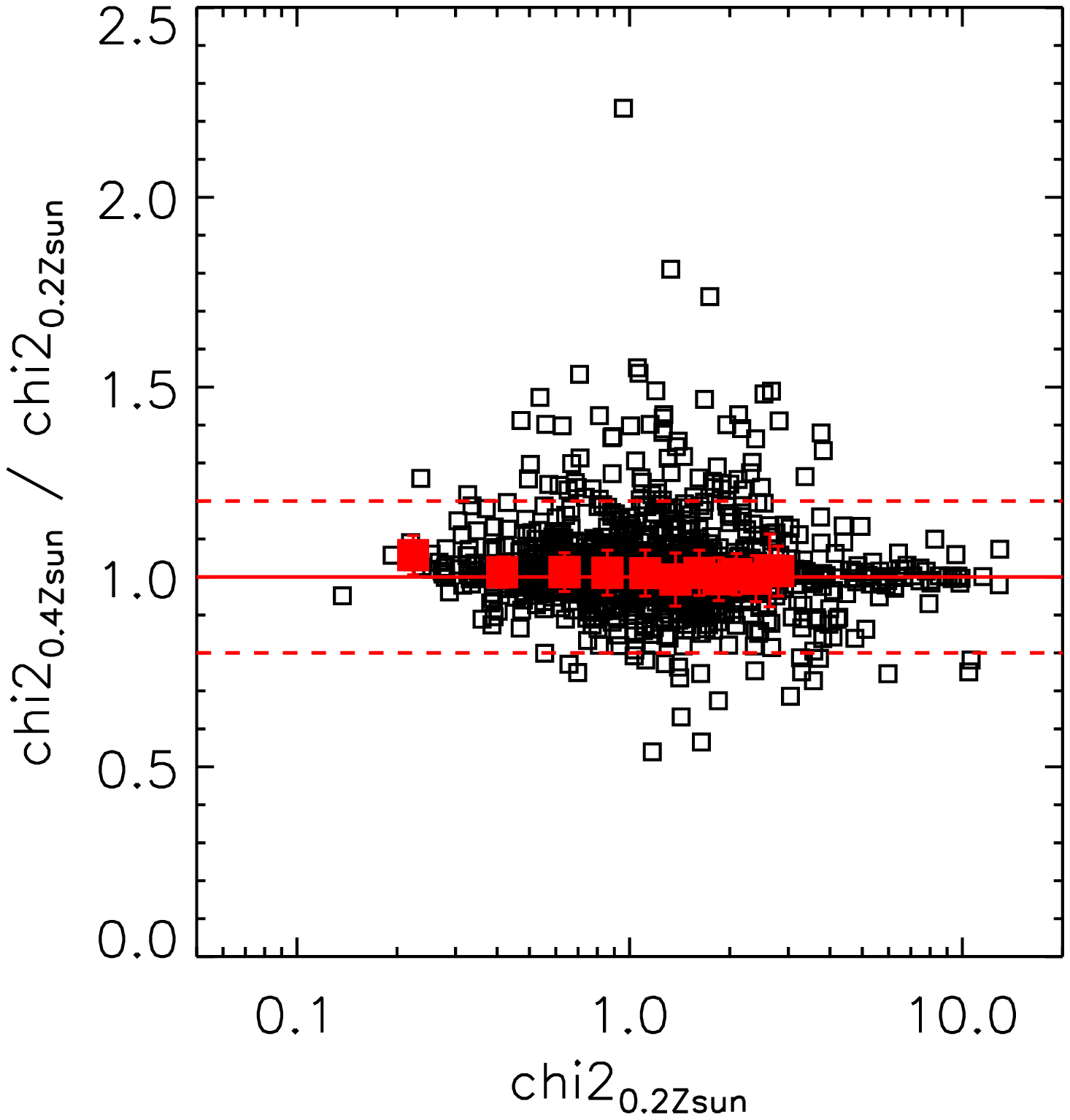}
\includegraphics[width=0.3\textwidth]{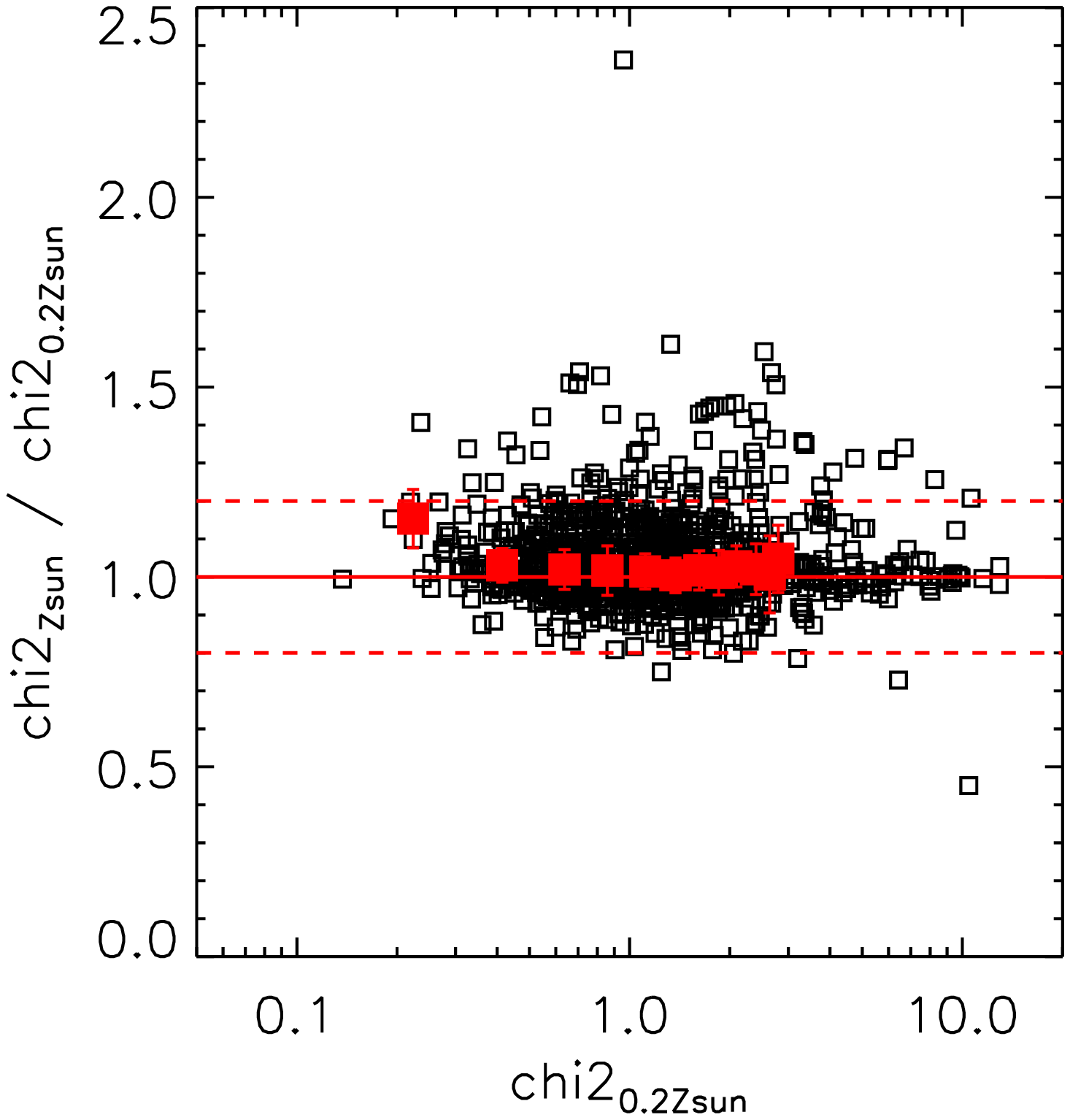}
\includegraphics[width=0.3\textwidth]{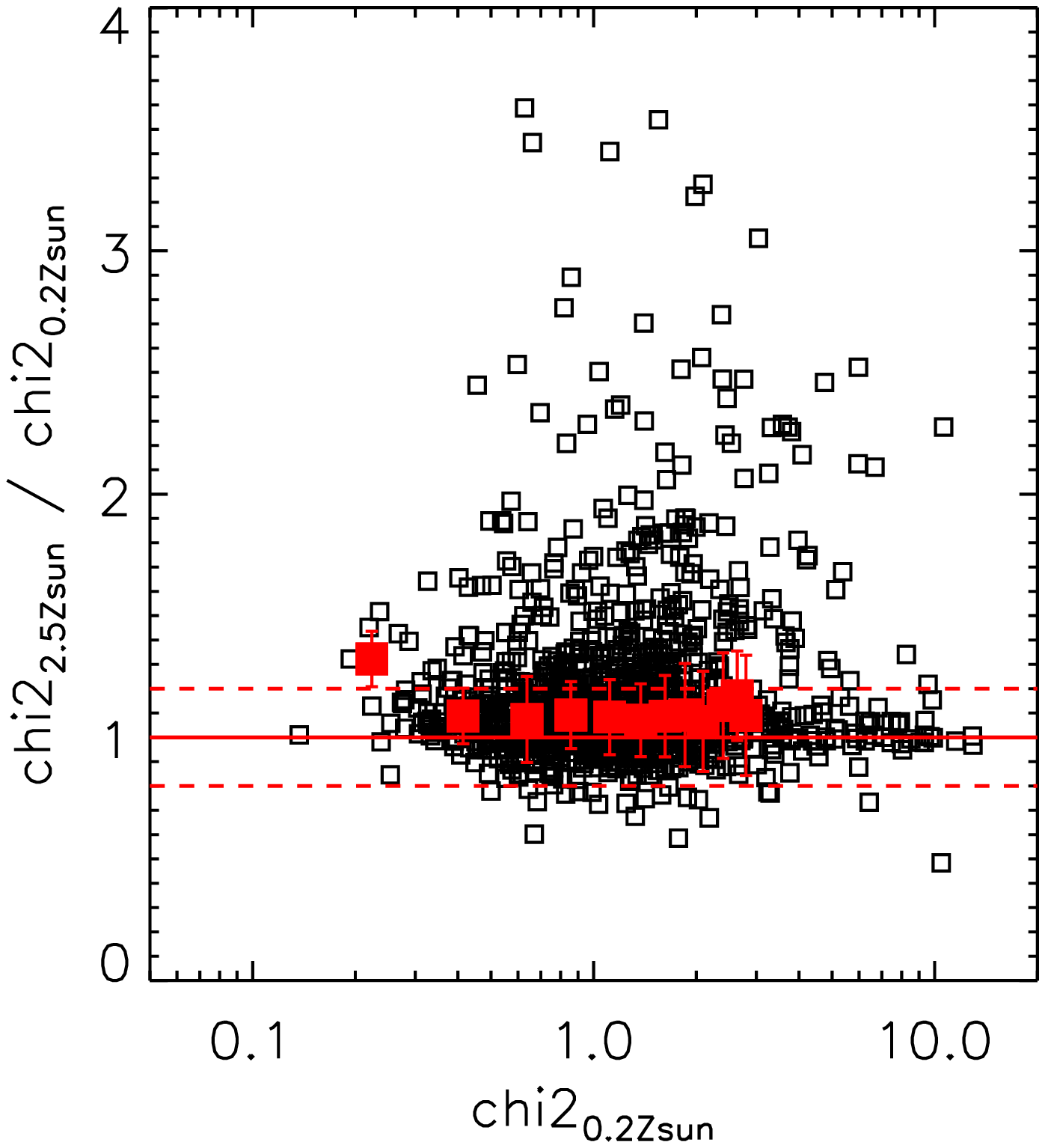}
\caption{Relation between the $\chi^2_r$ values of the SED-fitting results with \citet{Bruzual2003} templates built by assuming different metallicities. Red filled squares represent the the median $\chi^2_r$ ratios represented in the y-axis in different bins of the $\chi^2_r$ associated to SED-fitting results with metallicity $Z=0.2 Z_\odot$, the one adopted in this work for report the SED-derived properties of the studied galaxies. Red solid horizontal lines represent the one-to-one relation.
              }
\label{metal_Z1}
\end{figure*}


In its maximum likelihood mode, ZEBRA gives not only the best-fitted templates but also the probability that any of the others non-best-fitted templates can represent the photometric SED of a given galaxy. This probability can be used for deriving the uncertainties of the SED-derived parameters. To this aim, we define the weighted average ($WA$) of a given SED-derived physical property as: $WA=\sum_i^N P_i f_i/N$, where $P_i$ is the probability that a given template, $i$, can represent the observed SED of a given galaxy, $f_i$ is the value of one of the physical properties associated to the $i-th$ template, and $N$ is the number of templates. If the best-fitted template of a given galaxy has much higher probability to represent its observed SED than any of the other templates, the average WA of a given physical property would be quite similar to the best-fitted one and the uncertainty of that property should be low. On the other hand, if several templates associated to very different values of a given property have similar probability to represent the observed SED of a given galaxy, the WA would be dissimilar to the best-fitted value and the uncertainty should be high. Following this idea, we define the uncertainty of a given parameter as the difference between the best-fitted value and its corresponding weighted average.

\begin{figure*}
\centering
\includegraphics[width=0.24\textwidth]{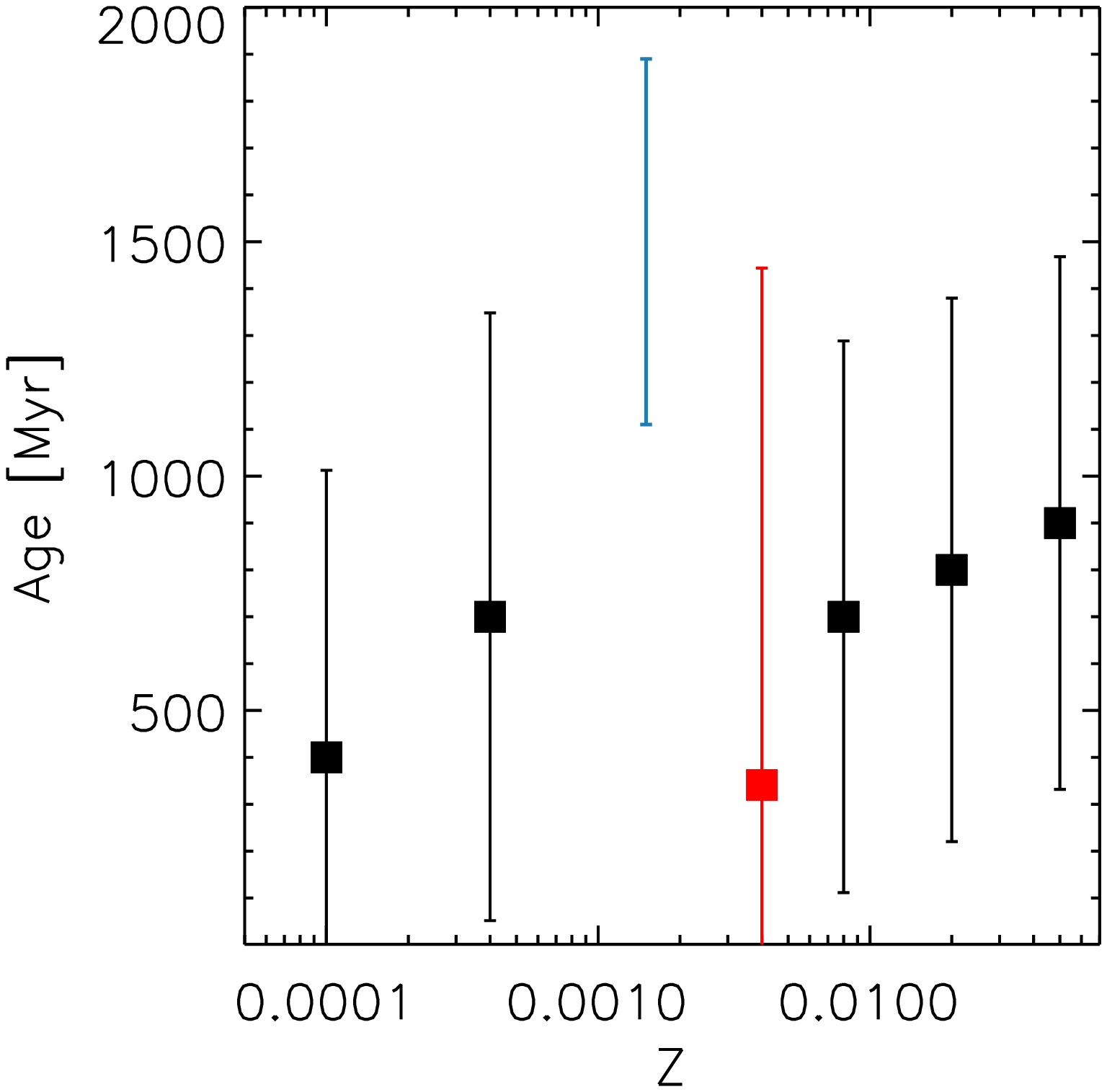}
\includegraphics[width=0.24\textwidth]{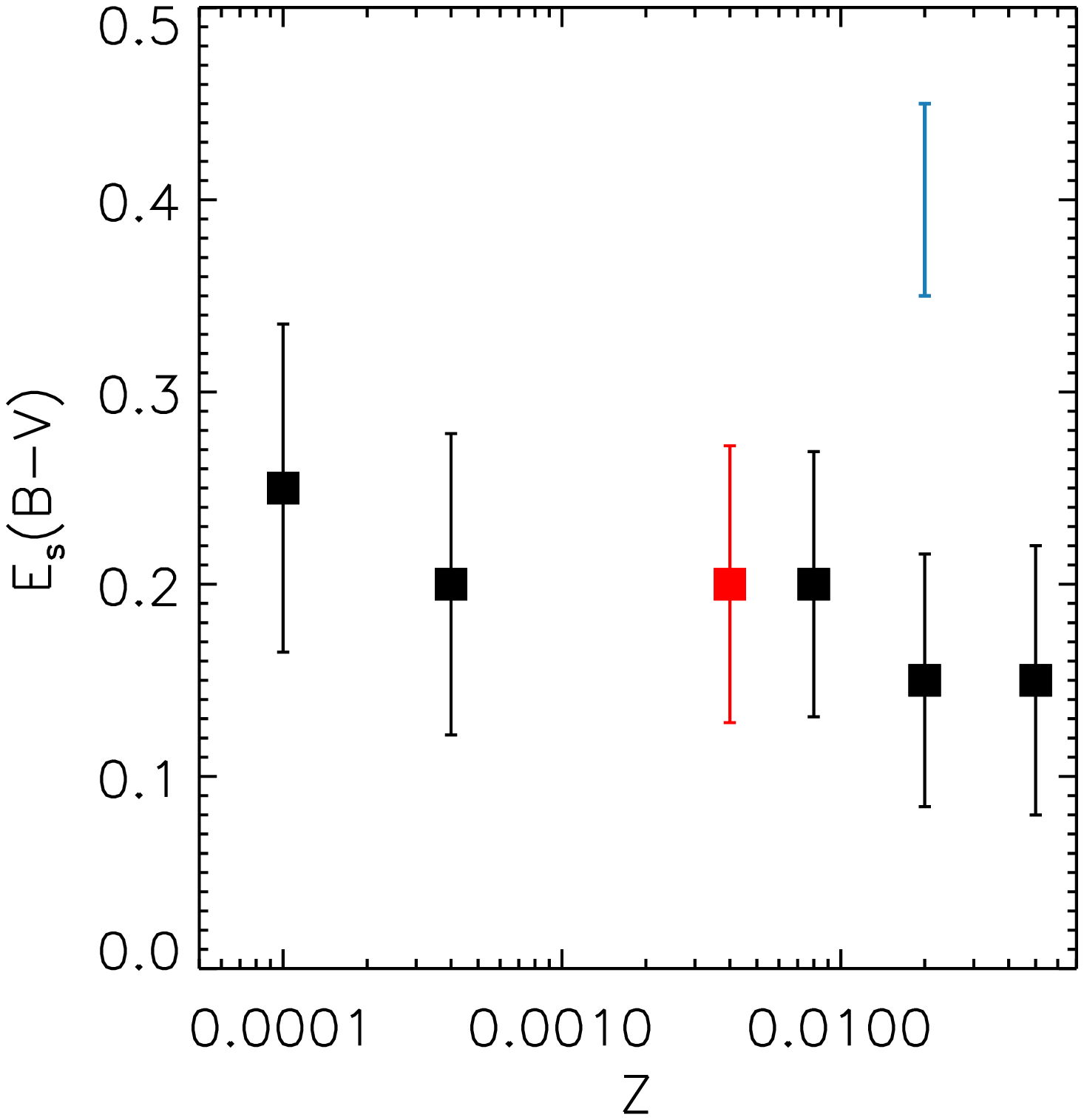}
\includegraphics[width=0.24\textwidth]{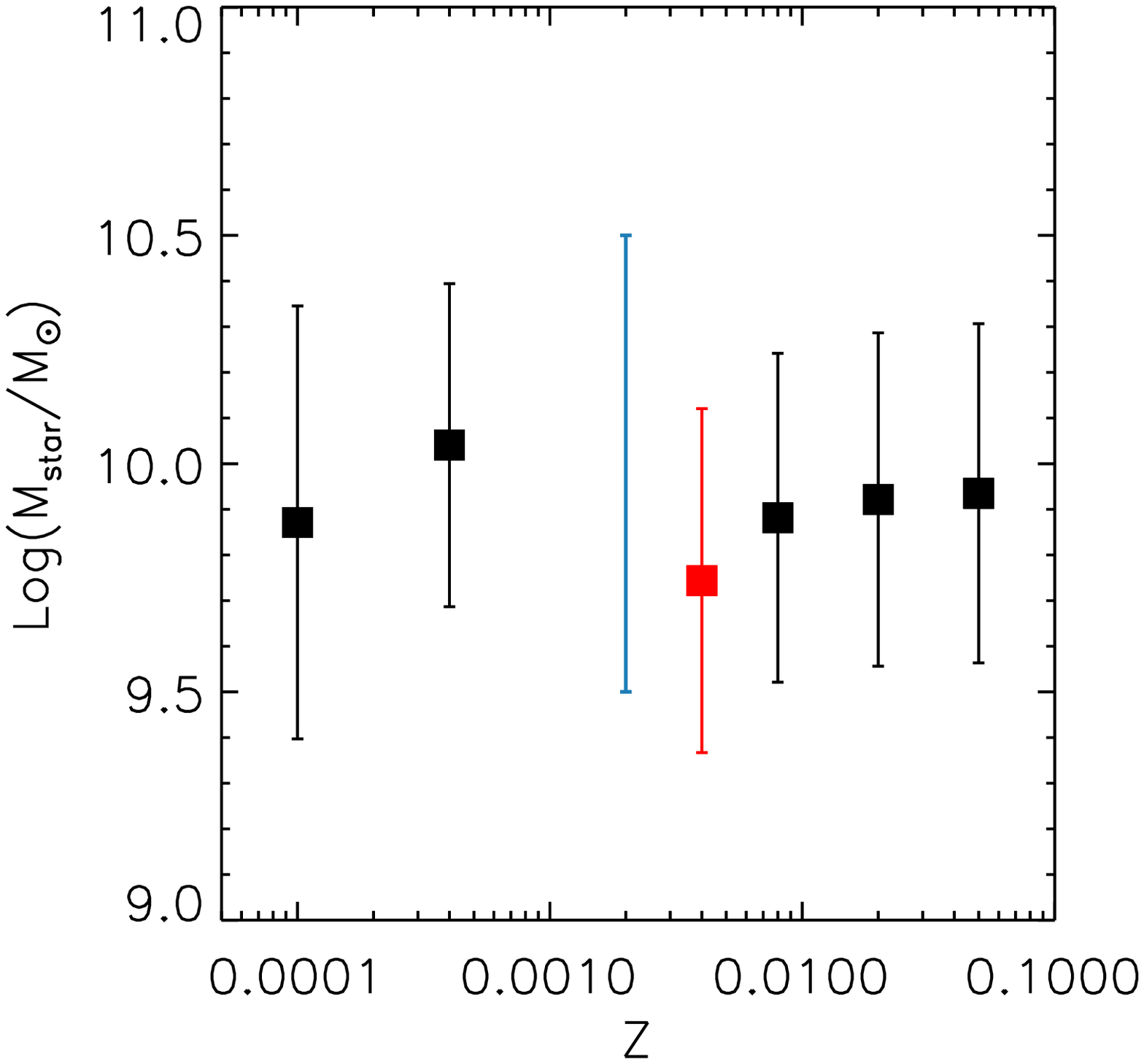}
\includegraphics[width=0.24\textwidth]{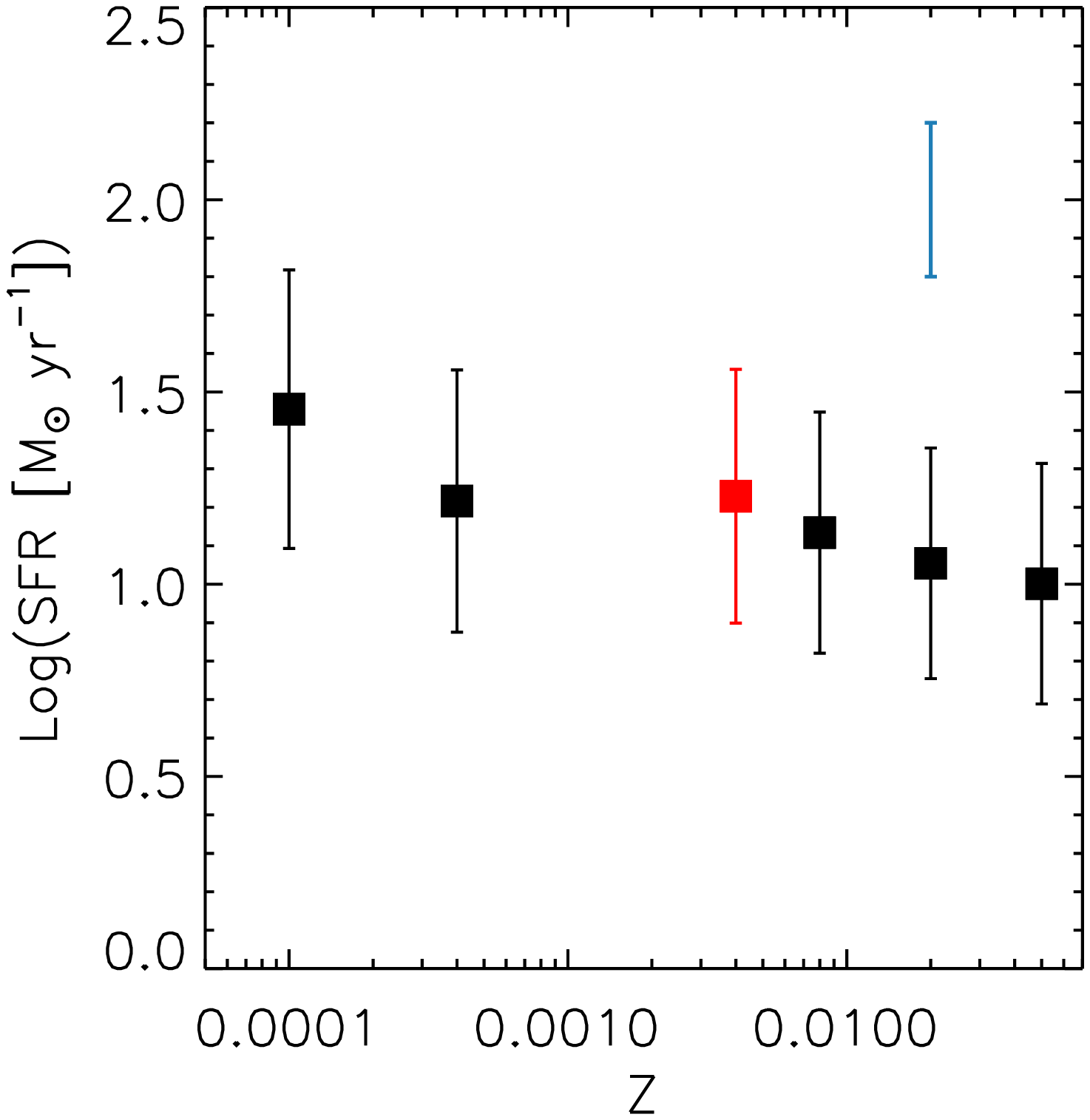}
\caption{Differences in the SED-derived age, dust attenuation, stellar mass, and dust-corrected total SFR when assuming \citet{Bruzual2003} templates associated to different metallicities. The dust-corrected total SFR is the one obtained by correcting the rest-frame UV luminosity with the SED-derived dust attenuation, $E_s(B-V)$. The squares represent the median values over the whole sample of GALEX-selected LBGs at $z \sim 1$ of each parameter for each value of the metallicity considered in the SED fits. The red symbols represent the results for the metallicity adopted in this work for reporting the SED-derived properties of our galaxies. Error bars indicate the width of the distributions of each parameter for each value of metallicity. The typical uncertainties associated to each parameter are shown with blue bars.
              }
\label{metal_Z2}
\end{figure*}

The physical parameters intrinsically related to the BC03 templates considered in this work are the age and the dust attenuation. The UV-derived dust-uncorrected SFR is obtained from the normalization of each observed template to the observed SED and the stellar mass is obtained from the values of age, dust attenuation, and dust-uncorrected SFR. This way, the procedure outlined above for deriving the uncertainties should be first applied to age and dust attenuation. As result, we obtain that the median value of the uncertainties of age and dust attenuation for our LBGs are $\Delta$Age = 390 Myr and $\Delta E_s(B-V) = 0.05$, respectively. The typical uncertainty of the SED-derived age is of the same order that the median age of our GALEX-selected LBGs. This implies that, even with the exceptional photometric coverage of the ALHAMBRA survey, which samples quite well the Balmer break of SF galaxies at $z \sim 1$, the age is a parameter difficult to determine accurately with an SED-fitting procedure. The uncertainty in the SFR are directly related to the uncertainties of the rest-frame UV luminosity, and thus to the photometric uncertainties of the sources. The rest-frame 1500 \AA\ is sampled by the NUV filter at $z \sim 1$. The typical uncertainties of the NUV magnitudes for the galaxies in our sample is 0.15 This value translates into a luminosity uncertainty of $\Delta \log{\left( L_{\rm UV} / L_\odot \right)} \sim 0.06$ and an UV-derived SFR uncertainty of $\Delta SFR = 0.15 M_\odot$/yr. Propagating the uncertainties of age, dust attenuation, and UV-derived SFR, and age we obtain typical uncertainties of 0.5 and 0.2 dex for $\log{M_*/M_\odot}$ and $\log{SFR_{\rm total}}$, respectively. The high uncertainty of the SED-derived stellar mass is a consequence that it comes from the combination of the uncertainties of age and dust attenuation. The uncertainties in the UV continuum slope are considered as those obtained in the power law fitting and turn out to have a median value of 0.15.


\cite{Nilsson2011_LBG} studied the optical SED-derived physical properties of a sample of 15 GALEX-selected LBGs at $z \sim 1$, taken from \cite{Burgarella2007}, by fitting with BC03 templates their photometric points derived from ACS slitless grism spectra. They obtain similar distributions for dust attenuation and stellar mass than those reported in the present work, but their ages tend to be lower.  In their work, \cite{Nilsson2011_LBG} employed BC03 templates which are a combination of two SSP models instead of the models with constant SFR assumed in the present work and the ages reported are the youngest ones associated to the two SSP models used in the fittings. This is likely the reason why the age distribution of \cite{Nilsson2011_LBG} is shifted towards lower values. \cite{Basu2011} studied the physical properties of a sample of 50 LBGs at $0.5 < z < 2$ selected with the \emph{SWIFT} ultraviolet/optical telescope. They reported lower values of stellar masses, $\langle \log{M_*/M_\odot} \rangle = 9.4 \pm 0.6$, than those obtained here and with a similar dust attenuation distribution. In this case, the differences are likely due to the fact that their sample contains galaxies at higher redshifts than our LBGs caused by the different red-ward and blue-ward filters adopted to segregate the galaxies.

\subsection{Dependance on the assumed metallicity}\label{metal}

In the elaboration of the BC03 templates employed to report the SED-derived physical properties of our GALEX-selected LBGs we adopted a fixed value of metallicity, $Z=0.2Z_\odot$. In this section, we explore the differences in the SED-derived physical properties of the studied galaxies when considering BC03 templates associated to different values of metallicity. Furthermore, we analyze whether it is possible to constrain the metallicity of our studied galaxies with an SED-fitting procedure with the exceptional photometric coverage that the combination of GALEX and ALHAMBRA data provides. To this aim, we built five more sets of BC03 templates associated to the other metallicities available in the software \verb+GALAXEV+ and redo the SED fits. In this process, we assume the same IMF, IGM absorption and consider the same sampling in age and dust attenuation than in Section \ref{combination}. We also employ a constant SFH. We then retained the values of $\chi^2_r$, age, dust attenuation, and stellar mass obtained with the fits for the different metallicities. We first compare the ratios between the $\chi^2_r$ associated to the SED-fitting results with set of templates of different metallicities for individual galaxies (see Figure \ref{metal_Z1}). It can be seen that the ratios in the $\chi^2_r$ are very similar to unity and, consequently, we cannot distinguish between different metallicities with our photometric SEDs built with GALEX and ALHAMBRA data. In order to examine the implications of this degeneracy in the SED-derived properties of our galaxies, we represent in Figure \ref{metal_Z2} the median age, dust attenuation, stellar mass, and dust-corrected total SFR of our galaxies as a function of the metallicity assumed in the SED fits. As can be seen, the difference in the SED-derived parameters are within their typical uncertainties for most of the cases. Since we cannot discriminate different values of metallicity, the difference between the median values shows the intrinsic uncertainties in the SED-fitting procedure.

\subsection{Dependance on the assumed SFH}\label{SFH}

\begin{figure*}
   \centering
   \includegraphics[width=0.33\textwidth]{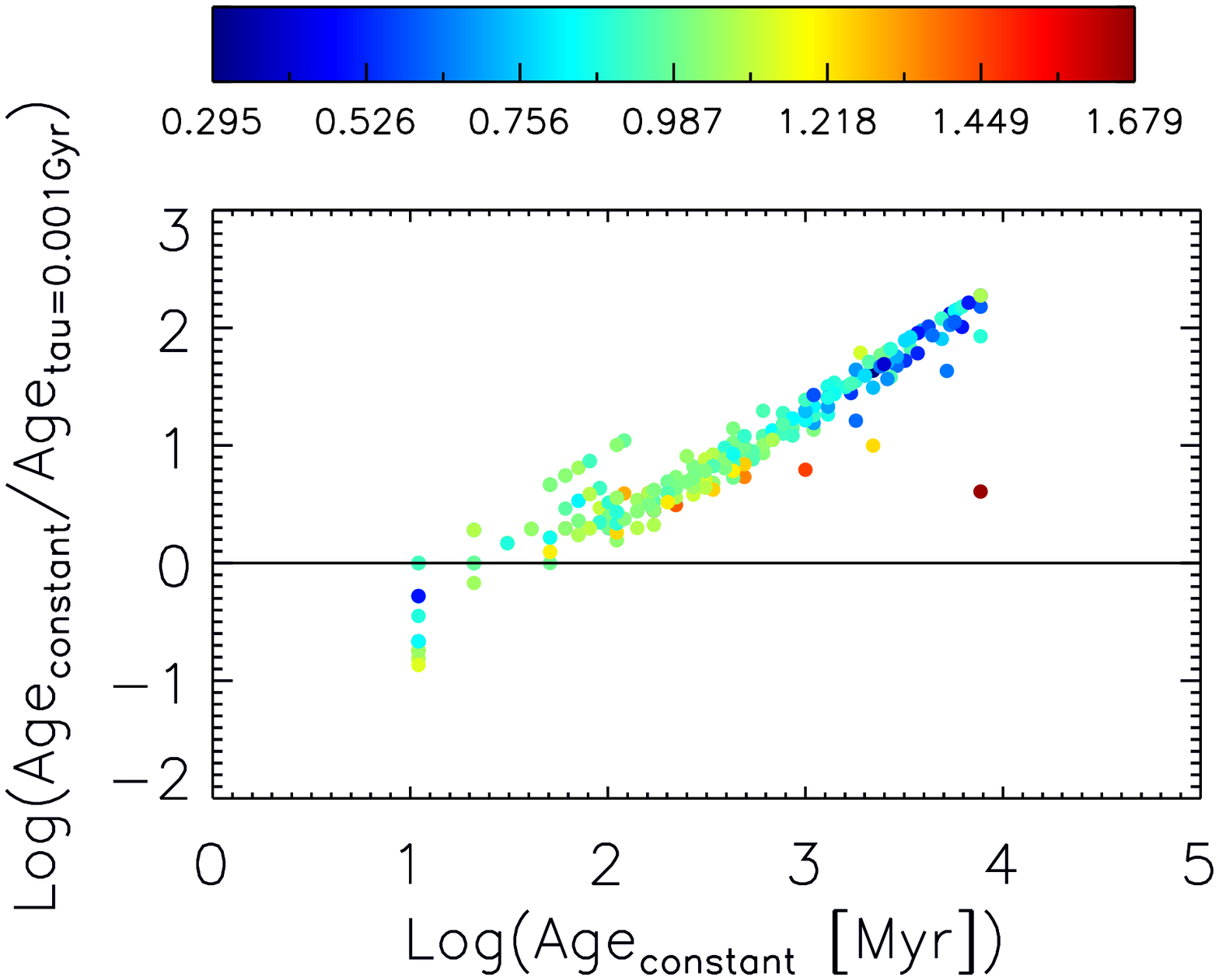}
   \includegraphics[width=0.33\textwidth]{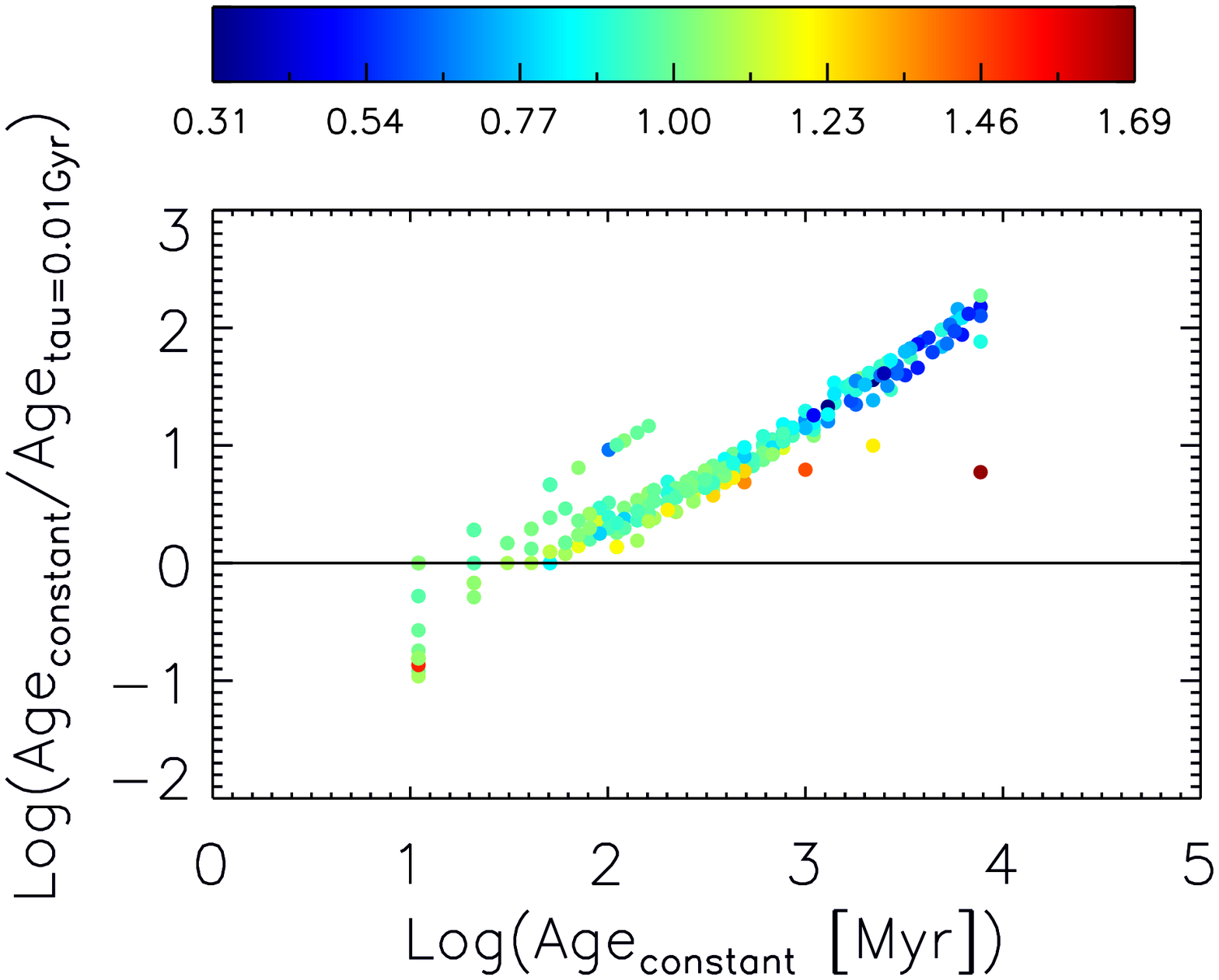}\\
   \includegraphics[width=0.33\textwidth]{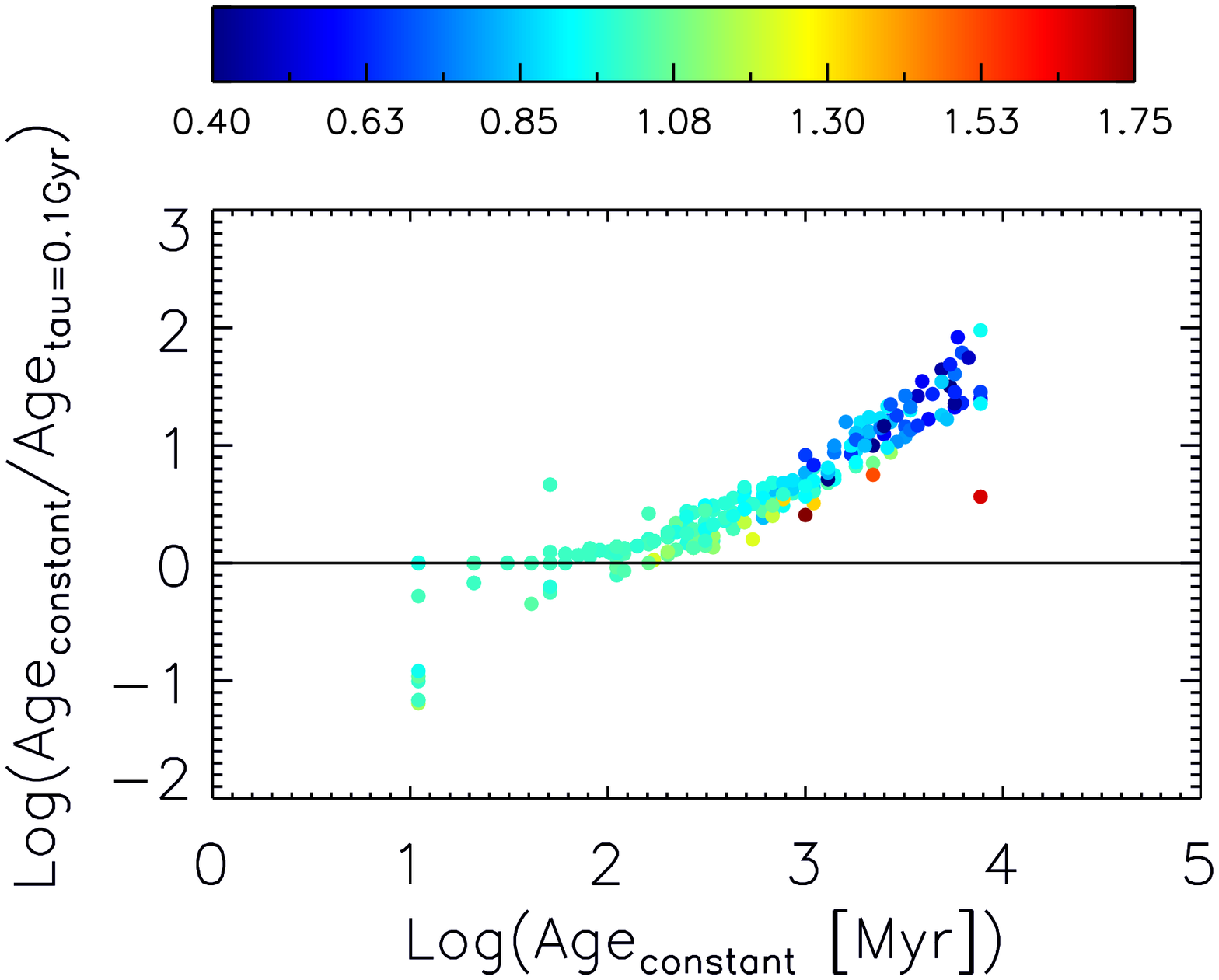}
   \includegraphics[width=0.33\textwidth]{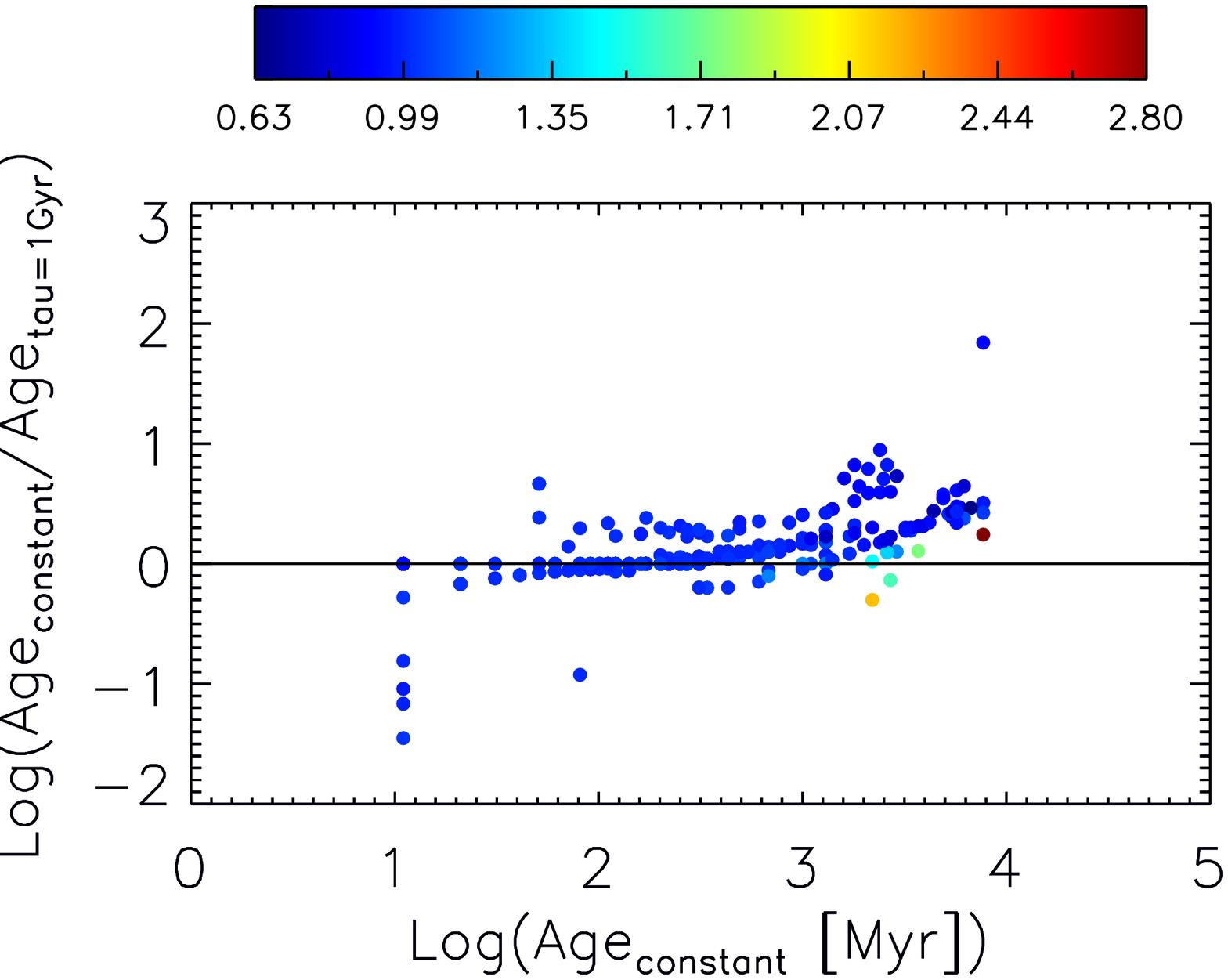}
   \includegraphics[width=0.33\textwidth]{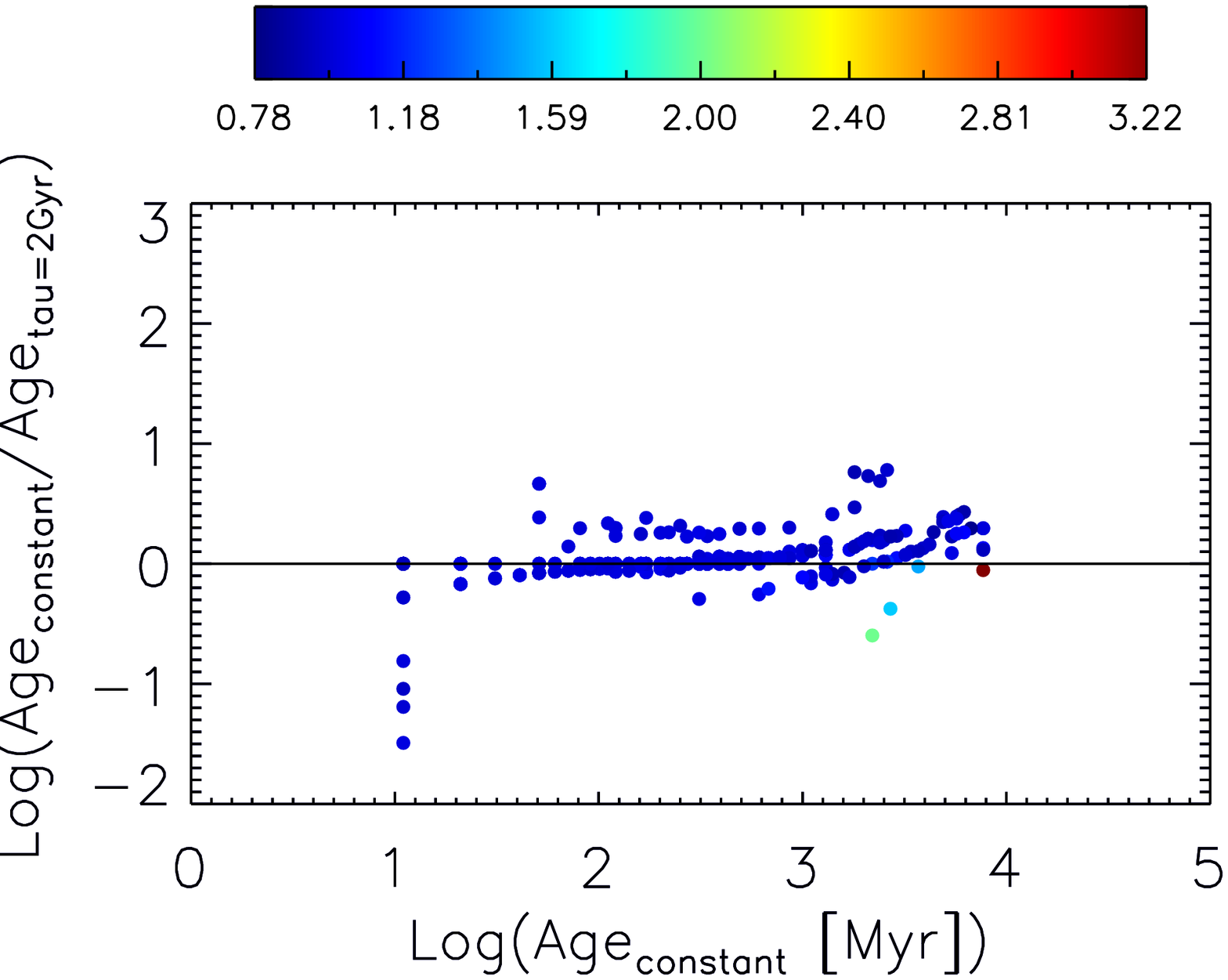}\\
   \includegraphics[width=0.33\textwidth]{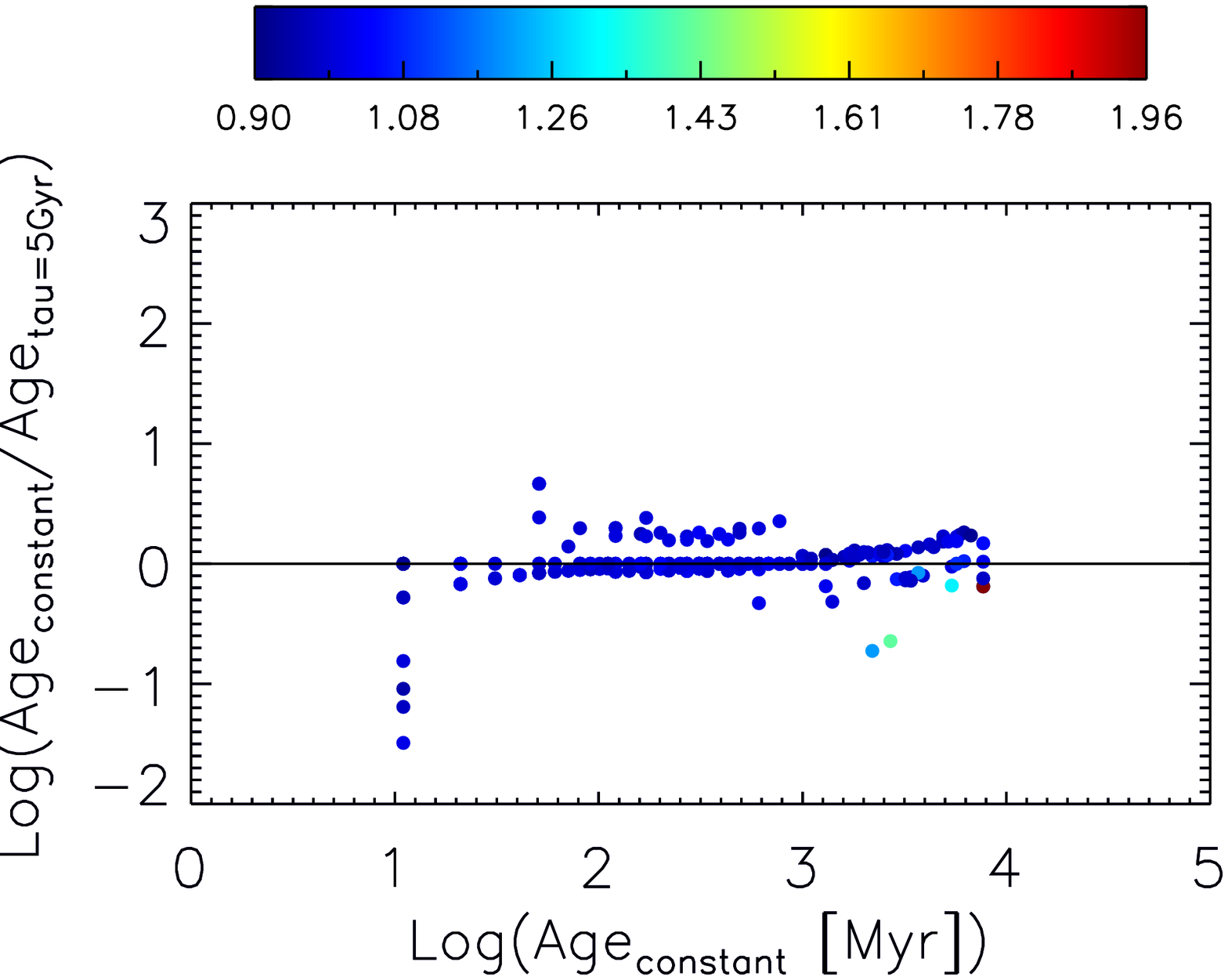}
   \includegraphics[width=0.33\textwidth]{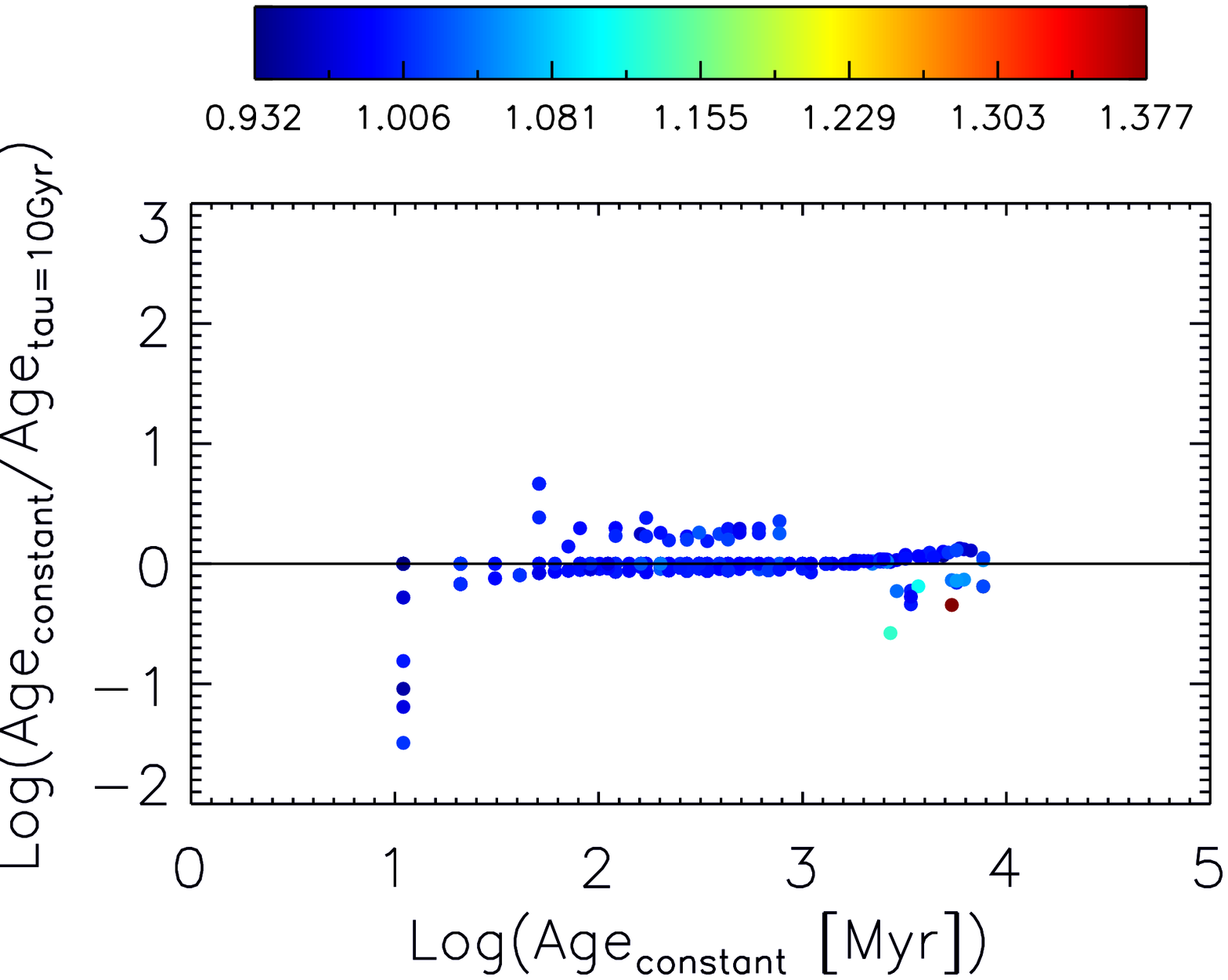}
   \includegraphics[width=0.33\textwidth]{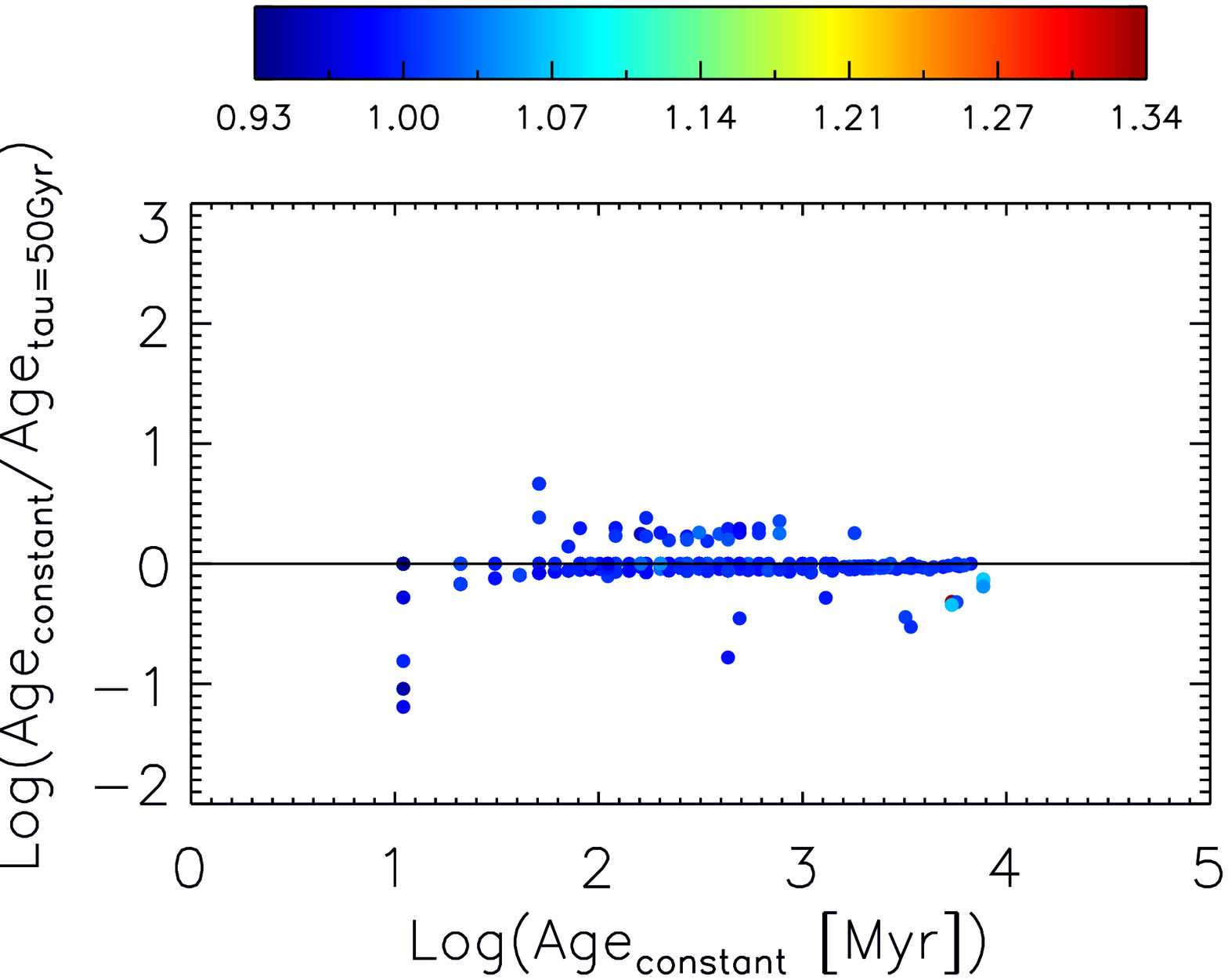}
\caption{Differences between the SED-derived age when performing SED fittings with BC03 templates associated to constant SFR and BC03 templates associated to exponentially declining SFH with finite values of the SFH time scale. The color of the points in each figure is related to the ratio in the $\chi^2_r$ values between the SED-fitting results when considering the different kinds of SFHs. The values of such $\chi^2_r$ ratios corresponding to each color are indicated by the color bars. Values close to one indicate that the templates associated to different SFHs fit the photometry with the same accuracy and, therefore, it is not possible  to distinguish between different kinds of SFHs. Each panel is associated to one value of the SFH time scale, as indicated in each vertical axis.
              }
\label{edad_edad}
\end{figure*}

\begin{figure*}
   \centering
   \includegraphics[width=0.33\textwidth]{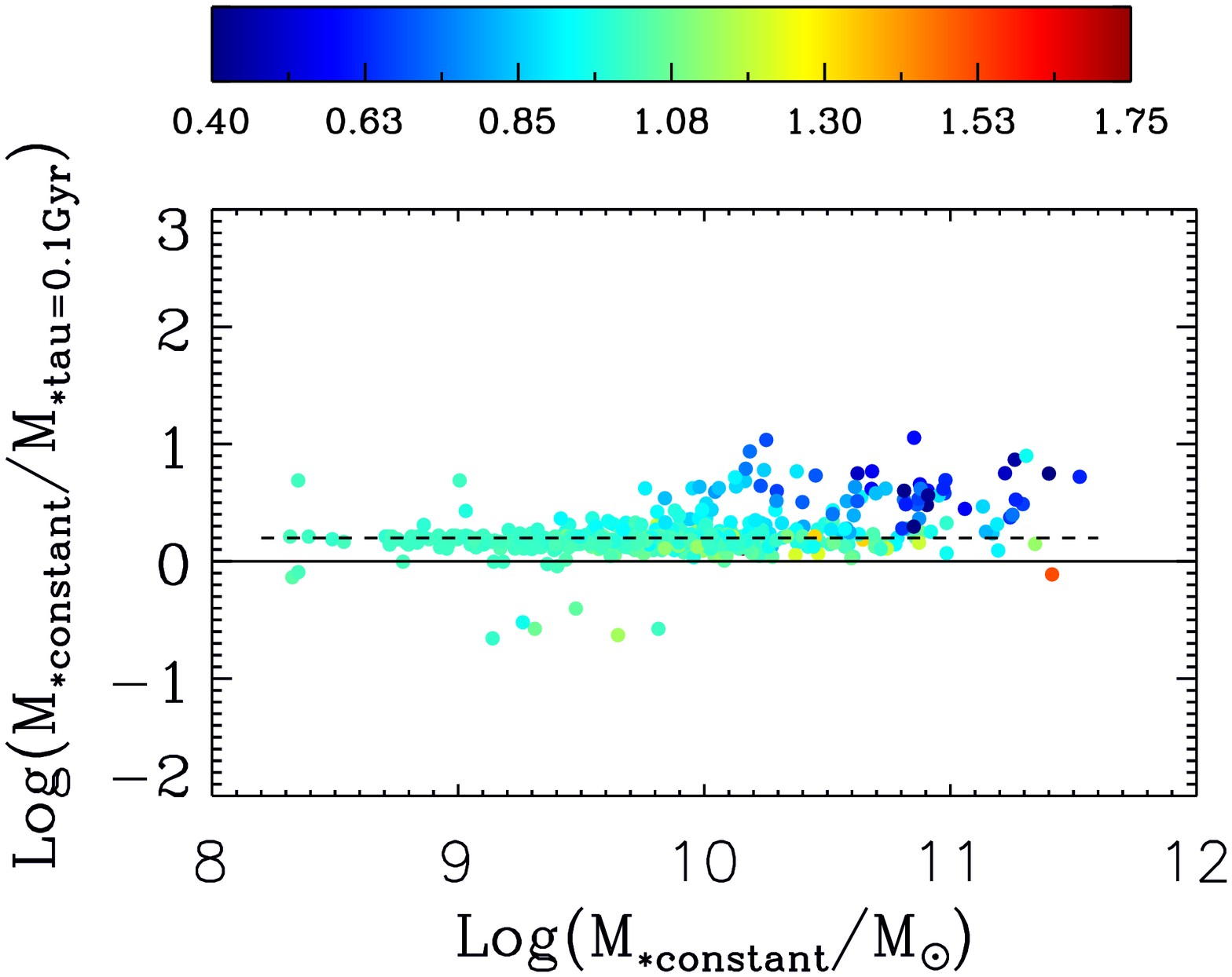}
   \includegraphics[width=0.33\textwidth]{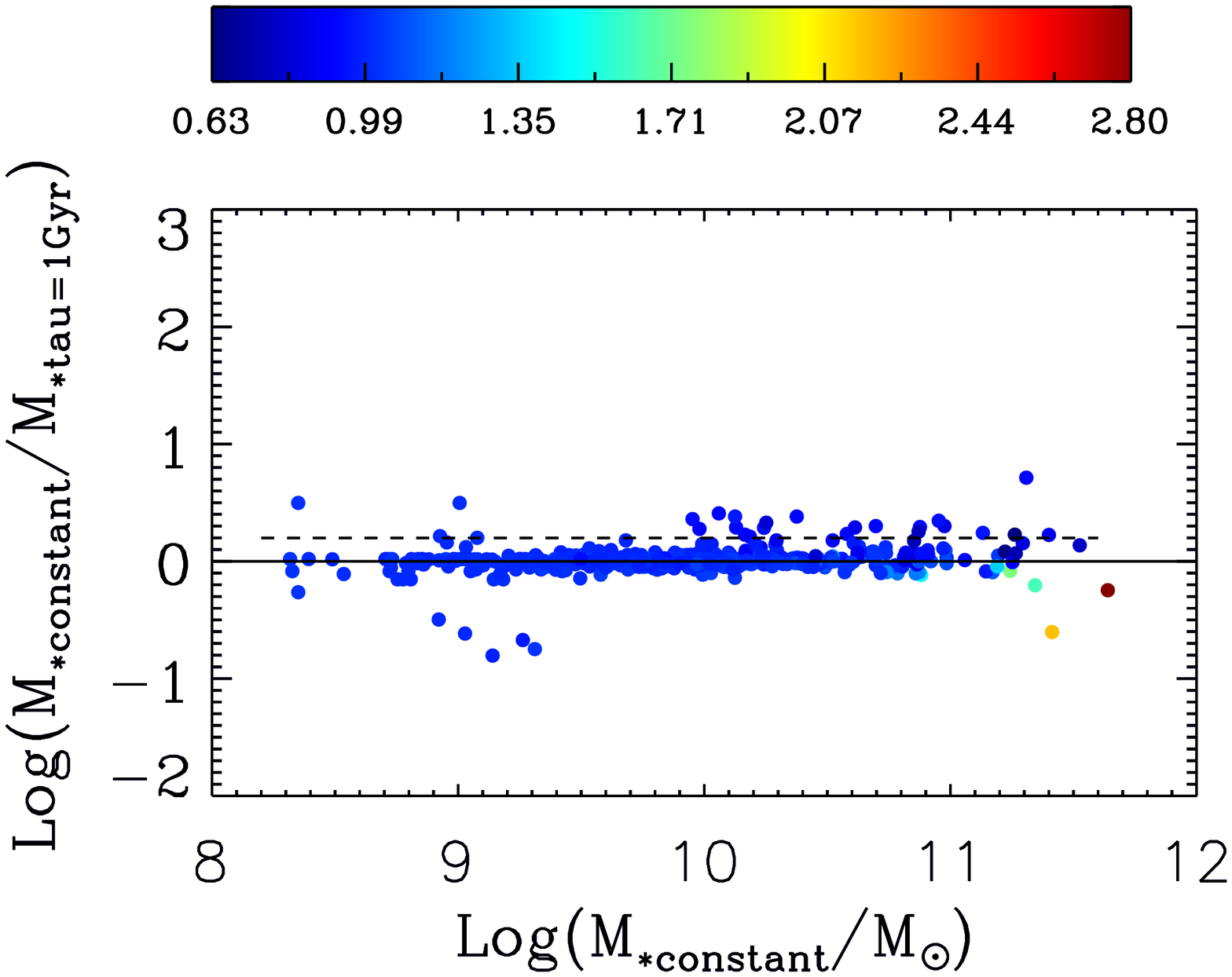}
   \includegraphics[width=0.33\textwidth]{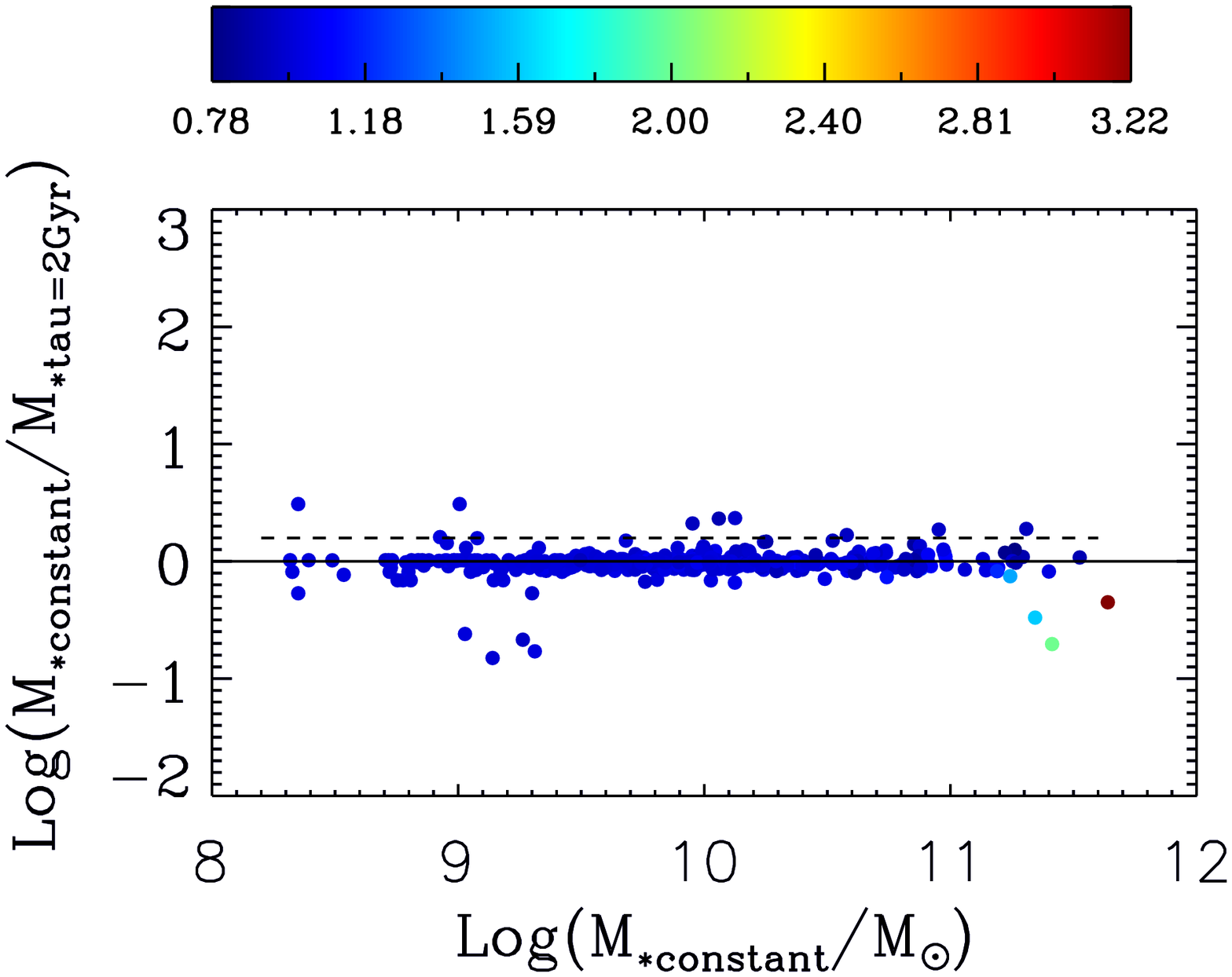}\\
   \includegraphics[width=0.33\textwidth]{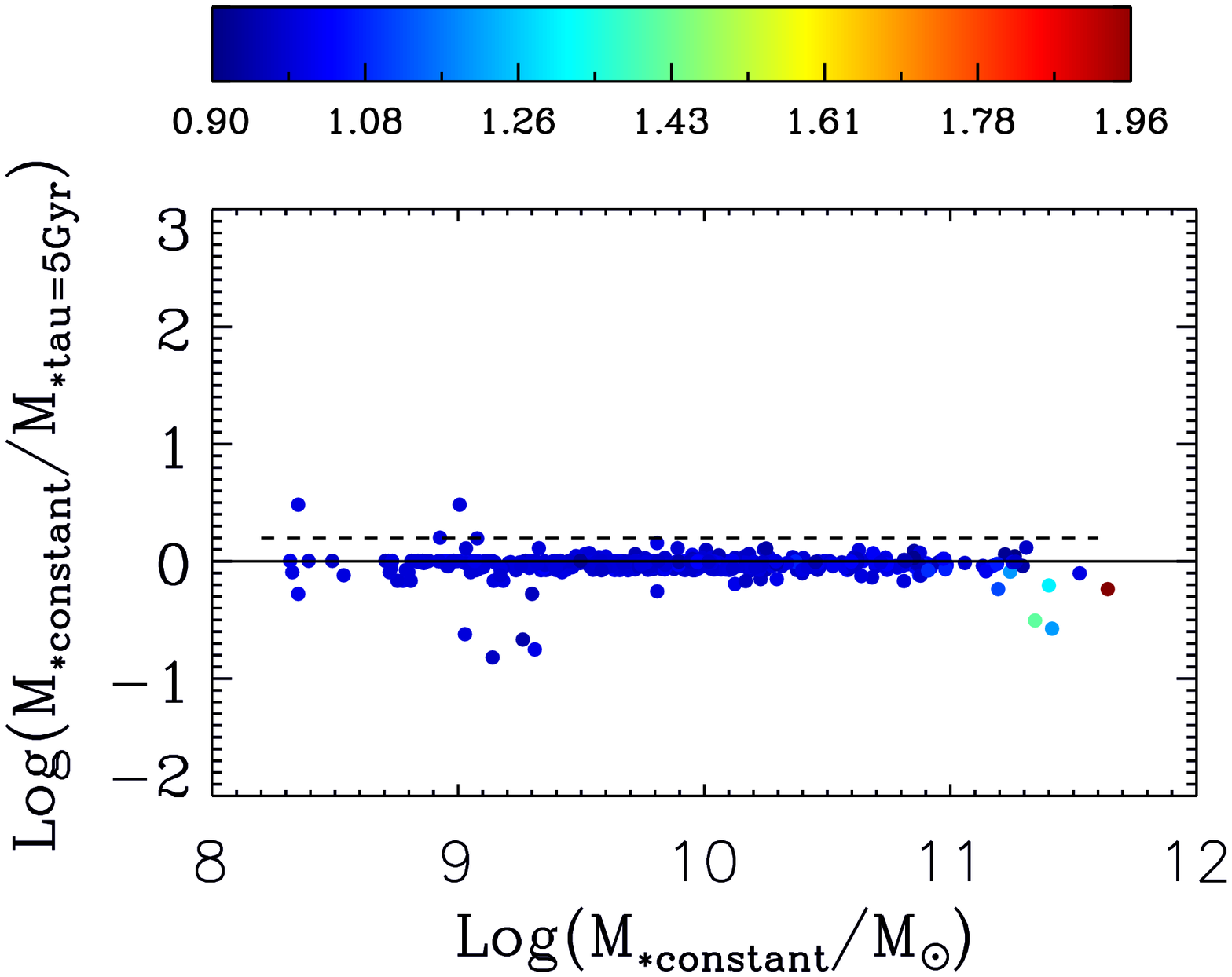}
   \includegraphics[width=0.33\textwidth]{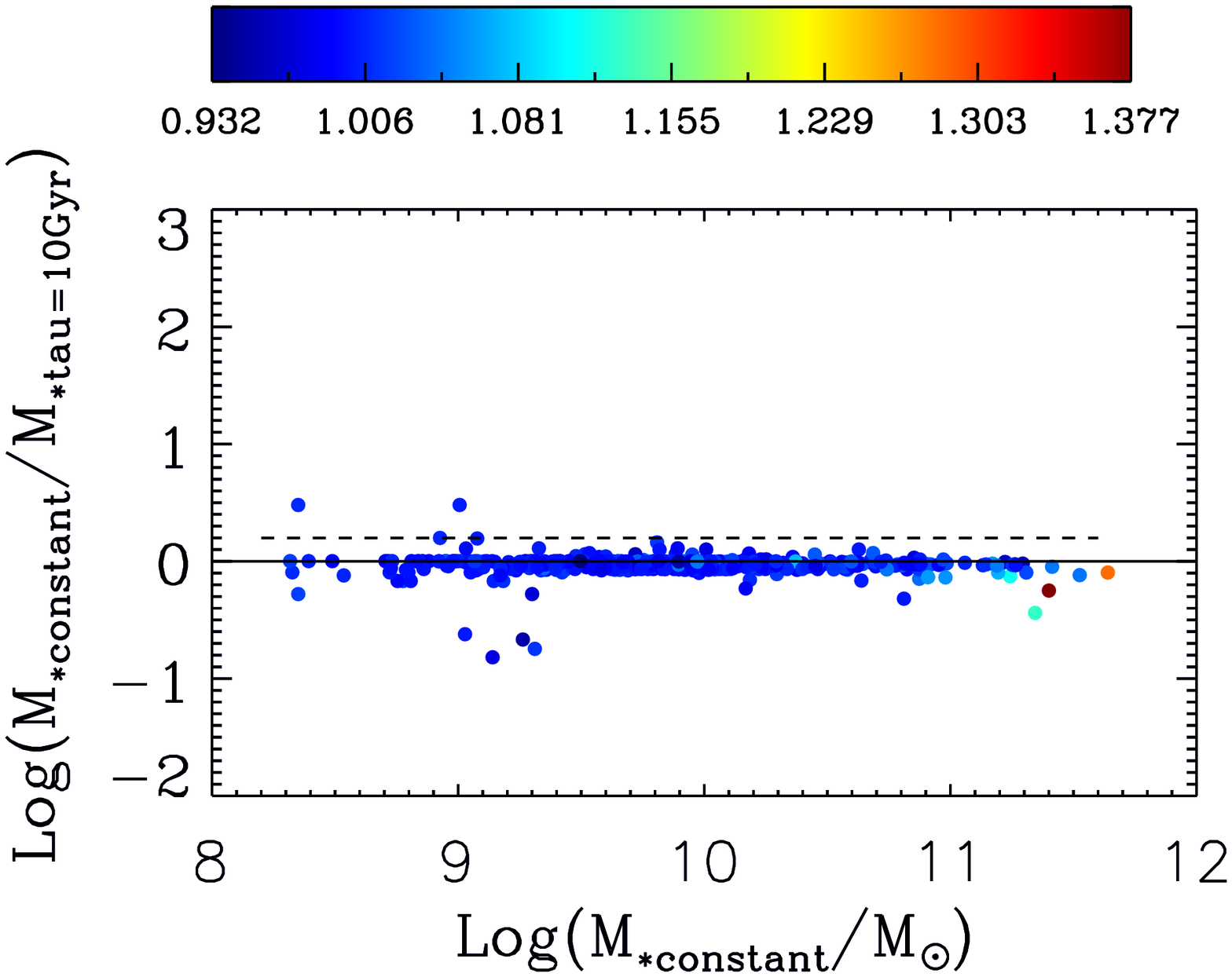}
   \includegraphics[width=0.33\textwidth]{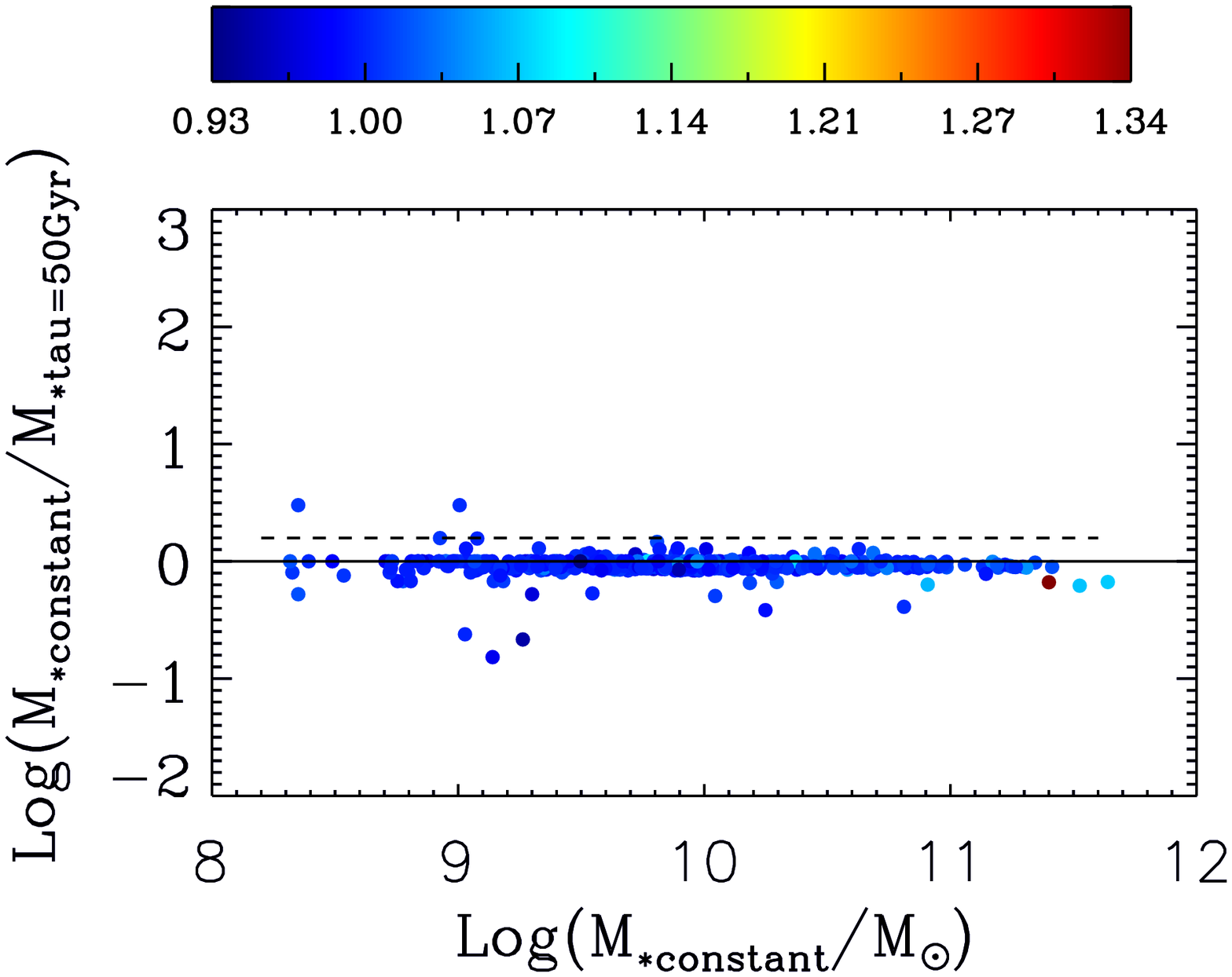}
\caption{Differences between the SED-derived stellar mass when performing SED fittings with BC03 templates associated to constant SFR and BC03 templates associated to exponentially declining SFH with finite values of the SFH time scale. The color of the points in each figure are related to the ratio in the $\chi^2_r$ values between the SED-fitting results when considering the different kinds of SFHs. The values of such $\chi^2_r$ ratios corresponding to each color are indicated by the color bars. Values close to one indicate that the templates associated to different SFHs fit the photometry with the same accuracy and, therefore, it is not possible  to distinguish between different kinds of SFHs. Each panel is associated to one value of the SFH time scale, as indicated in each vertical axis.
              }
\label{masa_masa}
\end{figure*}

The physical properties of our GALEX-selected LBGs that were discussed in the previous section were derived by carrying out an SED-fitting procedure with BC03 templates built by assuming  a constant SFR. Different works assume different SFH for the galaxies under study \citep{Nilsson2011_LBG,Basu2011,Haberzettl2012,Hathi2013}. One of the most used is that where the SFH varies exponentially with time. Analytically, it can be described by $SFH \propto \exp{\left( -t/\tau_{\rm SFH} \right)}$, where $\tau_{\rm SFH}$ is the SFH time scale. It should be noted that the constant SFR is a particular case of the exponentially declining SFH when the SFH time-scale tends to infinity. The usage of exponentially declining SFHs entails the fitting of the SFH time-scale, increasing the number of degrees of freedom. Furthermore, there is a degeneracy between the SED-derived age and the SFH time scale that produces that different combinations of age and $\tau_{\rm SFH}$ give templates with the same shape that might be undistinguishable with photometric SEDs. In this Section we analyze the differences in the SED-derived physical properties for our GALEX-selected LBGs when assuming SFH with different values of the SFH time-scale. Furthermore, we also analyze whether we can distinguish between different values of $\tau_{\rm SFH}$ and, therefore, constrain the SFH of our galaxies 

\begin{figure*}
   \centering
   \includegraphics[width=0.33\textwidth]{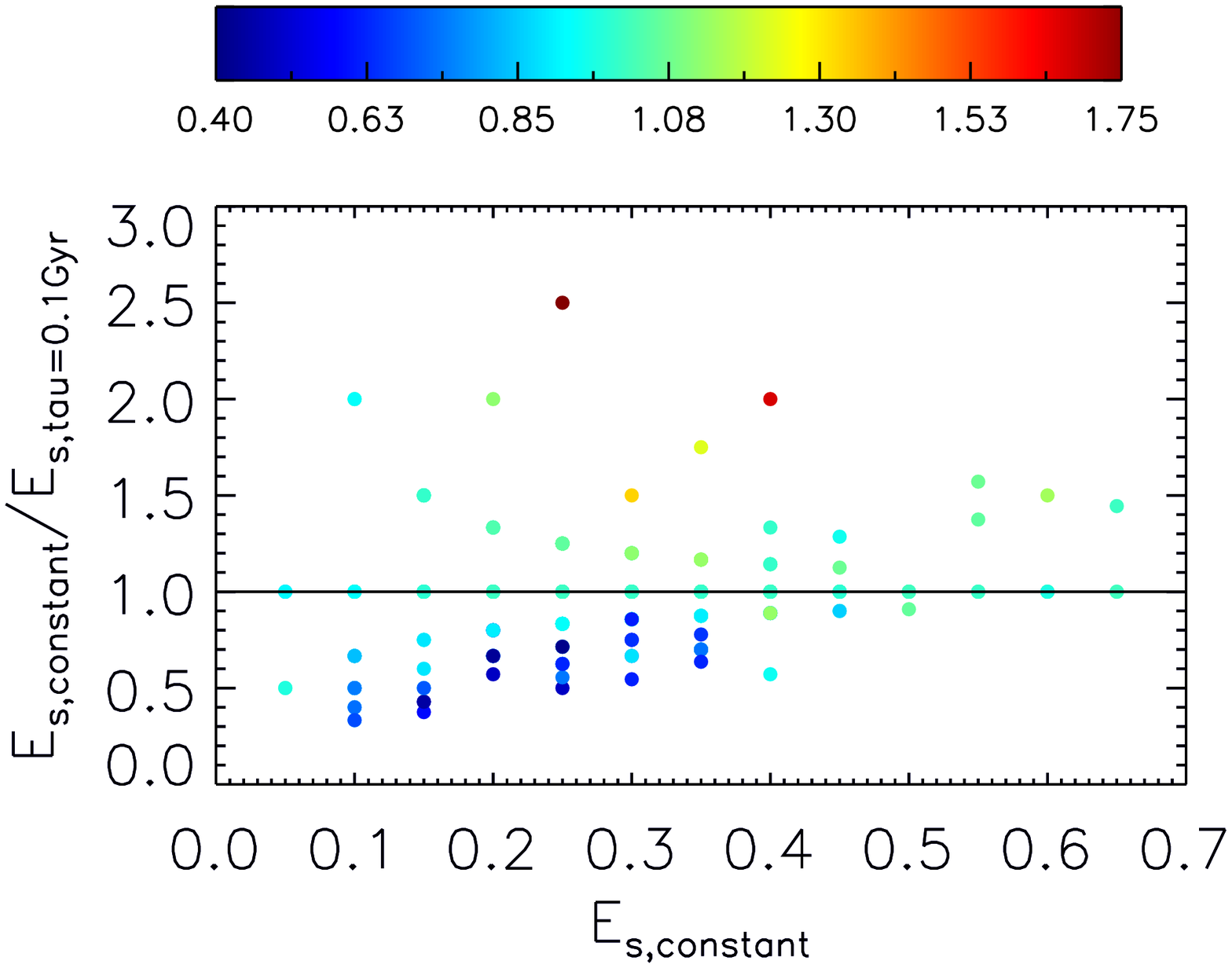}
   \includegraphics[width=0.33\textwidth]{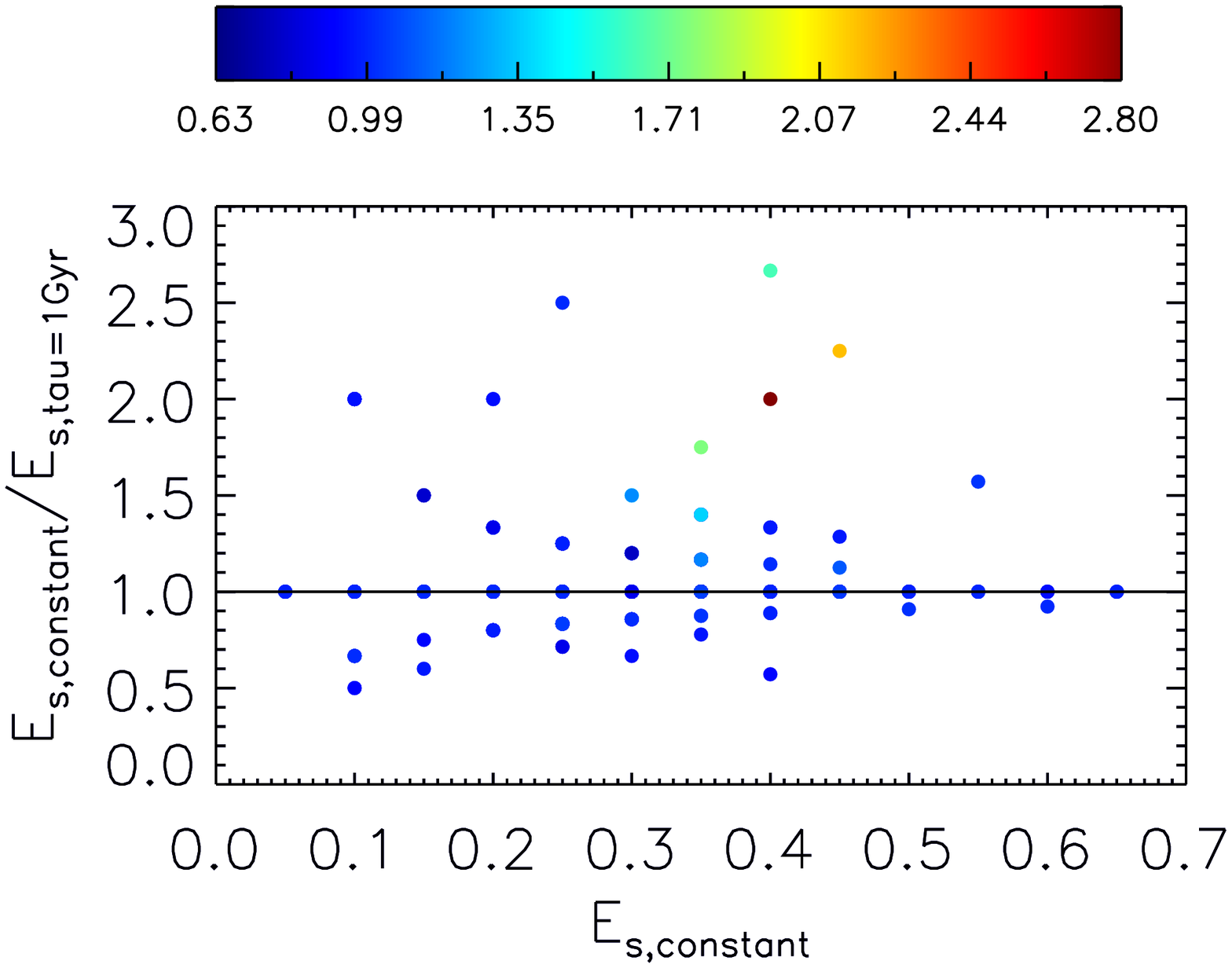}
   \includegraphics[width=0.33\textwidth]{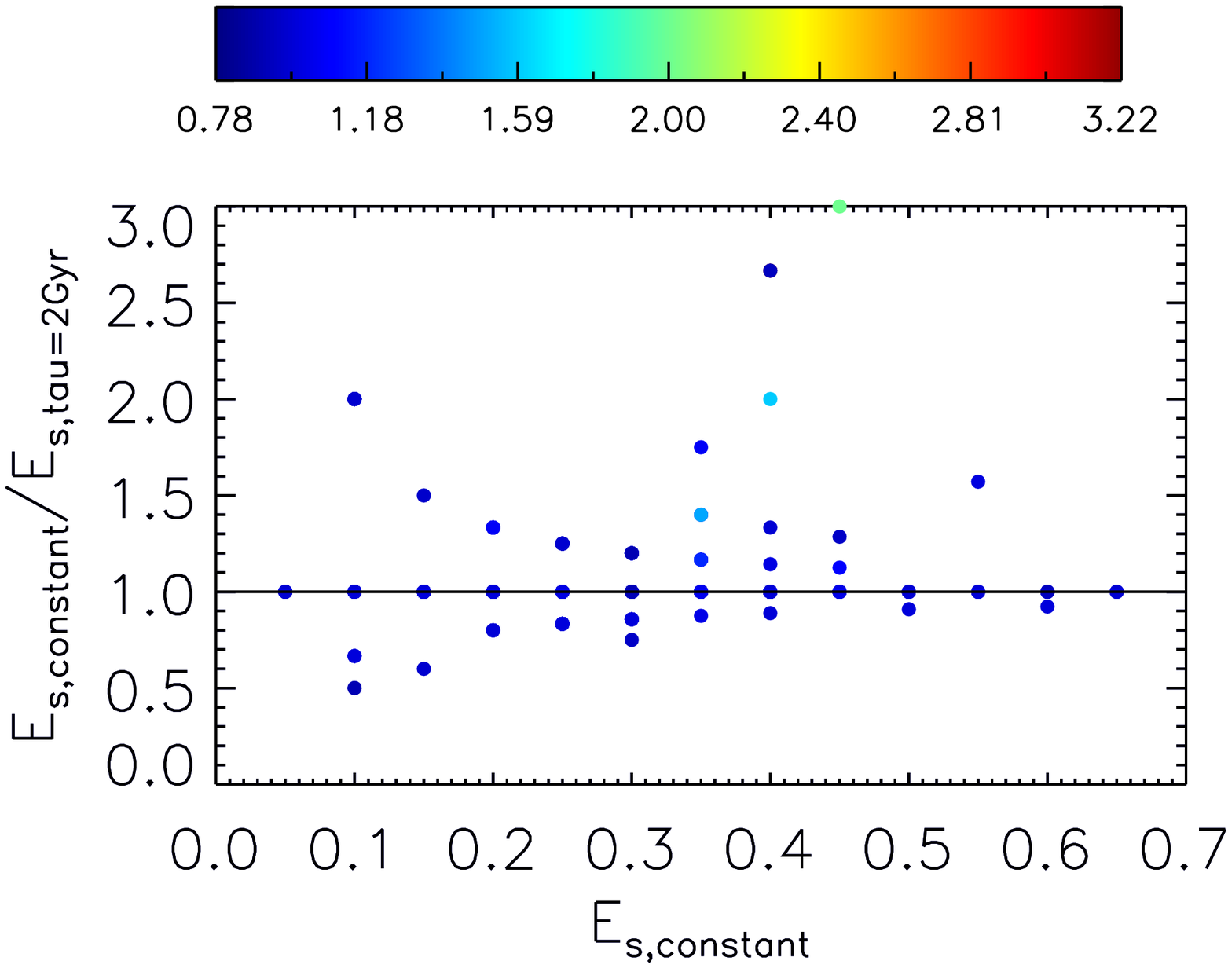}\\
   \includegraphics[width=0.33\textwidth]{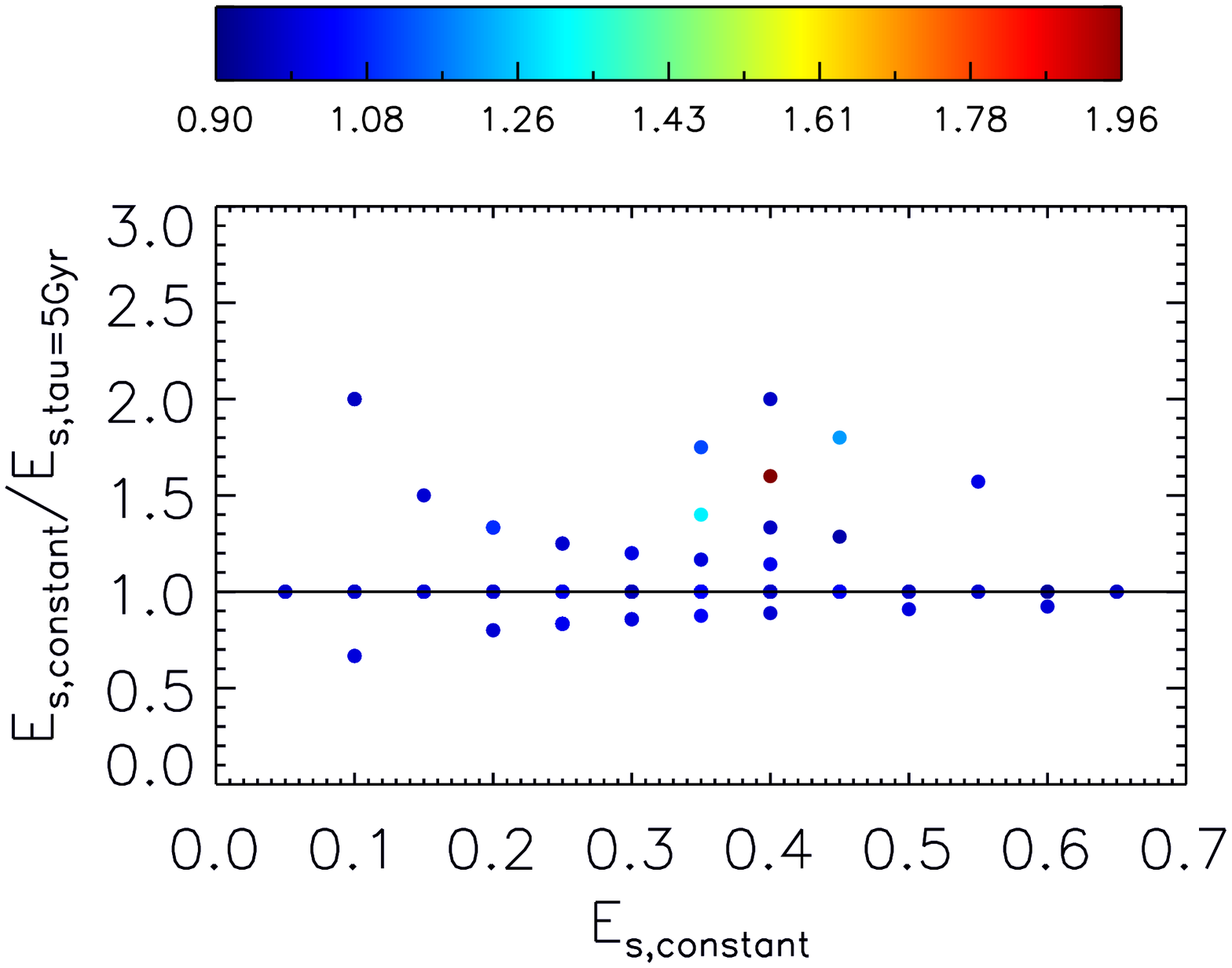}
   \includegraphics[width=0.33\textwidth]{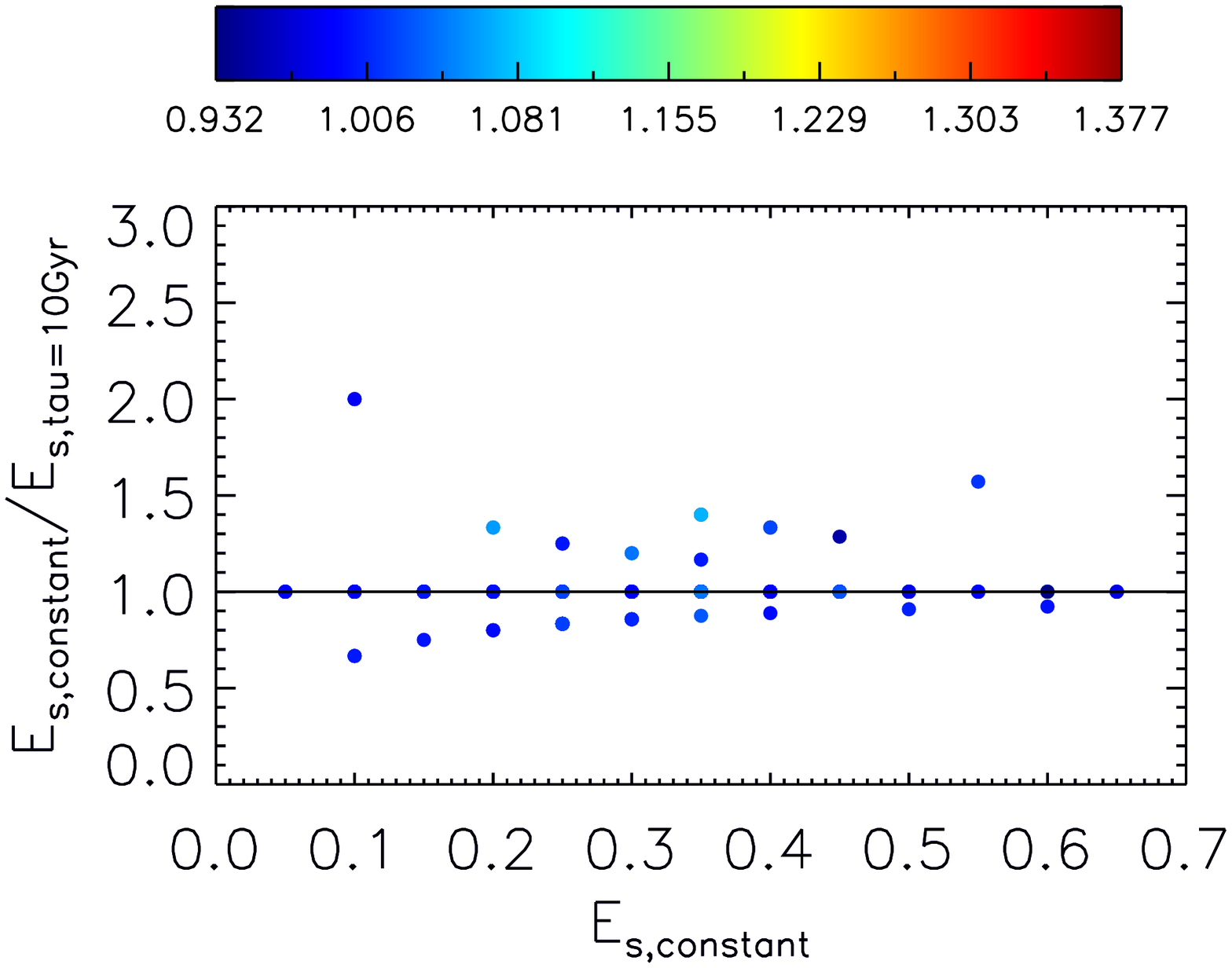}
   \includegraphics[width=0.33\textwidth]{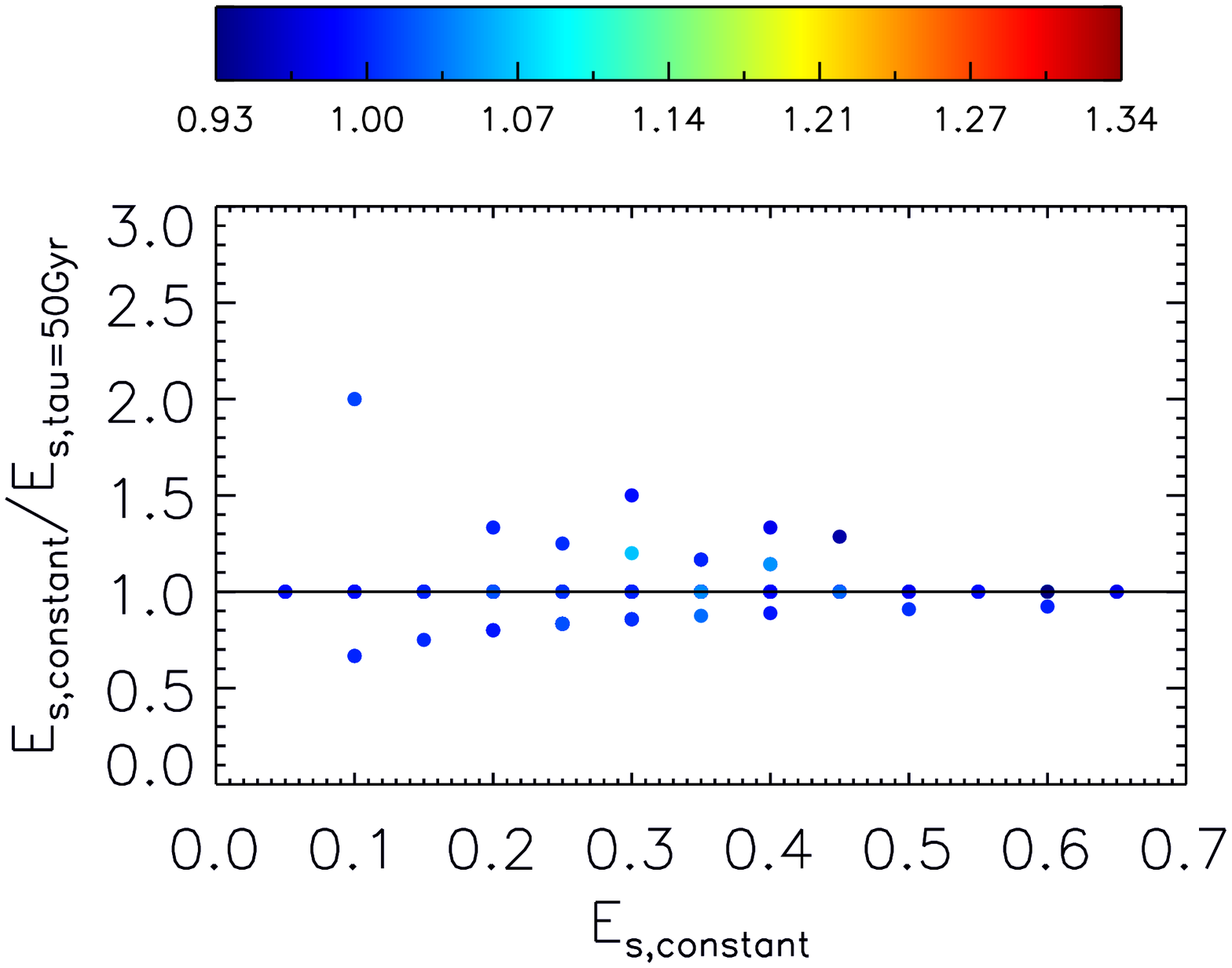}
\caption{Differences between the SED-derived dust attenuation when performing SED fittings with BC03 templates associated to constant SFR and BC03 templates associated to exponentially declining SFH with finite values of the SFH time scale. The color of the points in each figure are related to the ratio in the $\chi^2_r$ values between the SED-fitting results when considering the different kinds of SFHs. The values of such $\chi^2_r$ ratios corresponding to each color are indicated by the color bars. Values close to one indicate that the templates associated to different SFHs fit the photometry with the same accuracy and, therefore, it is not possible  to distinguish between different kinds of SFHs. Each panel is associated to one value of the SFH time scale, as indicated in each vertical axis.
              }
\label{polvo_polvo}
\end{figure*}

We build another set of BC03 templates considering that the SFH varies exponentially with time, with values $\tau_{SFH}=0.0001, 0.001, 0.01, 0.1, 1, 2, 5, 10$, and $50$ Gyr. In this process we adopt again a \cite{Salpeter1955} IMF distributing stars from 0.1 to 100 M$_\odot$ and select a fixed value of Z=0.2Z$_{\odot}$ for the metallicity. We select the same values of dust attenuation, $E_s(B-V)$ and age than those adopted in Section \ref{combination}. We also include intergalactic medium absorption adopting the prescription of \cite{Madau1995}. Figures \ref{edad_edad}, \ref{masa_masa}, and \ref{polvo_polvo} show the differences found for the SED-derived age, stellar mass, and dust attenuation, respectively, when assuming different SFHs. In those plots, the color code indicates the ratio of the $\chi^2_r$ values between the SED-fitting results with a given $\tau_{\rm SFH}$ and the case of SFR constant. In those cases where that ratio is close to one, the SED-fitting procedure would not be able to discriminate between different kinds of SFHs. Regarding age, Figure \ref{edad_edad} indicates that, as expected, there is a degeneracy between age and $\tau_{SFH}$. The SED-derived ages are systematically younger when adopting lower values of $\tau{\rm SFH}$.This tendency is more significant for galaxies whose SED-derived ages with constant SFR are higher. For $\tau_{SFH} \geq 1 Gyr$, the ratio of the $\chi^2_r$ values obtained with different SFHs are very similar to unity and, therefore, although the SED-derived ages are different for different values of $\tau_{SFH}$ (mostly for the oldest galaxies), the SED-fitting procedure is not able to distinguish between those SFHs. This produce an uncertainty in the age of the galaxies and prevents us from distinguishing between different temporal variation of their SFH. The ages derived with templates associated to constant SFH are typically higher than those obtained with templates of finite values of $\tau_{\rm SFH}$. In this way, the ages derived in Section \ref{properties} for our GALEX-selected LBGs might be considered as upper values. Considering diverse SFH scenarios, the differences in age found are typically comparable and higher in some cases (mostly for the lowest values of $\tau_{\rm SFH}$) than the typical uncertainty of the SED-derived age found in \ref{properties}. Therefore, both the uncertainties in the SED-fitting procedure and the degeneracy between age and SFH affect to the total uncertainty of the SED-derived age of our galaxies.

For $\tau_{\rm SFH} < 1$ Gyr, the ratio between the $\chi^2_r$ values is similar to one for the majority of the galaxies but there is a population of them (those for which the age derived with constant SFR is higher) whose $\chi^2_r$ indicate that the templates associated to an exponential variation of the SFH are worse fitted. Only in these cases we are able to distinguish between different kinds of SFHs and, thus, as a general trend, we obtained that the SEDs of our oldest GALEX-selected LBGs are better fitted with BC03 templates associated to SFH with $\tau_{SFH} \geq 1 Gyr$, including the case of constant SFR.

Figure \ref{masa_masa} indicates that templates with lower values of $\tau_{\rm SFH}$ tend to give lower values of the stellar mass than those associated to higher values of $\tau_{\rm SFH}$. In this way, the stellar masses derived with constant SFR and reported in the previous section for our GALEX-selected LBGs might be understood as upper limits. Again, as it happens for the age, even with the good photometric coverage of GALEX and ALHAMBRA we are not able to distinguish between the different SFH scenarios in most cases. In Figure \ref{masa_masa} we represent a horizontal line in each panel which corresponds to a deviation of 0.2 dex with respect to the case of constant SFR. This is the maximum median difference that we find and it is a measurement of the uncertainty of the SED-derived stellar mass when considering variuos values of $\tau_{\rm SFH}$.

Figure \ref{polvo_polvo} shows that there is also a degeneracy between the SED-derived dust attenuation and the SFH time-scale. In this case, the tendency between dust attenuation and $\tau_{\rm SFH}$ is not as clear as those for age and stellar mass. The differences between the results with constant SFR and finite values of $\tau_{\rm SFH}$ are within a factor of two, although for most galaxies both estimations agree.

Summarizing, even with the good photometric coverage of the combination of GALEX and ALHAMBRA data we are not able to distinguish between different SFHs for most of the cases. We can only conclude that older galaxies (in terms of the SED-derived age with templates associated to constant SFR) are better fitted with templates associated $\tau_{SFH} \geq 1 Gyr$. The differences between the SED-derived properties when assuming different kinds of SHFs imply additionally uncertainties in the SED-derived parameters to the previously studied in Section \ref{properties}.


\subsection{SFR-stellar mass plane}

\begin{figure*}
\centering
\includegraphics[width=0.49\textwidth]{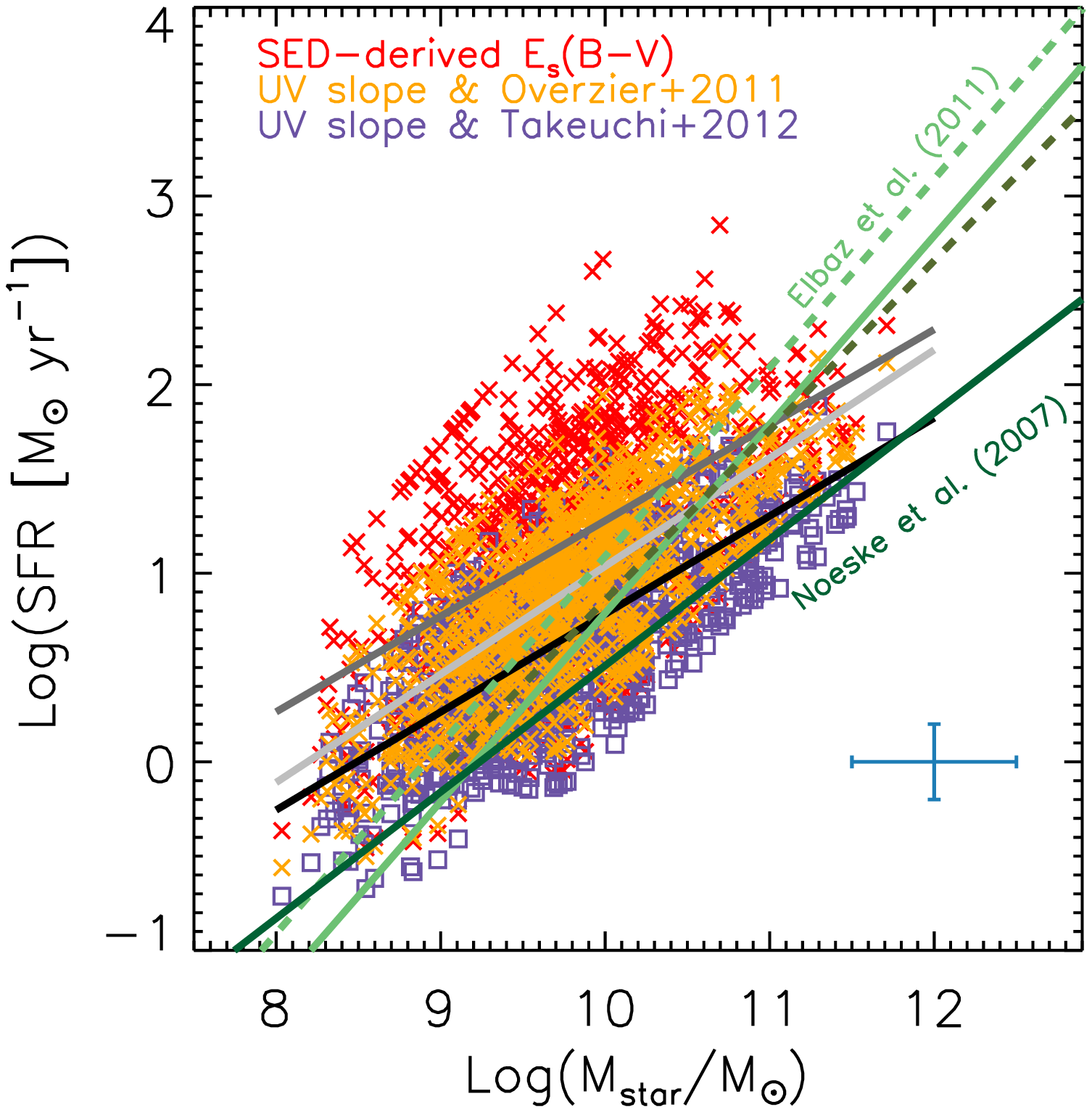}
\includegraphics[width=0.49\textwidth]{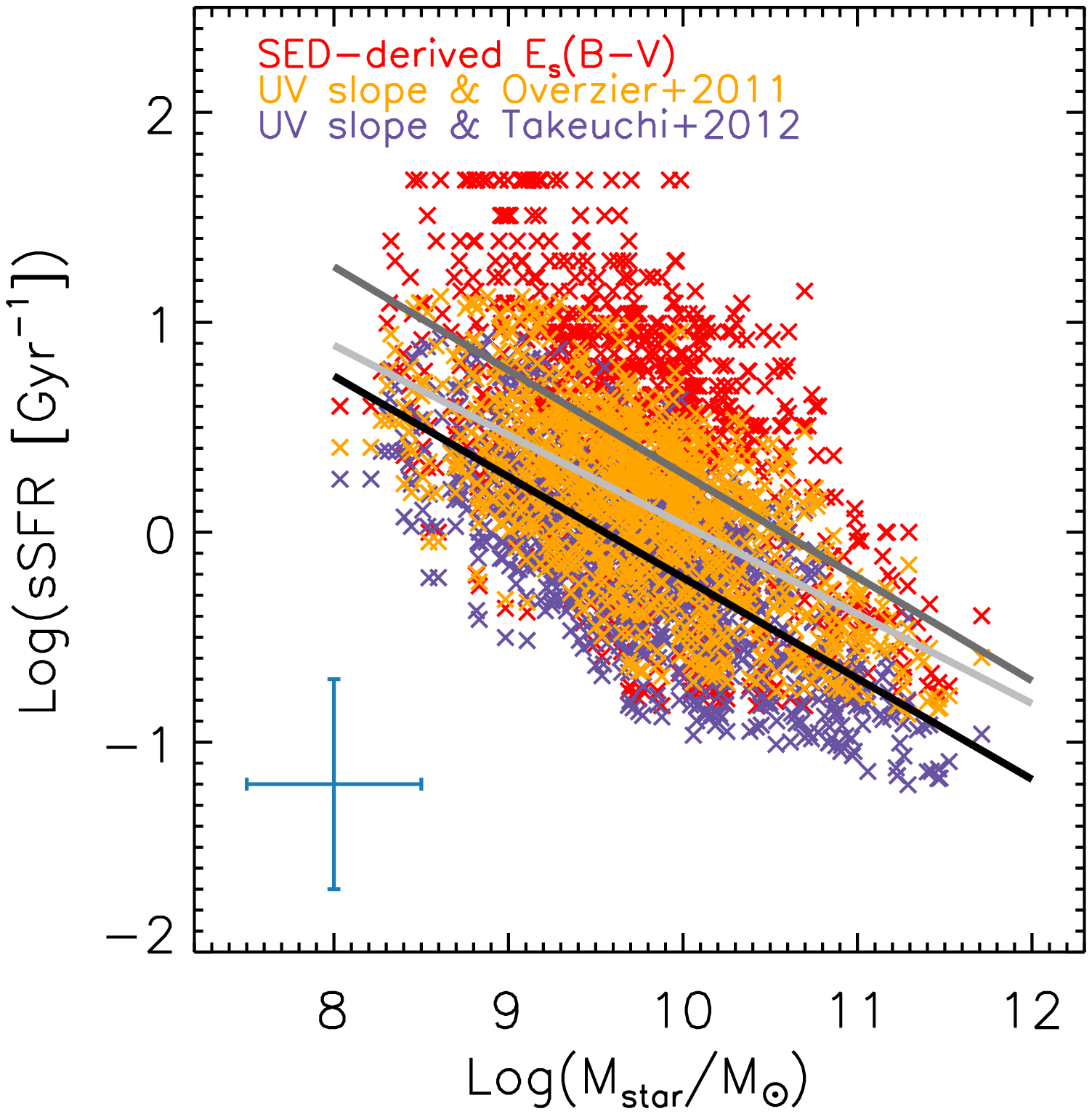}
\caption{\emph{Left}: Dust-corrected total SFR for our GALEX-selected LBGs as a function of their stellar mass. Red symbols correspond to the dust corrected total SFRs obtained with the SED-derived dust attenuation (as explained in Section \ref{combination}). Orange symbols represent the dust-corrected total SFRs obtained with the dust attenuation derived from the UV continuum slope using the \citet{Overzier2011} relation for Lyman break analogs. Purple symbols are obtained with the updated \citet{Meurer1999} relation given by \citet{Takeuchi2012} when correcting for the small aperture of IUE. Dark grey, clear grey, and black straight lines are the linear fits to the red, orange, and purple symbols, respectively. For comparison, we also show the SFR-mass relations for the main sequence of galaxies at $z \sim 1$ reported in \citet{Elbaz2011} (light green solid line), \citet{Elbaz2007} (blue solid line), and \citet{Noeske2007} (dark green continuous line) at a similar redshift range than that for our GALEX-selected LBGs. The dashed light green line represent twice the value of the main sequence of \citet{Elbaz2011}. \emph{Right}: Specific SFR of our GALEX-selected LBGs as a function of their stellar mass. The meaning of the colors and the fits is the same than those in the left panel. In both plots, the error bars represent the typical uncertainties in the SED-derived total SFR, steles mass, and sSFR (see Section \ref{properties}). The uncertainty of the SED-derived total SFR correspond to the values obtained when dust correcting with the SED-derived dust attenuation.}\label{sfr_mass}
\end{figure*}

One important parameter in the study of physical properties of galaxies and their evolution is the specific star formation rate (sSFR). This quantity is defined as the ratio between the star formation rate and the stellar mass, sSFR=SFR/M$_*$, and it is a measurement of the present over past star formation activity. A tight relation between the star formation rate and the stellar mass has been found at different redshifts in several works \citep{Salim2007,Elbaz2007,Noeske2007,Daddi2007,Pannella2009,Rodighiero2010,Gonzalez2010,Karim2011,Elbaz2011}. This has enabled the definition of a 'main sequence' (MS) of galaxies defined by a specific value of the sSFR. Galaxies in the MS are opposite to the idea of 'starburst galaxies', which are those galaxies whose nature makes them deviate from the MS towards higher values of the sSFR. The characteristic value of the sSFR for the MS of galaxies has been reported to change with redshift (see for example \cite{Elbaz2011} for a recent discussion of this evolution with deep \emph{Herschel} FIR data). Previous studies show that the $sSFR$ increases with increasing redshifts at all masses and that the $sSFR$ of massive galaxies is lower than that for less massive galaxies at any redshift \citep{Feulner2005,Erb2006,Dunne2009,Damen2009,Rodighiero2010}. Despite the number of studies analyzing the relation between stellar mass and SFR, there is still some controversy, mostly regarding the slope of the $sSFR$-$M_*$ relation.

In the left panel of Figure \ref{sfr_mass} we show the dust-corrected total SFR for our GALEX-selected LBG against their stellar mass. Along with the data points, the MS of galaxies at $z \sim 1$ taken from, \cite{Elbaz2007}, \cite{Elbaz2011} and \cite{Noeske2007} are also represented. In that plot, red symbols indicate the dust-corrected total SFR obtained following the procedure explained in Section \ref{combination}. Some other works, recover the dust attenuation of galaxies employing the UV continuum slope and a relation between the UV continuum slope and dust attenuation, as the \cite{Meurer1999} law. The values of dust attenuation obtained in this way can be also used for obtaining the dust-corrected total SFR. Recently, it has been found that the smaller \emph{IUE} apertures employed in \cite{Meurer1999} may have missed about half of the light. Consequently, for a given UV continuum slope, the newly found dust attenuations are lower than those predicted by the \cite{Meurer1999} relation \citep{Overzier2011,Takeuchi2012}. Following these results, we obtain the dust-corrected total SFR by correcting the UV luminosities with the dust attenuation derived from the UV continuum slope of our galaxies after the application of the \cite{Overzier2011} and \cite{Takeuchi2012} relations. Among the relations presented in \cite{Overzier2011} we employ the one for their Lyman break analogues. The results are also shown in the left panel of Figure \ref{sfr_mass}. It can be seen that whatever the dust-correction method applied, there is a clear relation between SFR and mass in the sense that more massive LBGs have also higher SFRs. We have fitted linear relations to the points obtained with the three dust-correction methods. Table \ref{fitting} summarizes the results. It can be seen that the slopes of the SFR-mass relation obtained with the three dust-correction methods are very similar and compatible within the uncertainties. The differences between the zero-points of the relations are due to the fact that, for a given stellar mass, the dust attenuation recovered by each method is slightly different.

\begin{table*}
\caption{\label{fitting}Results of a linear fitting to the SFR-mass and sSFR-mass relation for our GALEX-selected LBGs at $z \sim 1$. The data points are fitted to a relation in the form $\log{SFR} \, [M_\odot \, {\rm yr}^{-1}] = a + b \times \log{M_*/M_\odot}$ and $\log{sSFR} \, [{\rm Gyr}^{-1}] = a + b \times \log{M_*/M_\odot}$. The results for the three dust-correction method described in the text are included.}
\centering
\begin{tabular}{cccc}
\hline\hline
 SFR-mass & $E_s(B-V)$ & \cite{Overzier2011} & \cite{Takeuchi2012} \\
 $a$ & $-3.79 \pm 0.25$ & $-4.69 \pm 0.16$ & $-4.41 \pm 0.15$ \\
 $b$ & $0.50 \pm 0.03$ & $0.57 \pm 0.02$ & $0.51 \pm 0.02$ \\
\hline
 sSFR-mass & $E_s(B-V)$ & \cite{Overzier2011} & \cite{Takeuchi2012} \\
 $a$ &$5.21 \pm 0.25$ & $4.30 \pm 0.16$ & $4.59 \pm 0.15$ \\
 $b$ & $-0.49 \pm 0.03$ & $-0.42 \pm 0.02$ & $-0.48 \pm 0.02$ \\
\hline
\end{tabular}
\end{table*}


Right panel of Figure \ref{sfr_mass} shows the relation between the sSFR and the stellar mass of our GALEX-selected LBGs. Again, we represent the points associated to the dust-corrected total SFR obtained from the SED-derived dust attenuation and also those derived from the UV continuum slope and the \cite{Meurer1999} law. It can be seen that galaxies with larger masses tend to have lower values of the sSFR, independently of the dust-correction method employed. We have fitted a relation to the data points for the three dust attenuation methods. The best-fitted slope and zero-points are also shown in Table \ref{fitting}. It can be seen that the slope of the sSFR-mass relation is very similar for the three dust-correction methods and compatible within the uncertainties. Again, the differences between the zero-points is due to the slightly different dust attenuation obtained for a given stellar mass in each of the dust-correction methods.

\section{Sizes and morphology}\label{morfo}

In this section we study the morphology and the physical sizes of our GALEX-selected LBGs. To this aim, we use I-band ACS images taken from the HST observations of the COSMOS field, that for our redshift range correspond to a rest-frame band ranging from approximately 3500 to 4250 \AA. Out of the whole sample of LBGs, 897 have ACS information. We download 8''x8'' ACS cut-outs centered at the position of each source from the cut-outs service available on the NASA/IPAC Infrared Science Archive \footnote{http://irsa.ipac.caltech.edu/data/COSMOS/index\_cutouts.html}. In order to study the morphology of the galaxies we carry out a visual inspection of each cut-out and classify them into six groups: Disk-galaxies, compact galaxies, chain galaxies (CH), clump cluster (CC), interacting/merging galaxies and Irregular galaxies. For the definition of CH and CC we follow \cite{Elmegreen2009} and consider a galaxy as irregular when its morphology does not fit in any of the other groups. As a result of the visual classification, we find that the majority (69\%) of our GALEX-selected LBGs are disk-type galaxies, 11\% have indications of interactions or merging, 7\% are irregular, 6\% are compact galaxies and a minority are CH or CC galaxies, with a fraction of 5\% and 2\%, respectively. It should be noted that a visual classification is always a very subjective procedure, and therefore, the previous percentages are approximated. However, what it is quite clear is that the dominant morphology of our GALEX-selected LBGs at $z \sim 1$ is the disk-like class. \cite{Wolf2005} studied the contribution to the UV luminosity density of different morphologies of galaxies at $z \sim 0.7$ by combining high-resolution images from the GEMS survey \citep{Rix2004} with redshifts and SEDs from COMBO-17 survey \citep{Wolf2001,Wolf2004}. They report that seemingly normal disk galaxies are the largest contributors to the UV luminosity density. This is compatible with the result that we find in the present work, since LBGs are among the brightest UV galaxies and most of them have disk-like morphologies. Furthermore, the domination of disk-like galaxies morphology in LBGs at $z \sim 1$ is in agreement with the result of \cite{Burgarella2006} for LBGs at a similar redshift range than ours, who found that 75\% of LBGs are compatible with such morphology. However, our result contrasts with that found in \cite{Basu2011}, where most LBGs appear to have irregular morphologies and only a few have disk-like morphologies. This difference is likely due again (see Section \ref{stellar}) to the inclusion in their sample of galaxies at redshift higher than ours.

\begin{figure}
\centering
\includegraphics[width=0.23\textwidth]{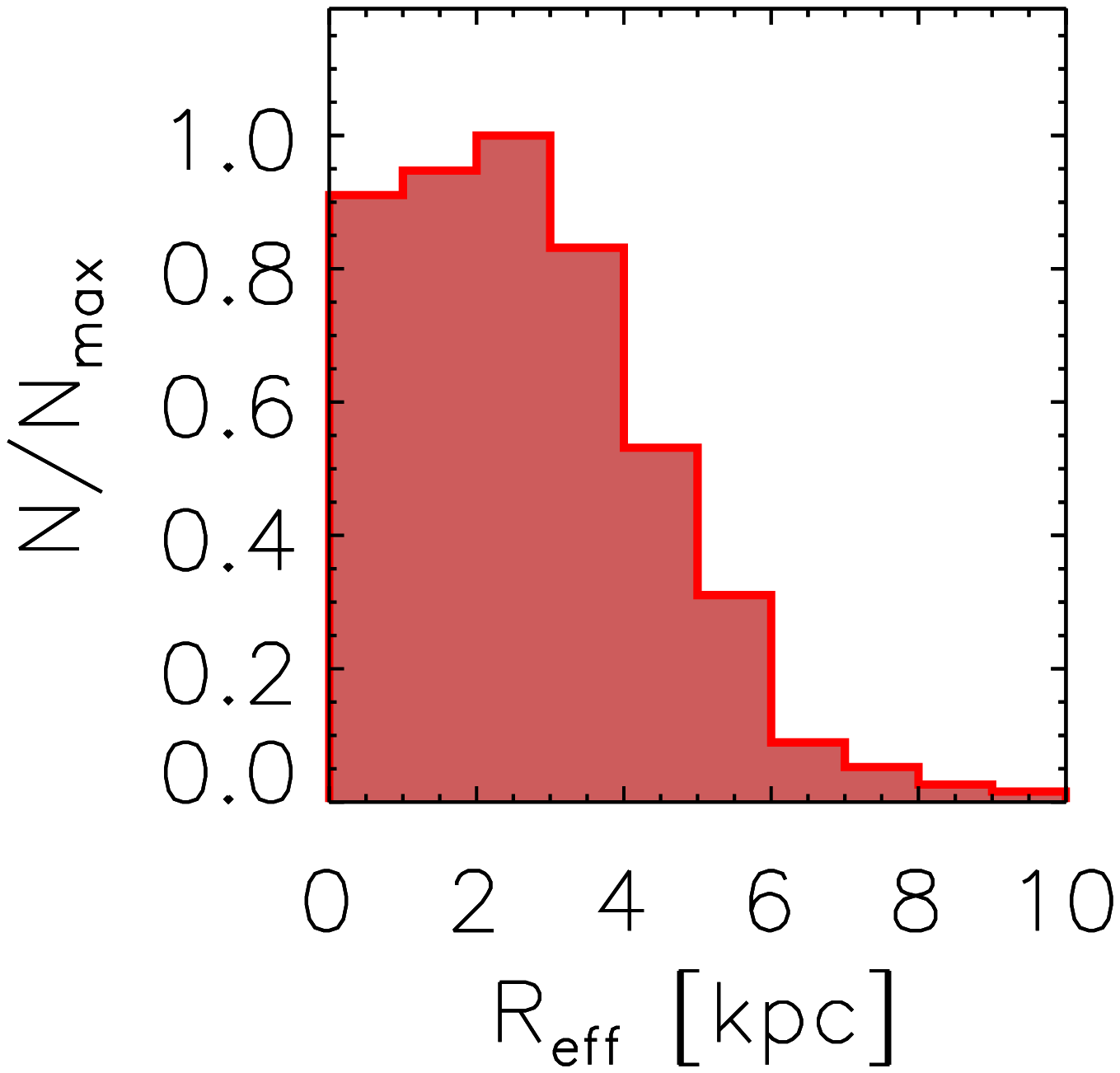}
\includegraphics[width=0.23\textwidth]{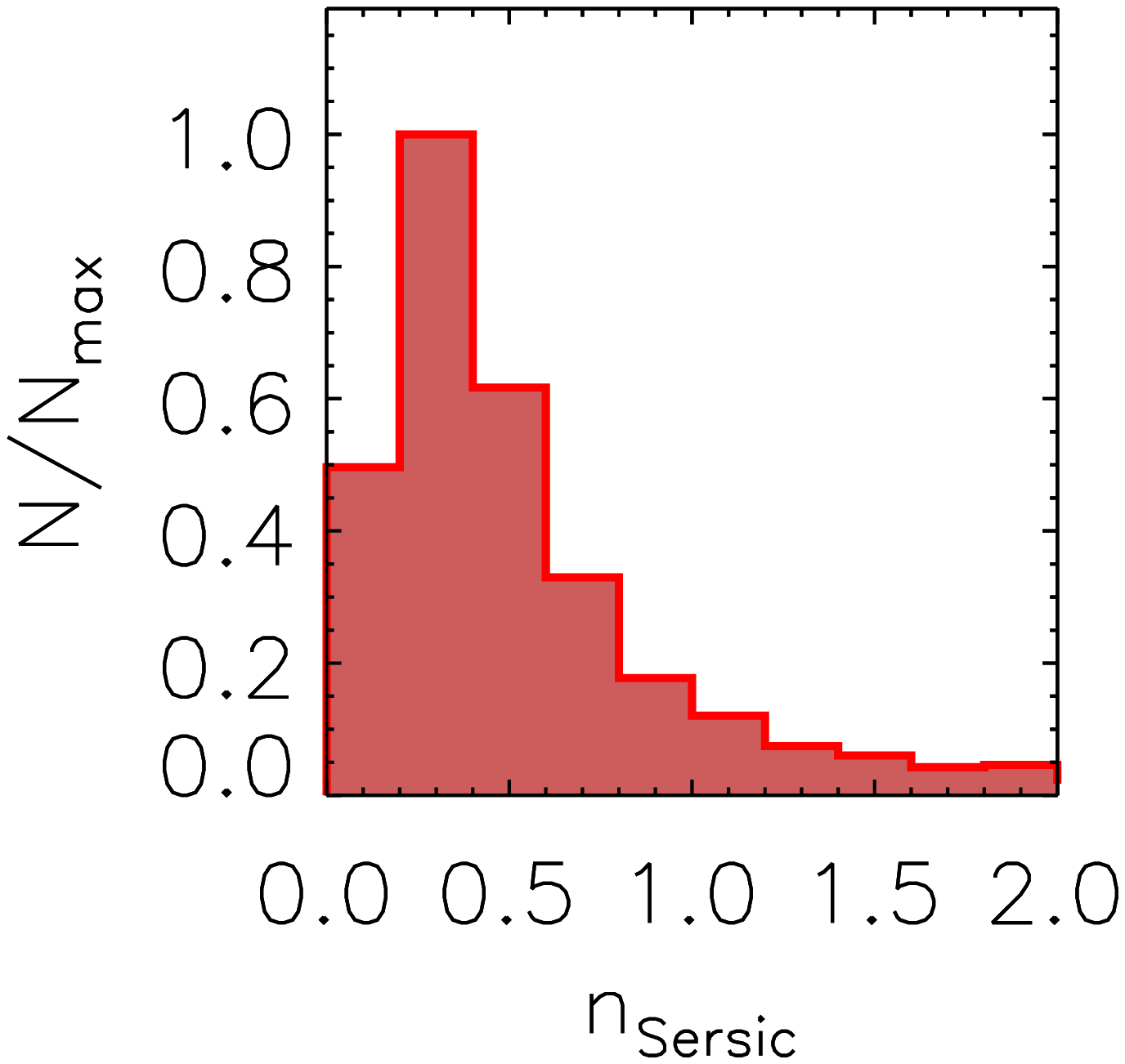}
\caption{Distribution of the effective radii (\emph{left}) and the Sersic indices (\emph{right}) for our GALEX-selected LBGs with available ACS images. Histograms have been normalized to their maxima in order to clarify the representations.
}
\label{radio_index}
\end{figure}

In order to obtain the physical sizes of our GALEX-selected LBGs we use the previous 8''x8'' I-band ACS cutouts and carry out fits to their radial light curves with \verb+GALFIT+ \citep{Peng2010}. In this step, we consider Sersic profiles \citep{Sersic1968}, which can be described as:

\begin{equation}
\Sigma (r) = \Sigma_e \exp{\left[ -\kappa \left(\left(\frac{r}{R_{\rm eff}}\right)^{1/n}-1\right) \right]}
\end{equation}


\noindent where $\Sigma_e$ is the pixel surface brightness at the effective radius $R_{\rm eff}$ and $n$ is the concentration parameter or Sersic index. The effective radius is the radius which encloses half the light of the galaxy. To make this definition true, the dependent variable $\kappa$ is coupled to $n$ \citep{Peng2010}. For each input galaxy, \verb+GALFIT+ provides its effective radius (in pixels) and Sersic index, along with their uncertainties. In order to convert the effective radius in pixels into the physical size in kpc we employ the ACS pixel scale and the assumed cosmology for calculating the ''/pix at the redshift of each galaxy. Figure \ref{radio_index} shows the distribution of the effective radius and the Sersic index for our GALEX-selected LBGs. The median effective radius for our LBGs is 2.48 kpc. The values of the Sersic indices for LBGs are compatible with most of them being disk-like galaxies. This is in agreement with the results of the visual morphological analysis.

\begin{figure*}
\centering
\includegraphics[width=0.49\textwidth]{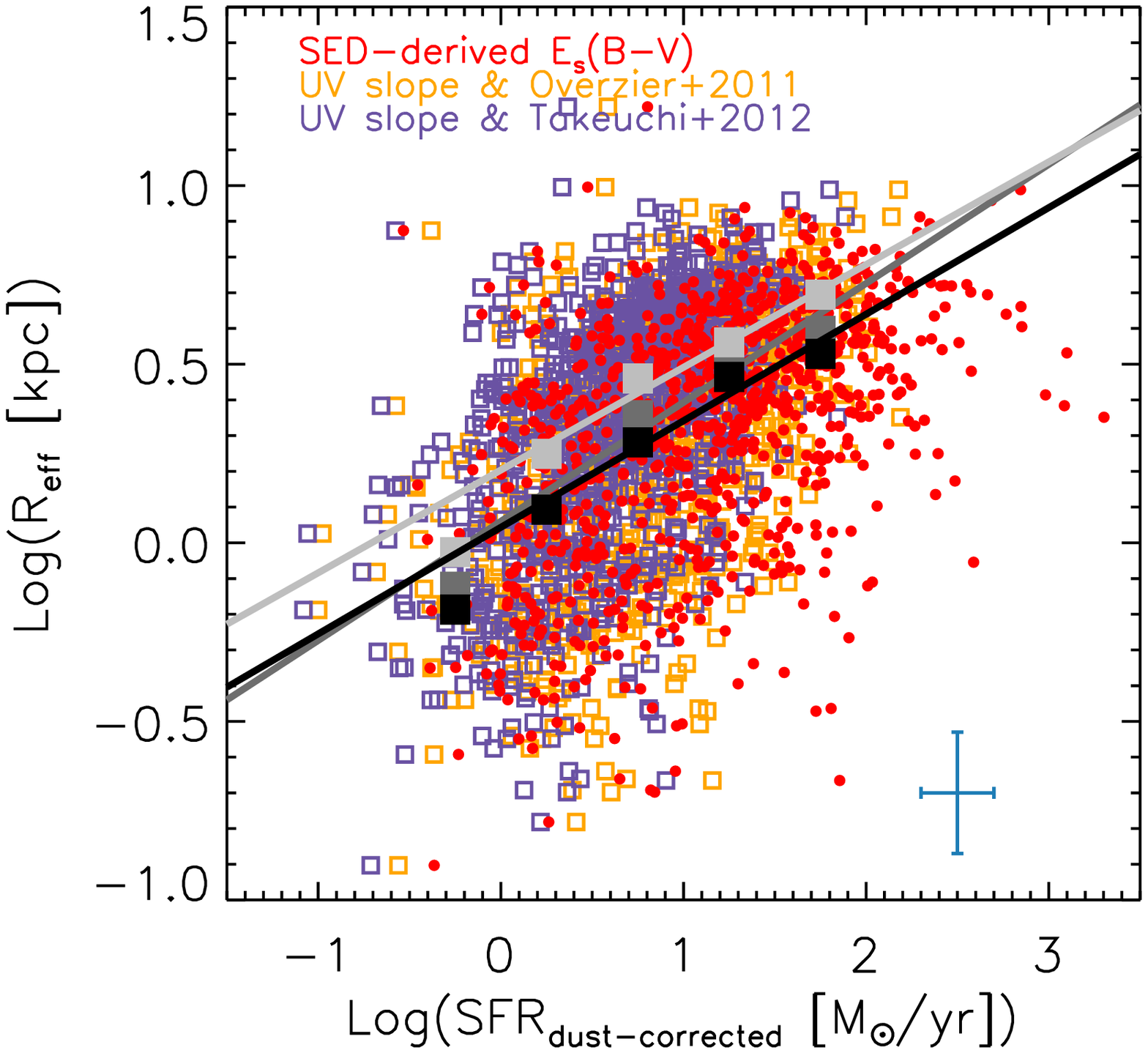}
\includegraphics[width=0.49\textwidth]{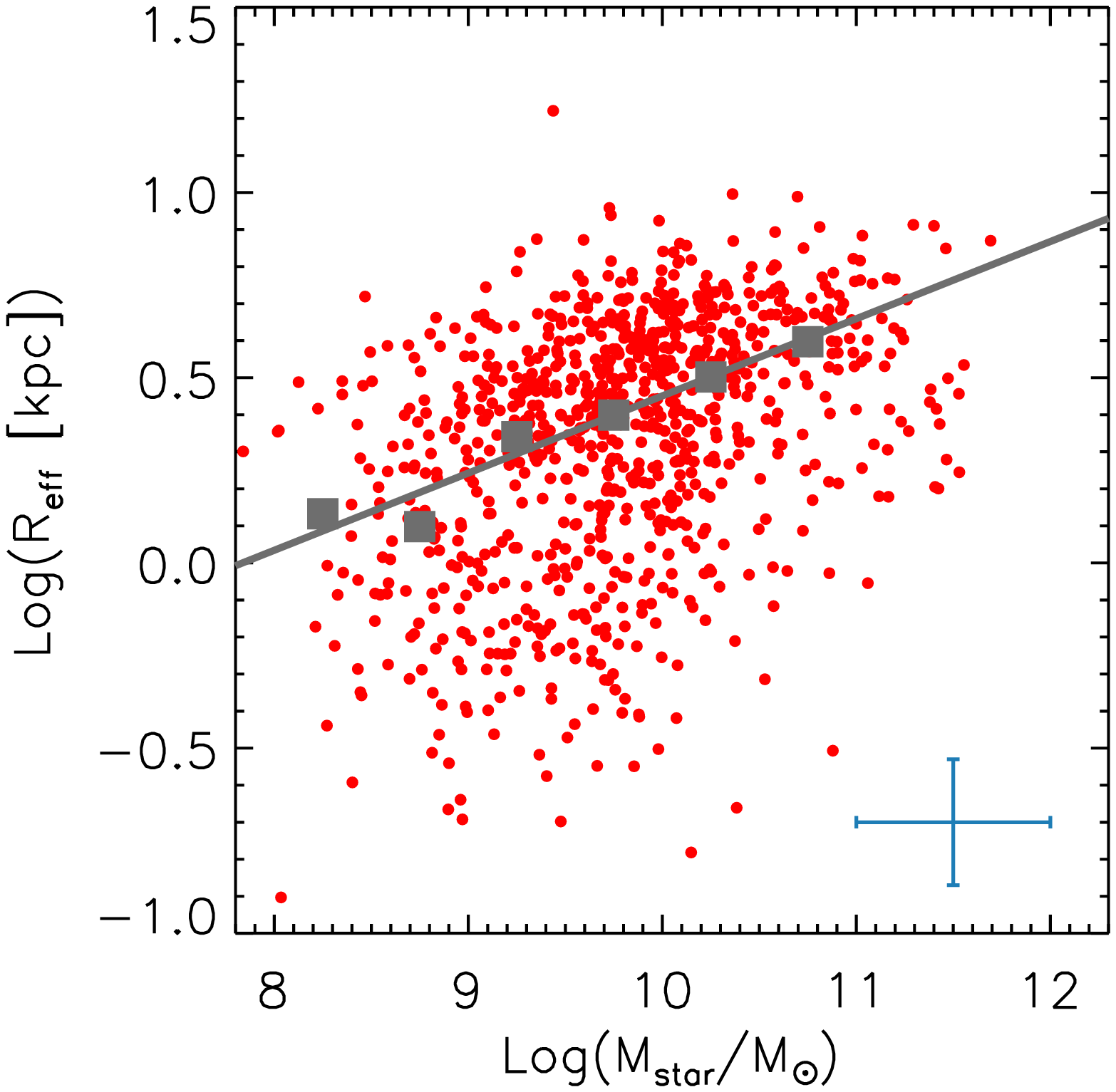}
\caption{\emph{Left}: Effective radius against the dust-corrected total SFR for our GALEX-selected LBGs at $z \sim 1$ with available ACS images. As indicated in the legend, we include the points associated to the total SFR obtained with the SED-derived dust attenuation (red symbols) and with the UV continuum slope and the application of the \citet{Overzier2011} (orange symbols) and \citet{Takeuchi2012} (purple symbols) laws. Black, dark grey, and light grey squares represent the median value of the dust-corrected total SFR obtained with the SED-derived dust attenuation, \citet{Overzier2011}, and \citet{Takeuchi2012} laws, respectively, for each considered bin of effective radius. \emph{Right}: Effective radius against the SED-derived stellar mass for our GALEX-selected LBGs at $z \sim 1$ (red dots) with available ACS images. Grey squares represent the median value of the SED-derived stellar mass for each considered bin of effective radius. Grey straight line represents a linear fits to the grey squares. In both plots, the blue error bars represent the typical uncertainty in the determinations of the SED-derived dust-corrected total SFR and stellar mass (see Section \ref{properties}) and effective radius.
}
\label{sfr_mass_radio}
\end{figure*}

Shown in Figure \ref{sfr_mass_radio} are the relations between the physical sizes of our GALEX-selected LBGs and their dust-corrected total SFR and stellar mass. It can be seen that there is a tendency between the physical size and both SFR and stellar mass: larger galaxies tend to form stars faster, and have higher stellar masses. The correlation between the total SFR and the effective radius exists whatever the dust attenuation method employed. The size-stellar mass relation has also reported to occur in LBGs at a similar and higher redshift ranges. \cite{Mosleh2012} find that the stellar mass-size for LBGs persists up to $z \sim 5$.

\section{Color-magnitude diagram}\label{color_mag}

\begin{figure*}
\centering
\includegraphics[width=0.49\textwidth]{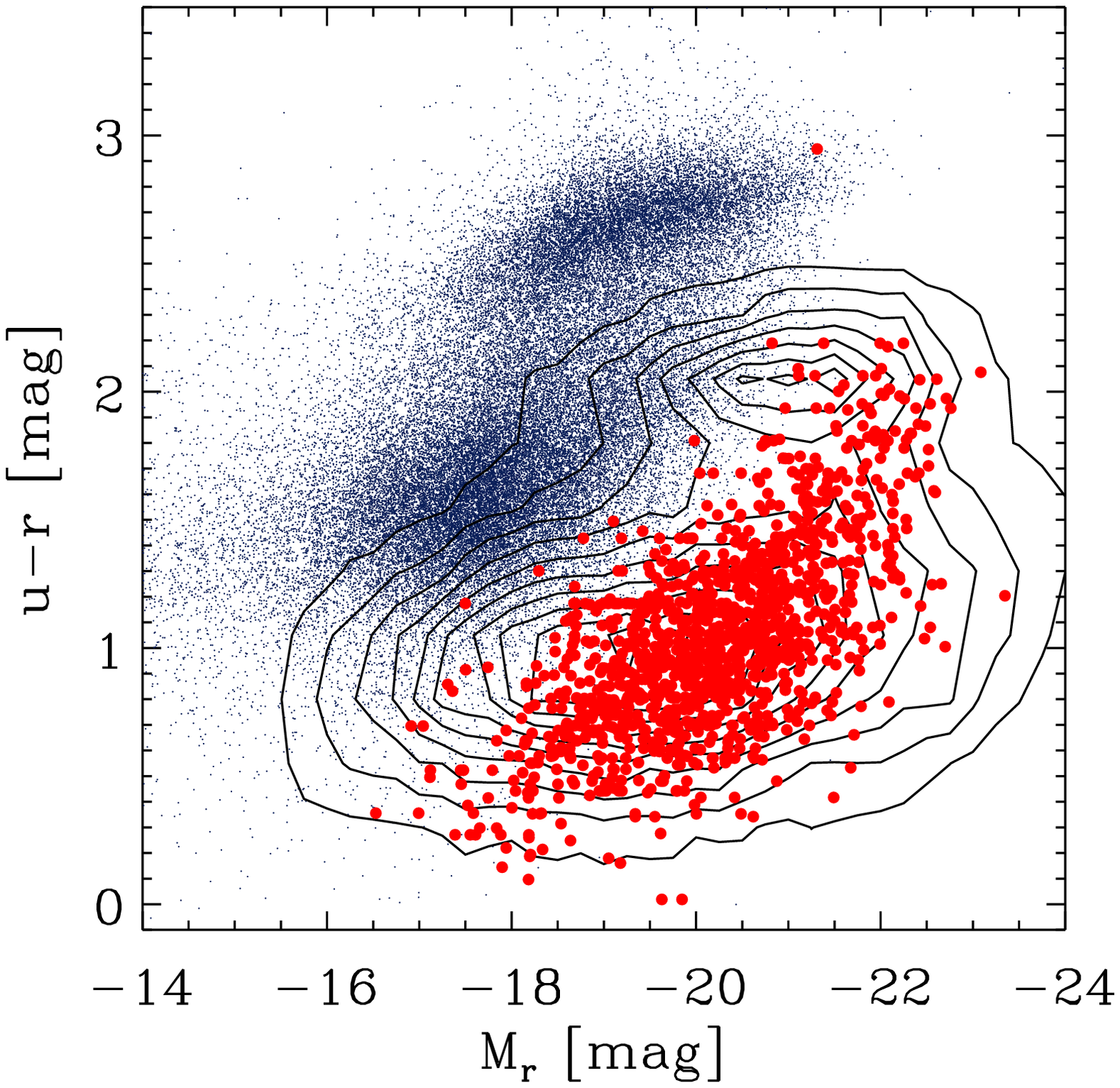}
\includegraphics[width=0.49\textwidth]{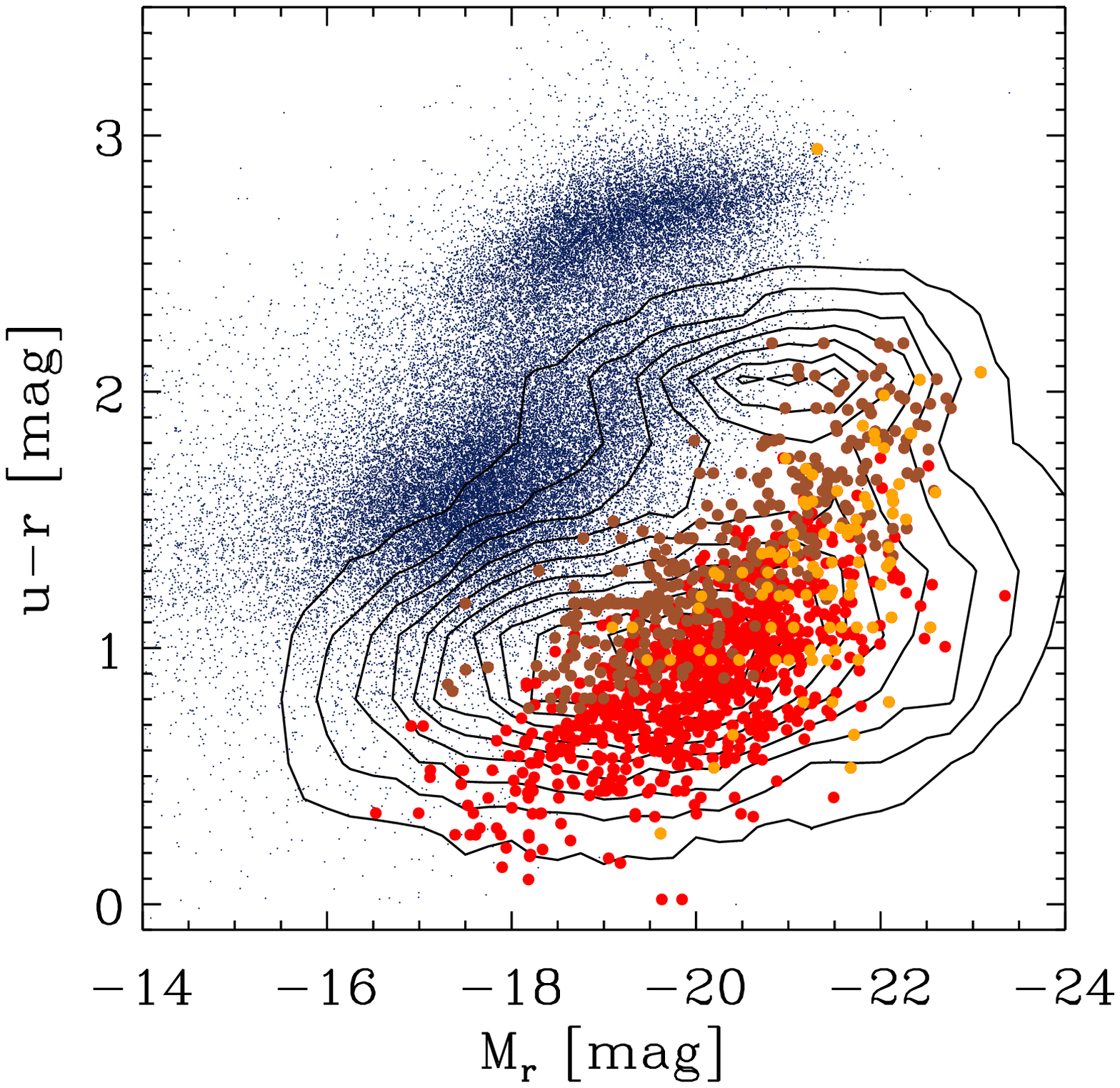}
\caption{\emph{Left:} Locus of our GALEX-selected LBGs (red dots) in a color-magnitude diagram. For comparison, we represent with blue dots the location in such a diagram of a sample of local galaxies selected from SDSS survey. Furthermore, we represent with black contours the typical CMD for galaxies at $z \sim 1$ obtained from a general population of galaxies at that redshift taken from the ALHAMBRA survey. These contours show the location of the blue cloud, green valley, and red sequence at that redshift and help in the discussions shown in the text. \emph{Right:} As in the left panel, but we segregate between old-LBGs (brown) and dusty-LBGs (orange). Old-LBGs are those LBGs whose ages are above 1200 Myr while dusty-LBGs are those LBGs whose dust attenuation are higher than $E_s(B-V)=0.4$.
}
\label{CMD}
\end{figure*}

An important tool to analyze the properties of our GALEX-selected LBGs is their location in a color-magnitude diagram. Traditionally, this kind of diagram has been used to separate local galaxies between non-SF galaxies earlier than the Sa morphological type and SF galaxies later than Sb in morphological type. In a color space, the former tend to populate a red sequence and the latter are located in the so-called blue cloud \citep{Hogg2002,Strateva2001}. This behavior translates into a bimodal distribution of the color of galaxies which allows either to study the nature of different samples of galaxies by looking at their position in the color space and to look for galaxies with different SF nature, by imposing conditions to their location in such a diagram. Furthermore, this bimodality in the local universe has been proven to apply at higher redshifts, at least up to $z \sim 1.6$ \citep{Bell2004,Nicol2011,Williams2009,Franzetti2007,Cirasuolo2007,Taylor2009,Weiner2005,Blanton2003}. Following this idea, we plot in Figure \ref{CMD} the location of our GALEX-selected LBGs in a color-magnitude diagram associated to the magnitudes in the $u$ and $r$ broad-band filters of the SDSS survey. The apparent $u$ and $r$ and absolute $r$ magnitudes are obtained by convolving the best-fitted template of each galaxy with the transmission of the $u$ and $r$ SDSS filters shifted in wavelength according the redshift of each source. Along with those points we also represent a sample of SDSS local galaxies taken from the DR7 \citep{Abazajian2009} and a sample of z-phot-selected galaxies at $z \sim 1$ taken from the ALHAMBRA survey. The sample of local galaxies comprises all the galaxies in the SDSS whose spectroscopic redshifts are below 0.035. In this case, the magnitudes plotted are those that we extract from the photometric catalogs of the SDSS survey. At such a low redshifts there is no need of K-correction. In order to build the sample of galaxies at $z \sim 1$ we select all the galaxies in the ALHAMBRA survey (in all the already observed fields) whose photometric redshifts are around that value and whose observed Ks-band magnitudes are brighter than 22 mag, similar to the limits employed in \cite{Williams2009} or \cite{Taylor2009}. By using the optical and near-IR photometry of the ALHAMBRA survey we fit their SEDs with BC03 templates and obtain their $u-r$ colors, and absolute r-band magnitude in the same way than for LBGs. It can be seen in the left panel of Figure \ref{CMD} that the bimodality which is seen in the local universe is also clearly present at $z \sim 1$. This result also indicates the power of the ALHAMBRA survey in characterizing the CMD of galaxies at different redshifts. 

The majority of our GALEX-selected LBGs are located in the blue-cloud of galaxies at their redshift, indicating that these kinds of galaxies are blue and active SF galaxies, as expected from their selection criteria in the UV. A minority of LBGs and are shifted toward the red sequence or are located between the blue cloud and the red sequence, the so-called green valley. This position does not necessarily indicates that these galaxies are non-SF. Actually, it can also occur that these galaxies have redder optical colors because of either a significant amount of dust which is attenuating their bluest emitted light, and/or there is an important contribution of old stellar populations in their SEDs, being more evolved systems. To clarify this issue we show in the right panel of Figure \ref{CMD} the position of our GALEX-selected LBGs in the CMD as a function of SED-derived dust attenuation and age. We arbitrarily consider two subclasses within the LBGs: Those with age $\gtrsim$ 1200 Myr (old-LBGs) and those with age $\lesssim$ 1200 Myr and $E_s(B-V) \gtrsim 0.4$ (dusty-LBGs). It can be clearly derived from Figure \ref{CMD} that those LBGs which are located over the green valley or near the red sequence are old-LBGs and dusty-LBGs, whereas those LBGs whose ages are bellow 1200 Myr and have low/intermediate ($E_s(B-V) < 0.4$) dust attenuation are located over the blue cloud.

\section{Comparison to high-redshift LBGs}\label{high_z}

In this section we analyze the differences/similarities between LBGs at $z \sim 1$ and $z \sim 3$ in order to study whether the Lyman break selection criterion selects different kind of galaxies at different redshifts. As commented in Section \ref{huh}, high-redshift LBGs tend to be intrinsically brighter than those studied in this work as a consequence of the usage of magnitude-limited observations. If we want to compare galaxies which are selected at different redshift with a similar selection criterion we must limit the rest-frame UV luminosities of the samples to the same range. We use the rest-frame UV luminosities since LBGs at any redshift are selected in the rest-frame UV. This sample was defined in Section \ref{huh} as UV-bright LBGs and it is formed by 65 galaxies. According to their SEDs, UV-bright LBGs are less dusty, have higher SFR, are more massive, and have bluer UV continuum slope than the whole population of LBGs at $z \sim 1$. However, there is no significant difference in their ages, which are mostly below 500 Myr for both populations.

\begin{figure}
\centering
\includegraphics[width=0.23\textwidth]{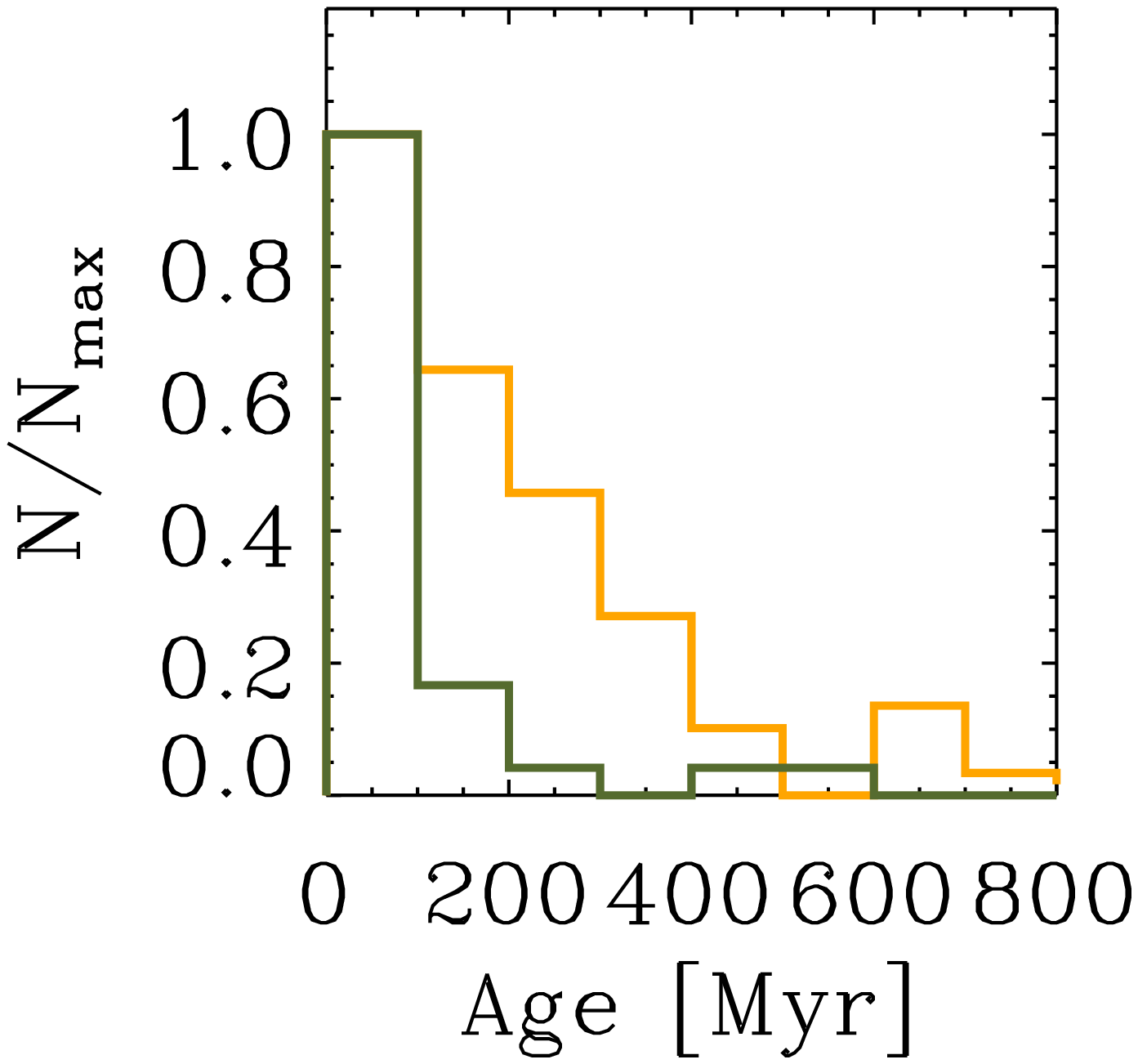}\\
\includegraphics[width=0.23\textwidth]{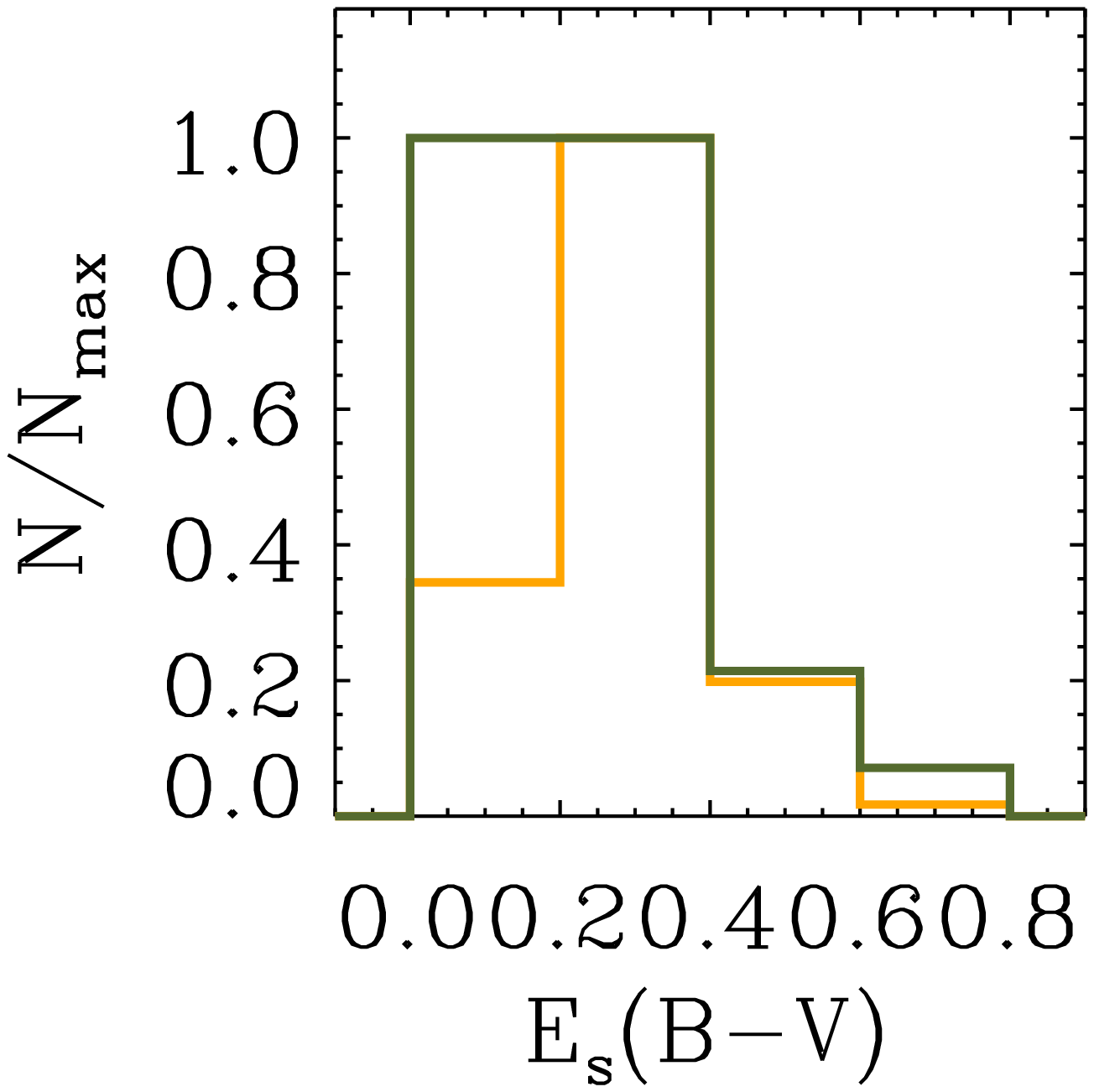}
\includegraphics[width=0.23\textwidth]{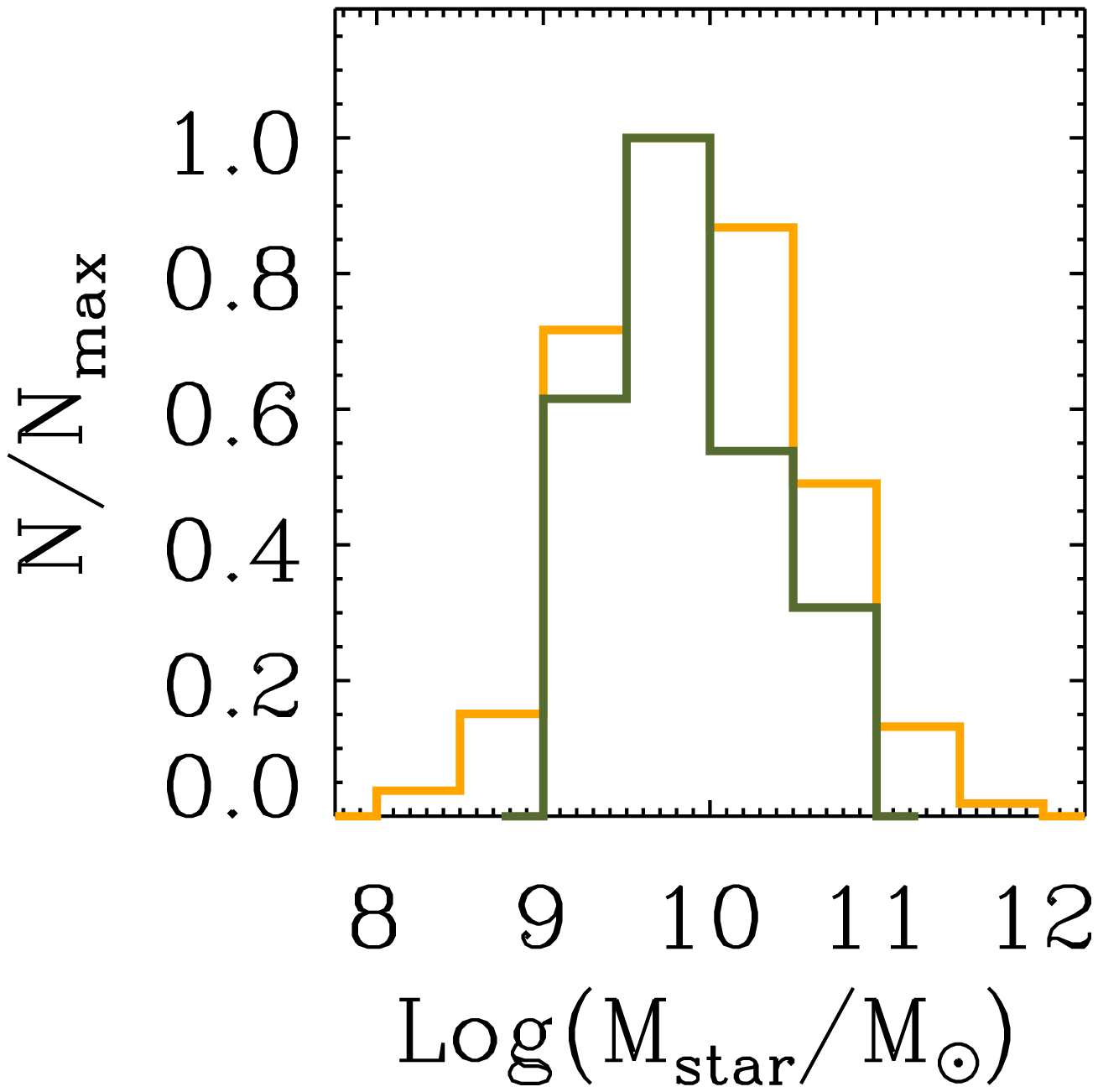}
\caption{Distribution of the SED-derived age, dust attenuation, and stellar mass for our UV-bright LBGs (orange histograms) and high-redshift LBGs (green histograms) taken from \citet{Papovich2001}. Histograms have been normalized to their maxima in order to clarify the representations.
}
\label{highz}
\end{figure}

Figure \ref{highz} shows the distribution of the SED-derived age, dust attenuation, and stellar mass for our UV-bright LBGs (orange histograms) and for a sample of LBGs at $z \sim 3$ studied in \cite{Papovich2001} (green histograms). UV-bright LBGs at $z \sim 1$ have ages mainly distributed between 1 and 400 Myr with a median value of 171 Myr, whereas LBGs at $z \sim 3$ are younger galaxies with a median age of 36 Myr. The difference in the median values is of the same order than the uncertainties of the SED-derived age at $z \sim 1$. However, as it can be seen in the histogram shown in Figure \ref{highz}, at $z \sim 1$ there is a presence of older stellar populations (with ages mainly between 150 Myr and 400 Myr) than at $z \sim 3$. A KS test applied to both histograms gives a very low probability that they represent similar distributions. This could indicate that the galaxies selected with the Lyman break criterion at $z \sim 1$ are in a later evolutionary stage than those at $z \sim 3$. It should be noted that, as pointed out in many previous works and in Section \ref{stellar}, the uncertainties of the SED-derived stellar age are usually high and, additionally, the age evolution found can be the consequence of diverse factors: a) using different SFHs in the analysis of the SED of the galaxies, b) employing photometric information with different wavelength coverage, c) the degeneracy between dust attenuation, metallicity, and age, etc. Therefore, the previous evidence of an evolution of the age of LBGs with redshift are not conclusive and should be confirmed in further studies with a more detailed study of the rest-frame UV-to-near-IR SED of these galaxies.

The dust attenuation distribution of LBGs at $z \sim 1$ seems to contain lower values than the distribution at high redshift, although both have median values of $E_s(B-V) = 0.25$. The typical uncertainty of the SED-derived dust attenuation in our work is $\Delta E_s(B-V) = 0.1$ (see Section \ref{stellar}). This value along with the similar median values of the distributions at $z \sim 1$ and $z \sim 3$ prevent us to constraining any evolution of the median values of dust attenuation of LBGs with redshifts. This can be an effect of the procedure employed. An SED-fitting technique is not precise enough to constrain an evolution of dust attenuation with redshift and others techniques should be employed. Direct measurements of dust emission of LBGs in the FIR could give clues for addressing this issue.

Regarding stellar masses, it can be seen that the distributions at $z \sim 1$ and $z \sim 3$ span within a similar range. The median values of the stellar mass of our UV-bright LBGs and LBGs at $z \sim 3$ are $\log{\left( M_*/M_\odot \right)}=10.0$ and $\log{\left( M_*/M_\odot \right)}=9.7$, respectively. This difference is similar to the typical uncertainties of the stellar mass determinations done in this work and, therefore, we cannot constrain any evolution in the stellar mass of LBGs with redshift either. The median value of the stellar mass found in the present work is between to those reported in \cite{Magdis2010_IRAC} for IRAC-8$\mu$m detected and IRAC-8$\mu$m faint LBGs at $z \sim 3$, $\langle\log{M_*/M_\odot}\rangle$=11, and $\langle\log{M_*/M_\odot}\rangle$=9, respectively. 

Regarding the UV continuum slope, UV-bright LBGs have a median value of $\beta = -1.44$. This value is larger (redder) than those reported at higher redshifts \citep{Lehnert2003,Bouwens2006,Hathi2008,Bouwens2009,Wilkins2011,Bouwens2012,Castellano2012}, indicating that LBGs at lower redshifts tend to be redder in the UV continuum than those at higher redshifts. The UV continuum slope is not a parameter associated directly to the BC03 templates, but it is obtained once the best-fitted BC03 template for each galaxy is known. Consequently, this parameter is more insensitive to the different SFH adopted for building the BC03 template and, therefore, is a good and accurate indicator of the evolution of LBG with redshift.

\cite{Mosleh2011} study the redshift evolution of the physical sizes of samples of LBGs and other UV/sub-mm-selected galaxies at different redshifts and find that their size increases with decreasing redshift. \cite{Mosleh2012} study the redshift evolution of LBGs from $z \sim 1$ up to $z \sim 7$ and find that the median size of LBGs at a given stellar mass increases toward lower redshifts. The median size of our sample of UV-bright LBGs is 2.92 kpc. \cite{Mosleh2012} study the redshift evolution of LBGs considering galaxies in two bins of stellar mass, $8.6 < \log{\left( M_*/M_\odot \right)} < 9.5$ and $9.5 < \log{\left( M_*/M_\odot \right)} < 10.4$. As indicated before, the median value of the stellar mass of our UV-bright LBGs is $\log{\left( M_*/M_\odot \right)}=10.0$. The median value of the size of UV-bright LBGs is slightly lower than that presented in \cite{Mosleh2012} and \cite{Mosleh2011} for the corresponding stellar mass range. This small difference is likely due that, although they also work with GALEX-selected LBGs, they consider galaxies located within $0.6 < z < 1.4$, whereas we limit the redshift of our sample to $z > 0.8$. The inclusion of galaxies at lower redshifts might increase the median value of the size, explaining the difference found between the two works. 

It is important to remark again that there is a fundamental difference in the selection criteria of LBGs at different redshifts. Given that the combination of GALEX and ALHAMBRA (including IRAC for a subsample) provides very accurate photometric redshifts of our galaxies, we do not need any extra condition to rule out interlopers. However, this is not the case at high redshifts. At $z \gtrsim 2$ the photometric redshifts of the sources are not accurate enough to ensure a proper cleaning of the sample from interlopers. As a consequence, additional criteria should be applied. These extra criteria usually involve limits in the observed optical colors of the samples. For example, as it is discussed in \cite{Madau1996} for their sample of F300W dropouts, the application of extra criteria rules out interlopers which are at another redshift range, but they also missed galaxies at the proper redshift at the same time. The missing galaxies tend to be redder than those included in the final sample, either because they are older or are attenuated by dust. Therefore, it is clear than the additional criteria employed at high redshift get rid of certain kinds of subclasses of galaxies at each redshift. In contrast, in our work, as in \cite{Burgarella2006} and \cite{Burgarella2007}, we do not apply any extra observed optical color criteria and therefore we include in the sample all kind of galaxies which have a break between FUV and NUV filters, regardless their age, dust attenuation or optical colors. 

All previous results seem to indicate that LBGs at $z \sim 1$ tend to be older, bluer in their UV continuum, and have larger sizes than those at higher redshifts. Therefore, SED fitting and morphological studies indicate that LBGs at lower redshifts are in a later evolutionary stage than those at higher redshifts. It should be considered that these differences between LBGs at different redshifts have been found by comparing the results obtained in this work with other from previous studies done by other authors. Therefore, this comparison might suffer from uncertainties coming from the different methods employed in each work or slightly different selection criteria. For example, the use of BC03 templates associated to different SFHs or different photometric coverage of the SED of galaxies at different redshifts. Then, in order to properly characterize the evolution of LBGs with redshift a more precise work should be done where the photometric SEDs of the galaxies are sampled over the same rest-frame wavelength range and the selection criteria and the procedures employed for the analysis of the physical properties of the galaxies are as similar as possible.

\section{Conclusions}\label{conclu}

In this work we have analyzed the SED-fitting derived physical properties of a sample of 1225 GALEX-selected LBGs at $0.8 \lesssim z \lesssim 1.2$ by using a combination of UV and optical/near-IR data coming from GALEX observations and the ALHAMBRA survey, respectively. ALHAMBRA uses a set of 20 medium-band (width $\sim 300\AA$) optical and the three classical near-IR JHKs filters for covering the observed optical SED of galaxies in an unprecedented way. This provides a good sampling of both the UV continuum slope and the Balmer break, increasing significantly the accuracy of the results of the SED-fitting technique. We defined LBGs as those galaxies that have a difference of color greater than 1.5 magnitudes between the FUV and NUV filters of the GALEX satellite. Our main conclusions are as follows:

\begin{enumerate}

\item According to SED fitting with BC03 templates built assuming a constant SFR, Salpeter IMF, and metallicity $Z=0.2Z_\odot$, GALEX-selected LBGs at $z \sim 1$ are young galaxies with ages mostly below 300 Myr, with median dust attenuation of $E_s(B-V)$ of 0.20, and a median UV continuum slope of -1.53. As a consequence of the selection criteria used they are UV-bright objects with UV-uncorrected SFR of about 2.0 M$_\odot$/yr. When dust-correcting their rest-frame UV luminosity, their total SFR turns out to have a median value of 46.4 M$_\odot$/yr. Combining the total SFRs and ages, we find that GALEX-selected LBGs have a median stellar mass of $\log{(M_*/M_\odot)}=9.74$. Only 2\% of the galaxies selection with the Lyman break selection criterion have an AGN according to their X-ray emission.


\item LBGs with higher stellar masses have higher total SFRs and lower values of the specific SFR. The anti-correlation between the specific SFR and stellar mass supports the downsizing scenario, where more massive galaxies have formed their stars earlier and faster than galaxies with lower stellar mass. 

\item Morphologically, LBGs at $z \sim 1$ are mostly disk-like galaxies (about 69\%), while the remaining are interacting, compact or irregular systems in much lower percentages. This is confirmed by their Sersic indices, which are typically below 0.5. The median effective radius for our GALEX-selected LBGs at $z \sim 1$ is 2.48 kpc. Bigger galaxies tend to have higher total SFR and stellar masses.

\item In a color-magnitude diagram, most GALEX-selected LBGs are located over the blue cloud at their redshift, which indicates that they are active SF galaxies. Some LBGs are located over the green valley or near the red sequence. They turn out to be the dustiest and/or oldest galaxies in the sample, as signs of more evolved systems.

\item Comparing with their high-redshift analogs, we find that the galaxies selected through the Lyman break criterion at $z \sim 1$ seem to be in a later evolutionary stage than those at high-redshift. However, the uncertainties in the SED-derived age are typically significant and, consequently, the age evolution should be confirmed with a more detailed study of the rest-frame UV-to-near-IR SEDs of LBGs at different redshifts. We do not find any significant difference in the distributions of stellar mass or dust attenuation for LBGs at high and intermediate redshift. LBGs at lower redshifts are bigger, have more contribution of older stellar population to their SEDs, and are redder in their UV continuum than their high-redshift analogues. 

\end{enumerate}

\section*{Acknowledgments}

The authors would like to thank the referee for the careful reading of the manuscripts and for a valuable feedback that has improved the presentation of our results. I. Oteo would also like to thank Professor Tsutomu T. Takeuchi for kindly providing useful comments. This research has been supported by the Spanish Ministerio de Econom'a y Competitividad (MINECO) under the grant AYA2011-29517-C03-01. Some/all of the data presented in this paper were obtained from the Multimission Archive at the Space Telescope Science Institute (MAST). STScI is operated by the Association of Universities for Research in Astronomy, Inc., under NASA contract NAS5-26555. Support for MAST for non-HST data is provided by the NASA Office of Space Science via grant NNX09AF08G and by other grants and contracts. Based on observations made with the European Southern Observatory telescopes obtained from the ESO/ST-ECF Science Archive Facility. Based on zCOSMOS observations carried out using the Very Large Telescope at the ESO Paranal Observatory under Programme ID: LP175.A-0839. 

Funding for the SDSS and SDSS-II has been provided by the Alfred P. Sloan Foundation, the Participating Institutions, the National Science Foundation, the U.S. Department of Energy, the National Aeronautics and Space Administration, the Japanese Monbukagakusho, the Max Planck Society, and the Higher Education Funding Council for England. The SDSS Web Site is http://www.sdss.org/.

The SDSS is managed by the Astrophysical Research Consortium for the Participating Institutions. The Participating Institutions are the American Museum of Natural History, Astrophysical Institute Potsdam, University of Basel, University of Cambridge, Case Western Reserve University, University of Chicago, Drexel University, Fermilab, the Institute for Advanced Study, the Japan Participation Group, Johns Hopkins University, the Joint Institute for Nuclear Astrophysics, the Kavli Institute for Particle Astrophysics and Cosmology, the Korean Scientist Group, the Chinese Academy of Sciences (LAMOST), Los Alamos National Laboratory, the Max-Planck-Institute for Astronomy (MPIA), the Max-Planck-Institute for Astrophysics (MPA), New Mexico State University, Ohio State University, University of Pittsburgh, University of Portsmouth, Princeton University, the United States Naval Observatory, and the University of Washington. Financial support from the Spanish grant AYA2010-15169 and from the Junta de Andalucia through TIC-114 and the Excellence Project P08-TIC-03531 is acknowledged.

\bibliographystyle{mn2e}
\bibliography{ioteo_biblio}

\bsp

\label{lastpage}

\end{document}